\pdfoutput=1
\documentclass[
nofootinbib,
superscriptaddress,amsfonts,amssymb,amsmath, twocolumn,
tightenlines
]{revtex4-2}

\allowdisplaybreaks
\usepackage{tabularx}
\usepackage{graphicx}

\usepackage{sastyle} 
\graphicspath{ {images/} }
\usepackage{slashed}
\usepackage{tablefootnote}

\newcommand{\beq}{\begin{equation}}
\newcommand{\eeq}{\end{equation}}

\newcommand{\qed}{QED\(_3\)\,}
\newcommand{\Nf}{\(N_f\)\,}

\def\note[#1]#2\noteend{%
{\bfseries \textcolor{orange}{#1}: \textcolor{blue}{#2}}
}
\def\ub{\widebar{u}}
\newcommand{\psb}{\widebar\psi}
\usepackage{makecell}

\renewcommand{\<}{\langle} 
\renewcommand{\>}{\rangle} 
\newcommand{\colvec}[1]{
	\left(\begin{array}{c}
	#1
	\end{array}
	\right)
}
\newcommand{\pr}[1]{
	\left(#1\right)
}
\renewcommand{\d}{\partial}

\newcommand{\CO}{\mathcal{O}}
\newcommand{\CM}{\mathcal{M}}

\definecolor{pearlypurple}{RGB}{183, 104, 162}

\renewcommand{\geq}{\geqslant}
\renewcommand{\leq}{\leqslant}

\begin{document}
\title{Bootstrapping $N_f=4$ conformal QED$_3$}
\author{Soner Albayrak}
\affiliation{Department of Physics, Yale University, New Haven, CT 06511, USA}
\affiliation{Institute of Physics, University of Amsterdam, Amsterdam, 1098 XH, The Netherlands}
\author{Rajeev S. Erramilli}
\affiliation{Department of Physics, Yale University, New Haven, CT 06511, USA}
\author{Zhijin Li}
\affiliation{Department of Physics, Yale University, New Haven, CT 06511, USA}
\author{David Poland}
\affiliation{Department of Physics, Yale University, New Haven, CT 06511, USA}
\author{Yuan Xin}
\affiliation{Department of Physics, Yale University, New Haven, CT 06511, USA}

	\begin{abstract}
We present the results of a conformal bootstrap study of the presumed unitary IR fixed point of quantum electrodynamics in three dimensions (QED$_3$) coupled to $N_f=4$ two-component Dirac fermions. Specifically, we study the four-point correlators of the $SU(4)$ adjoint fermion bilinear $r$ and the monopole of lowest topological charge $\mathcal{M}_{1/2}$. Most notably, the scaling dimensions of the fermion bilinear $r$ and the monopole $\mathcal{M}_{1/2}$ are found to be constrained into a closed island with a combination of spectrum assumptions inspired by the $1/N_f$ perturbative results as well as a novel interval positivity constraint on the next-lowest-charge monopole $\mathcal{M}_1$. Bounds in this island on the $SU(4)$ and topological $U(1)_t$ conserved current central charges $c_J$, $c_J^t$, as well as on the stress tensor central charge $c_T$, are comfortably consistent with the perturbative results. Together with the scaling dimensions, this suggests that a part of estimates from the $1/N_f$ expansion --- even at $N_f=4$ --- provide a self-consistent solution to the bootstrap crossing relations, despite some of our assumptions not being strictly justified.
	\end{abstract}
\date{\today}
\maketitle
	
	
	\tableofcontents
	
	
\section{Introduction}

Quantum electrodynamics in three dimensions (QED$_3$) has been extensively studied over the past decades, partially motivated by qualitative similarities with four dimensional quantum chromodynamics. The gauge coupling in QED$_3$ has positive mass dimension, and so the theory is asymptotically free and strongly coupled in the infrared (IR) limit. The IR phase of QED$_3$ depends on the number of electrons $N_f$.\footnote{In this work, the flavor number refers to $N_f$ two-component Dirac fermions, and we will assume $N_f$ is even to avoid the parity anomaly.}
In the large $N_f$ limit, the theory can be solved using a $1/N_f$ expansion, which suggests a renormalization group flow to an IR fixed point \cite{Appelquist:1988sr, Nash:1989xx}. The pure $U(1)$ gauge theory with $N_f=0$ is expected to confine in the IR due to a proliferation of monopoles \cite{Polyakov:1975rs, Polyakov:1976fu}.  Schwinger-Dyson equation analysis \cite{Pisarski, Appelquist:1986fd} and some lattice simulations \cite{Dagotto:1989td} suggest that at small $N_f$, the IR phase of QED$_3$ has its chiral symmetry spontaneously broken ($\chi$SB),
\begin{equation*}
SU(N_f) \rightarrow SU(N_f/2)\times SU(N_f/2)\times U(1) \,,
\end{equation*}
due to the dynamical generation of a fermion mass. It is expected that there is a critical flavor number $N_f^*$ which separates the conformal phase from the $\chi$SB phase.  

QED\(_3\) also has various fundamental applications in condensed matter physics. In particular, $N_f=4$ QED$_3$ has been utilized to describe high-temperature superconductors, or more generally Dirac spin liquids \cite{Rantner:2000wer, Rantner:2002zz, Herbut:2002yq, Franz:2002qy, Hermele_2005,Wen2008}. $N_f=2$ QED$_3$ has been proposed to be part of the 3d fermion-boson duality web and is an effective theory for the deconfined quantum critical point, see \cite{Senthil_2019} for a comprehensive review.

A crucial unanswered question in these studies is the value of the critical flavor number $N_f^*$ of QED$_3$.
Various approaches have been used to estimate $N_f^*$ \cite{Appelquist:1988sr, Nash:1989xx, Maris:1996zg, Aitchison:1997ua, Appelquist:1999hr, Kubota:2001kk,  Appelquist:2004ib, Franz_2003, Fischer:2004nq, Kotikov:2016wrb,  Kaveh_2005, Giombi:2015haa, DiPietro:2015taa, Giombi:2016fct, DiPietro:2017kcd, Zerf:2018csr, Herbut:2016ide, Gusynin:2016som, Benvenuti:2018cwd, Christofi:2007ye, Janssen:2012pq, Hands:2020itv, Braun:2014wja, Gukov:2016tnp, Hands:2002qt, Hands:2002dv, Hands:2004bh, Strouthos:2008kc, Karthik:2015sgq, Karthik:2016ppr,Karthik:2019mrr, Karthik:2020shl, Li:2021emd};\footnote{Some of these studies focused on QED$_3$ with a noncompact gauge group $\R$, in which the monopole contributions have been suppressed. At small $N_f$ its low energy dynamics may be different from compact QED$_3$.} however, there is no general consensus to what the actual value should be. Estimates range from $0$ all the way up to $10$.\footnote{See \cite{Gukov:2016tnp} for more details on this discrepancy.}
The problem is made worse by the fact that the theory is actually strongly coupled near $N_f^*$, rendering the estimations of perturbative approaches unreliable.
Lattice simulations do offer a nonperturbative approach, but their results remain inconsistent between each other. In particular, some lattice simulations indicate that there is no $\chi$SB for any $N_f>0$ \cite{Karthik:2015sgq, Karthik:2016ppr,Karthik:2019mrr, Karthik:2020shl} and that the low energy limit of QED$_3$ coupled with massless fermions is always conformal. This assertion stands in contrast to other lattice results which observed $\chi$SB at $N_f=2$ and a conformal phase at $N_f\geqslant4$ \cite{Hands:2002qt, Hands:2002dv, Hands:2004bh, Strouthos:2008kc}. A subtle issue in the study of  \(N_f=2\) QED$_3$ by lattice simulations is the violation of conformality by a small non-unitary factor, as this could not be distinguished from the standard conformal phase due to the finiteness of practical lattice simulations. For instance, \cite{Karthik:2019mrr} measures the scaling dimension of the monopole with lowest unit of topological charge in $N_f=2$ QED$_3$, and according to the bootstrap result \cite{Li:2021emd}, their data requires a weakly relevant singlet scalar, indicating that the theory is slightly below the conformal window in the so-called merger and annihilation scenario for the loss of conformality in QED$_3$ \cite{Kubota:2001kk, Kaveh_2005, Gies:2005as, Kaplan:2009kr, Giombi:2015haa, Gorbenko:2018ncu}.\footnote{
In the merger and annihilation scenario for the loss of conformality in QED$_3$, we expect the fixed point of QED$_3$ to merge with another fixed point as we continuously vary $N_f$ down to $N_f^*$ from above, and these points annihilate one another below $N_f^*$. A candidate theory for the other fixed point is the so-called \emph{QED Gross Neveu Yukawa} (QED$_3$-GNY) fixed point.
}
 
The modern conformal bootstrap \cite{Rattazzi:2008pe, Poland:2018epd} provides a powerful non-perturbative approach to study conformal QED$_3$, free of the subtleties of the perturbative and lattice computations, and poised to be able to answer puzzles such as the value of the critical \(N_f^*\). 
Bootstrap studies of \qed have been initiated in \cite{Chester:2016wrc,Chester:2017vdh}  by focusing on the monopole operators in QED$_3$. In 3d, $U(1)$ gauge theories have a unique property of admitting a topological symmetry $U(1)_t$, whose non-trivial representations are constructed by the monopole operators.  From bootstrap point of view, the power of the monopole operators is that they let us distinguish QED\(_3\) from e.g. the $SU(N_c)$ QCD\(_3\). Moreover,
monopole operators are known to play important roles in QED$_3$ with small $N_f$. For instance, in $N_f=2$ QED$_3$ which is a part of the 3d boson-fermion duality web  \cite{Senthil_2019}, the monopoles provide dual descriptions of the gauge invariant composite operators made from elementary fermions.
In \cite{Chester:2016wrc}, the authors obtained bootstrap bounds on the scaling dimensions of the leading charge \(q=1/2\) and \(q=1\) monopoles close to saturation, but these bounds were quite sensitive to the gap assumptions, especially to what the authors refer to \(\Delta_2\) (which we will refer to as \(\Delta_{S_{(220)}}\)), which will also play an important role in our study.

Other encouraging results towards bootstrapping conformal QED$_3$ have been obtained by bootstrapping $SU(N_f)$ adjoint fermion bilinear scalars \cite{Li:2018lyb}; these operators are the leading gauge-invariant operators with a nontrivial $SU(4)$ representation, and therefore can give us a view into the flavor symmetry of this theory. The study \cite{Li:2018lyb} found bootstrap bounds with sharp kinks for $N_f>2$: for large $N_f$, the location of the kink approaches free fermion theory; for large but finite $N_f$, the location is close to the perturbative predictions of conformal QED$_3$; and for sufficiently small \(N_f\) the kink disappears, implying some critical $N_f^*\in(2,3)$.
 The lowest singlet operator approaches marginality condition near $N_f^*$, consistent with the merger and annihilation mechanism \cite{Kubota:2001kk, Kaveh_2005, Gies:2005as, Kaplan:2009kr, Giombi:2015haa, Gorbenko:2018ncu} for the loss of conformality in QED$_3$.
However, it has been proved in \cite{Li:2020bnb} that the kinks in the singlet bounds are wholly $SO(N)$ symmetric and can not literally be identified with conformal QED$_3$, while they may correspond to the conformal QED$_3$ through $SO(N)$ symmetry enhancement in the bootstrap bounds \cite{Li:2020bnb, Li:2020tsm}. 
Another set of studies focused on the $SU(N)$ adjoint bilinears in scalar \qed~\cite{He:2021xvg,Manenti:2021elk}, with similarly promising results (including isolated regions at large $N_f$ or in $d = 2 + \epsilon$ which may contain the scalar \qed solution~\cite{He:2021xvg}).

A natural next step would be to bootstrap crossing equations of both the monopoles and the $SU(4)$ adjoint fermion bilinears; this was recently pursued in \cite{He:2021sto}. The authors make assumptions inspired by the constraints of lattice implementations, based on which they obtain lower bounds on the dimension of the leading monopole \(\Delta_{\cM_{1/2}}\) in order to reach the IR fixed point of $N_f=4$ \qed on a triangular lattice \(\Delta_{\cM_{1/2}}>1.046\) or kagome lattice \(\Delta_{\cM_{1/2}}>1.105\). The bounds are consistent with recent Monte Carlo estimates \cite{Karthik:2015sgq,Xu:2018wyg,Karthik:2019mrr} but they exclude the large \(N_f\) expansion prediction \(\Delta_{\cM_{1/2}}\approx 1.022\). 

In this work we will provide a more comprehensive bootstrap study for $N_f=4$ conformal QED$_3$. An important element of our analysis is that the crossing equations of single correlators with the $SU(4)$-adjoint fermion bilinear operator $r$ and the monopole operator with lowest unit of topological charge $\cM_{1/2}$ have enhanced $SO(15)$ and $SO(12)$ symmetry, respectively. A direct consequence of the $SO(N)$ symmetry enhancement of the crossing equations is that suitable gap assumptions are necessary to obtain bootstrap results for non-$SO(N)$ symmetric theories, e.g., conformal QED$_3$. We will use the fermion bilinear bootstrap to demonstrate the gap-dependence of the bootstrap bounds, and show that interesting results for $N_f=4$ conformal QED$_3$ can be obtained after introducing gap assumptions inspired by the perturbative results. 
Our most interesting results are obtained from the monopole bootstrap, presented in \secref{\ref{islands}}, in which the scaling dimensions of operators $r$ and $\cM_{1/2}$ are restricted into a closed island after introducing an {\it interval positivity} assumption, along with some input about gaps in the monopole spectrum. Parity symmetry also plays a critical role in generating the monopole bootstrap results. Our bootstrap results suggest that part of the perturbative CFT data of $N_f=4$ conformal QED$_3$ provides a consistent solutions to the crossing equations.  

The paper is organized as follows. In Section \ref{sec:perturbative} we briefly review the perturbative results on conformal QED$_3$, which provide useful guides for our bootstrap studies. In Section \ref{sec: Fermion bilinear scalar bootstrap} we explain the gap-dependence of the $SU(4)$ adjoint bootstrap bounds caused by the $SO(15)$ symmetry enhancement in the crossing equations and show that interesting results can be obtained after introducing gap assumptions inspired by the perturbative results. In Section \ref{sec:monopole} we revisit the monopole bootstrap. We explain the $SO(12)$ symmetry enhancement in the crossing equations and show that parity symmetry can help to restrict the CFT data in a closed region consistent with the results from $1/N_f$ expansions.
In Section \ref{sec:mixed} we study the mixed correlator bootstrap with fermion bilinear operator $r$ and the monopole $\cM_{1/2}$. Conclusions and discussions are given in Section \ref{sec:conclusion}. Technical details related to the bootstrap studies are provided in Appendices.

\section{Perturbative results for conformal \texorpdfstring{QED$_3$}{QED3}}\label{sec:perturbative}

\qed can be understood pertubatively in the large \Nf limit, where one can identify a conformal fixed point and solve conformal \qed analytically in a 1/\Nf expansion. At small \Nf this expansion breaks down and the theory becomes strongly coupled, making perturbative estimates of the theory as well as the critical flavor number \(N_f^*\) harder to calculate. One of the main objectives of this work is to test whether the results from perturbative computations can be consistent with constraints from the conformal bootstrap. 
 
In Euclidean signature, the QED$_3$ action, i.e. the action of a $U(1)$ gauge theory coupled to $N_f$ massless charged two-component Dirac fermions, is
\be
\cS=\int d^3x\left(\frac{1}{4e^2}F^{\mu\nu}F_{\mu\nu} - \sum_{i=1}^{N_f}   \widebar{\psi}_i \sigma^\mu (\partial_\mu+i A_\mu)\psi^i\right),
\ee
where $e$ is the $U(1)$ gauge coupling constant, $A_\mu$ is the gauge field with field strength $F_{\mu\nu}=\partial_\mu A_{\nu}-\partial_\nu A_{\mu}$, and $\psi^i$ are the $N_f$ fermions in the fundamental representation of the flavor symmetry $SU(N_f)$.
The gamma matrices associated with two-component Dirac fermions are given by the Pauli matrices $\sigma^\mu, ~\mu=1,2,3$. Besides the flavor symmetry $SU(N_f),$ the theory also has a $U(1)_t$ global symmetry associated with a conserved current
\be
J_\mu^t=\frac{1}{4\pi}\epsilon_{\mu\nu\rho}F^{\nu\rho}.
\ee
The current $J_\mu^t$ is conserved  due to the Bianchi identity of the $U(1)$ gauge field, i.e. $dF=0$. The local operators charged under $U(1)_t$ are the monopole operators corresponding to the non-trivial topology of the $U(1)$ gauge field. The $U(1)_t$ charges $q$ of the monopole operators are quantized according to the Dirac quantization condition. We will follow the normalization of monopole operators in \cite{Chester:2016wrc, Chester:2017vdh}, in which $2q\in \Z$.

Due to the contributions from fermionic zero modes in the topological gauge field configurations, the monopole operators also construct nontrivial representations of the flavor symmetry $SU(N_f)$.
According to their charges under topological $U(1)_t$, the local gauge invariant operators in QED$_3$ can be separated into two parts: the $U(1)_t$ charged monopole operators, and the composite operators made from products of fundamental fields which are neutral under topological $U(1)_t$.  

\subsection{Scaling dimensions of low-lying gauge invariant operators with \texorpdfstring{$U(1)_t$}{U(1)t} charge \texorpdfstring{$q=0$}{q=0}}
In large $N_f$ QED$_3$, a set of  local gauge invariant operators can be constructed out of the fundamental fields $\psi_i$, $A_\mu$ and their derivatives. These operators do not correspond to any nontrivial topology of $U(1)$ gauge field and are neutral ($q=0$) under the topological $U(1)_t$; however, they form non-trivial representations of the flavor symmetry $SU(N_f)$. In this work, we will be interested in the fermion bilinear operator $r\equiv \psb_i\psi^j-\frac{1}{N_f} \delta_i^j\psb_k\psi^k$ with $N_f=4$, which forms an $SU(4)$ adjoint  representation. The OPE of $r\times r$ can be decomposed into $SU(N_f)$ irreducible representations (irreps):
\begin{multline}
(211) \bigotimes (211) =  (000)^+\bigoplus (211)^+\bigoplus (211)^- \bigoplus (220)^+\\\bigoplus(310)^-\bigoplus(332)^- \bigoplus(422)^+, \label{OPE}
\end{multline}
where the $i-$th number in the vector $(abc)$ denotes the number of boxes in the $i-$th line in the Young diagram of the representation, e.g. $(211)$ is the adjoint representation. The superscripts $+/-$ denote even/odd spin selection rules. Since $r$ forms a real representation of $SU(4)$, only real representations can appear in the right hand side (RHS) of above equation; for instance, only the real combination of $(310)$ and $(332)$ can appear in the $r\times r$ OPE, which will be denoted by $(310)_R$ throughout this paper.

Another important fact to take into account is the parity symmetry. The fermion bilinear scalar is parity odd, and so all the operators in the RHS of (\ref{OPE}) are parity even. The lowest parity-even operators in these sectors are constructed from fermion quadrilinear operators or their mixing with the gauge kinetic operator $F^2$. These four-fermion operators play important roles in solving the conformal QED$_3$ crossing equations. The scaling dimensions of these operators have been computed using $1/N_f$ expansion in previous studies \cite{Appelquist:1988sr, Nash:1989xx, Gracey:1993iu, Gracey:1993sn, Rantner_2002, Xu_2008, Hermele_2005, Kaul2008, Chester:2016ref}, which we now summarize.

\begin{subequations}
\label{eq: perturbative results}
The scaling dimension of the parity odd $SU(N_f)$ adjoint fermion bilinear scalar has been computed to the order $1/N_f^2$ \cite{Gracey:1993sn}:
\be 
\Delta_{(211)}=2-\frac{64}{3\pi^2N_f}+\frac{256(28-3\pi^2)}{9\pi^4N_f^2}\;. \label{fb}
\ee 
The $SU(N_f)$ singlet four-fermion operator $(\psb_i\psi^i)^2$ has scaling dimension 4 at tree level, identical to the $U(1)$ gauge kinetic term $F_{\mu\nu}F^{\mu\nu}$. They can mix with each other through quantum loop corrections; the scaling dimensions of the resultant two operators at order $1/N_f$ are
\be
\Delta_{(000)}^{\pm}= 4+\frac{64(2\pm \sqrt{7})}{3\pi^2N_f}\;.
\ee
We expect that the singlet operator with negative anomalous dimension $\Delta_S<4$ plays an important role in the loss of conformality in QED$_3$. For sufficiently small $N_f$, $\Delta_S$ approaches the marginality condition $\Delta_S=3$ from above and eventually generates an RG flow, dissolving the IR fixed point of QED$_3$ below $N_f^*$. Above $N_f^*$, the singlet four-fermion coupling can also generate a UV fixed point, whose UV completion is given by the QED$_3$-Gross-Neveu-Yukawa model. In this work, we will only focus on the QED$_3$ IR fixed point, and assume that $N_f^*<4$, as indicated by previous bootstrap studies and some lattice simulations.

The scaling dimension of the lowest scalar in the parity even $(220)$ sector has been computed in \cite{Xu_2008, Chester:2016ref} at the order $1/N_f$ to be
\be
\Delta_{(220)}=4-\frac{64}{\pi^2N_f}\;;
\ee
this operator will play an important role in bootstrap computations. Meanwhile, the scaling dimension of the parity even adjoint scalar\footnote{We remind the reader that \equref{adjeven} is for the \emph{parity even} scalar, whereas the result in \equref{fb} is for the \emph{parity odd} operator $r$, which is also in the $(211)$ sector.} is
\be
\Delta_{(211)}=4+\frac{8(25\pm \sqrt{2317})}{3\pi^2N_f}\;, \label{adjeven}
\ee
where the two operators differ by the contraction of the flavor indices at tree level \cite{Chester:2016ref}. Note that with $N_f=4$ the two operators have scaling dimensions about 2.44 and 8.94, respectively! Clearly, these first order corrections to the scaling dimensions of these four-fermion operators are significant, and so these results should be taken cautiously: it would be interesting to know if the higher order corrections can improve the behavior of these perturbative expansions. In \cite{Xu_2008} the author also computed the scaling dimension of lowest parity even scalar in $(422)$ sector 
\be
\Delta_{(422)}=4+\frac{64}{3\pi^2N_f}\;.
\ee
\end{subequations}

We would like to briefly comment on the convergence of the perturbative results in \equref{eq: perturbative results}. 
For the parity odd fermion bilinear adjoint operator, the second order correction is quite small, being only $5.4\%$ of the first order correction. 
Meanwhile, for the lowest scalars in the parity even $(220)$ and $(211)$ sectors, the first order corrections at $N_f=4$ are nearly $40\%$ of the tree level results. The $1/N_f$ perturbative results obtained in \cite{Chester:2016ref} suggest that the convergence becomes worse for composite operators with more fermions. It is currently unclear how well the leading order results can estimate scaling dimensions of four fermion operators in these sectors: as noted previously, it would be quite useful to compute higher order corrections to clarify this issue. For the $SU(N_f)$ adjoint fermion bilinear scalar, Monte Carlo simulations in \cite{Karthik:2015sgq, Xu:2018wyg} have computed the scaling dimension of $r$, which is consistent with the leading order results even for $N_f=4$; however, there are significant error bars in the estimates, which cannot exclude a potentially notable correction to the current result.

\subsection{Conserved charges in conformal \texorpdfstring{QED$_3$}{QED3}}

Conserved currents play fundamental roles in the study of CFTs. In  conformal QED$_3$,  there are three such currents: the stress tensor $T_{\mu\nu}$ and two global symmetry currents, $J_{\mu,i}^j$ and $J^t_\mu$, the latter of which are associated with the $SU(N_f)$ flavor symmetry and the topological $U(1)_t$  symmetry, respectively. The two-point functions of these conserved currents (in the normalization of \cite{Chester:2016ref}) are
\small 
\bea
\< T_{\mu\nu}(x_1) T_{\lambda\rho}(x_2)\> &=c_T \frac{3}{16\pi^2\left(x_{12}^2\right)^3}  I_{\mu\nu,\lambda\rho}(x_{12})\;,  \label{cT}\\
\< J_{\mu,i}^j(x_1) J_{\nu, k}^l(x_2)\> &=c_J \frac{1}{8\pi^2\left(x_{12}^2\right)^2}  I_{\mu\nu}(x_{12})\left( \delta_i^l\delta_k^j-\frac{1}{N_f} \delta_i^j\delta_k^l\right),   \label{cJ}\\
\< J^t_{\mu}(x_1) J^t_{\nu}(x_2)\> &=c_J^t  \frac{1}{8\pi^2\left(x_{12}^2\right)^2} I_{\mu\nu}(x_{12})\;, \label{cJt}  
\eea
\normalsize
where $c_x$ are the central charges and the tensor structures are defined through
\small 
\bea
I_{\mu\nu}(x) &\equiv\delta_{\mu\nu}-2\frac{x_\mu x_\nu}{x^2}\;, \\ 
I_{\mu\nu,\lambda\rho}(x)& \equiv \frac{1}{2}\left(   I_{\mu\lambda}(x) I_{\nu\rho}(x)+I_{\mu\rho}(x) I_{\nu\lambda}(x) \right)-\frac{1}{3}\delta_{\mu\nu}\delta_{\lambda\rho}\;.  
\eea
\normalsize
for convenience. The above central charges have been computed to sub-leading order in the $1/N_f$ expansion in \cite{Giombi:2016fct}:\footnote{These central charges  have also been studied in \cite{Huh:2013vga, Huh:2014eea}.}
\bea[eq:centralcharges]
c_T &=c_{T0} \left(1+\frac{0.7193}{N_f}+O(1/N_f^2) \right),   \\
c_J &=c_{J0} \left(1+\frac{0.1429}{N_f}+O(1/N_f^2)\right),   \\
c_J^t &= \frac{6.4846}{N_f} \left(1-\frac{{0.1429}}{N_f} +O(1/N_f^2)\right).
\eea
Here $c_{T0}$ and $c_{J0}$ are the contributions from the free fermions to the central charges, which are equal to $1$ in our normalization.  

It is worth mentioning one other result from \cite{Giombi:2016fct}, on $c_T$ and $c_J$ in QCD$_3$ with an $SU(N_c)$ Yang-Mills gauge field coupled with quarks in the fundamental representation of the color group:
\bea
c_T&=N_c c_{T0}\left(1+\frac{0.7193}{N_f}\frac{N_c^2-1}{N_c} +O(1/N_f^2)\right), \\
c_J&=N_c c_{J0}\left(1+\frac{0.1429}{N_f}\frac{N_c^2-1}{N_c} +O(1/N_f^2)\right).
\eea
Compared with QED$_3$, $c_T$ and $c_J$ in QCD$_3$ with gauge group $SU(N_c)$ have an additional factor of $N_c$, due to the color degrees of freedom carried by the fermions. The non-Abelian gauge fields also contain more degrees of freedom than the $U(1)$ gauge field, which increase the subleading order corrections in $c_T$ and $c_J$.
This provides a key differentiation between QED$_3$ and QCD$_3$, which otherwise might be hard to distinguish in bootstrap studies just by looking at their low-lying spectrum. 

\subsection{Large \texorpdfstring{$N_f$}{Nf} expansion of the monopole spectrum in \texorpdfstring{QED$_3$}{QED3}  }
 
Monopole operators in  QED$_3$ have been studied in various works \cite{Borokhov:2002ib, Pufu:2013vpa, Dyer:2013fja, Chester:2017vdh, Dupuis:2021flq}. 
Their quantum numbers $(\Delta_i, q_i, R_i)$ consist of their scaling dimension $\Delta_i$, their topological charge $q_i$ under $U(1)_t$ symmetry, and their $SU(4)$ representation $R_i$. 
We will be particularly interested in the monopoles $\cM_{1/2}$ and $\cM_1$ carrying the lowest topological charges $q=1/2$ and $q=1$,\footnote{In this work we will follow the conventions and the normalization used in \cite{Pufu:2013vpa,Dyer:2013fja}.}
which respectively sit in $(110)$ and $(220)$ representations of $SU(4)$. The scaling dimensions of these monopole operators were computed in \cite{Dyer:2013fja, Dupuis:2021flq} to subleading order in the large $N_f$ expansion. The authors computed the free energy on $S^2\times \mathbb{R}$ in the presence of a monopole flux in the IR limit $e^2 N_f\rightarrow \infty$. The scaling dimensions of the monopole operators on $\mathbb{R}^3$ are then given by the energies of the monopole states on $S^2\times \mathbb{R}$ through the state-operator correspondence. For the monopoles $\cM_{1/2}$ and $\cM_1$, their scaling dimensions are \cite{Dupuis:2021flq}
\bea 
\Delta_{\cM_{1/2}}=&0.26510N_f - 0.038138(5)+O(1/N_f)\;,\\
\Delta_{\cM_1}=&0.67315N_f-0.19340(3)+O(1/N_f)\;. 
\eea 
At $N_f=4$, the above formulas give $\Delta_{\cM_{1/2}}\simeq1.022$, $\Delta_{\cM_1}\simeq2.499$. The subleading corrections are fairly small compared with the leading terms, even at small $N_f=4$.  

The OPE of the monopole operators $\cM_{1/2}\times \cM_{1/2}$ plays a key role in our bootstrap study. There are an infinity family of operators  with topological charge $q=1$ appearing in this OPE. Like the monopoles $\cM_{1/2}$ and $\cM_1$, these operators can be constructed by applying fermionic creation operators on the monopole vacuum with $4\pi$ background magnetic flux. Our bootstrap study will make important use of the topological charge 1 spectrum appearing in the $\cM_{1/2}\times \cM_{1/2}$ OPE, which we discuss in more detail below.

States or operators with topological charge $q$ can be explicitly constructed in the free theory limit $e^2 N_f\rightarrow 0$ using a formalism developed in \cite{Chester:2017vdh}. To construct these states, one first chooses a monopole vacuum with background magnetic flux $4\pi q$ uniformly distributed in the 2D sphere of the Lorentizan spacetime $S^2\times \mathbb{R}$. Then the spectrum with topological charge $q$ can be obtained by constructing the gauge invariant states of free massless fermions $\psi^i$ in this background. The building blocks of a generic state are the fermionic modes in the classical solutions of the fermion field, which can be obtained by solving the Dirac equation  $(i\slashed{\nabla}+\mathcal{A})\psi=0$ in the monopole vacuum, giving a mode expansion
\be 
\label{eq:largeNModeExpansion}
\psi^i(t,x) = &\sum_{m=1/2-q}^{q-1/2} 
c_{q-1/2,m}^{i,\dagger} C_{q,q-1/2,m}(x)
\\+\sum_{j>q-1/2,m}&
\left(
a_{jm}^{i,\dagger} A_{qjm}(x)e^{i\lambda_j t}
+
b_{jm}^{i} B_{qjm}(x)e^{-i\lambda_j t}
\right) \, ,
\ee 
where $q$ is an overall label of the family of states on the same monopole background, and each fermion mode is labeled by the flavor indices $i$ and total angular momentum quantum numbers $j$ and $m$. 
The operators $a_{jm}^{i,\dagger}$, $b_{jm,i}^{\dagger}$, and $c_{q-1/2,m}^{i,\dagger}$ are fermion creation operators, and their corresponding coefficients $A_{qjm}$, $B_{qjm}$, and $C_{q,q-1/2,m}$, are spinor spherical harmonics. Specifically,  $a_{jm}^{i,\dagger}$ ( $b_{jm,i}^{\dagger}$) corresponds to (anti-)particles whereas $c_{q-1/2,m}^{i,\dagger}$ corresponds to fermion zero modes; furthermore, each
$a_{jm}^{i,\dagger}$ and $c_{q-1/2,m}^{i,\dagger}$ ($b_{jm,i}^{\dagger}$) mode transforms in the (anti-)fundamental representation of the $SU(N)$ group, and carry gauge $U(1)$ charge $+1$ ($-1$). 
The quantum numbers of the creation operators are given in Table \ref{QEDmodes}.\footnote{The energy of the bare monopole is the Casimir energy of the fermion fields
$$
\Delta_{\rm bare} = -N \sum_{j=q-1/2}^{\infty} d_j \lambda_j\, ,
$$
where $d_j = 2j+1$ is the degeneracy. The infinite sum is treated by $\zeta$-function regularization to give a finite answer.} See \cite{Chester:2017vdh} for more details on the monopole vacuum and fermionic creation operators.

\begin{table}
	\centering 
	\caption{Quantum numbers of the bare monopole with topological charge $q$ and the fermionic creation operators, adapted from \cite{Chester:2017vdh}. \label{QEDmodes}}
	\begin{tabularx}{.47\textwidth}{ccccc}
		\hline\hline\\[-1em]
	 	&  \textbf{energy/scaling dim.}   & \; \textbf{spin} \; & \; $\begin{aligned}
	 		\textbf{gauge}\\\textbf{charge}
	 	\end{aligned}$ \; & \;$\begin{aligned}
	 	SU(N)\\\textbf{irrep} 
 	\end{aligned}$
		\\[.1em]\hline\\[-1em]
		 \;$a^{i,\dagger}_{j m}$ & $\sqrt{(j+1/2)^2-q^2}$ & $j\;(\geqslant q+1/2)$ & $+1$ & $\mathbf{N}$
		\\[.1em]\hline\\[-1em]
		 \;$b^{i,\dagger}_{j m,i}$ & $\sqrt{(j+1/2)^2-q^2}$ & $j\;(\geqslant q+1/2)$ & $-1$& $\overline{\mathbf{N}}$ 
		\\[.1em]\hline\\[-1em]
	     \;$c^{i,\dagger}_{q-1/2, m}$ & $0$ & $q-1/2$ & $+1$ & $\mathbf{N}$
		\\[.1em]\hline\\[-1em]
		 \;$M_\text{bare}$ & $\Delta_{\rm bare}$  & $0$ & $-qN$ & $\mathbf{1}$
		 \\[.1em]\hline\hline
	\end{tabularx}
\end{table}

In principle the fermionic creation operators in Table \ref{QEDmodes}  allow us to construct any states or operators in the topological charge $q$ sector. 
There is a subtle issue that the above microstate construction is based on the free fermions in the UV limit $e^2N_f\rightarrow 0$ of QED$_3$, while the theory we are interested in corresponds to its IR fixed point, which relates to the $e^2N_f\rightarrow \infty$ limit. Nevertheless, there is evidence from the thermal computation which suggests that the states have significant overlaps between the two different limits \cite{Chester:2017vdh}.

We then set out to construct as completely as possible the low-lying states of $N_f=4$ QED$_3$. Our strategy is the following:
\begin{enumerate}
\item We first set a maximum energy threshold $\Delta_{\rm max}$, and exhaust all possible combinations of creation operators $a_{jm}^{i,\dagger}$, $b_{jm,i}^{\dagger}$, and $c_{q-1/2,m}^{i,\dagger}$, with the constraint that the net gauge charge is zero.
\item We decompose states created by each sequence of $a_{jm}^{i,\dagger}$, $b_{jm,i}^{\dagger}$, and $c_{q-1/2,m}^{i,\dagger}$ operators into irreps of the product group of spin and flavor symmetries $SU(2)\times SU(4)$. 
\item Within the sectors of the same $SU(2)\times SU(4)$ irreps, we anti-symmetrize the fermion creation operators, and collect the linearly independent states. 
\item After obtaining all possible states created by the fermion modes, it is straightforward to get the scaling dimension, spin, $SU(4)$ irreps and parity of the corresponding operators.
\end{enumerate}
More details of our procedure can be found in Appendix~\ref{sec: large n mode construction}, and we present the results in Table~\ref{tab:largeNSpectrum}. Here, we would like to briefly comment on the data in Table~\ref{tab:largeNSpectrum}: it describes the low-lying spectrum predicted by the large $N_f$ mode expansion, where some entries are improved wherever results about subleading corrections in $1/N_f$ are available. Additionally, there is a possible caveat of the above procedure that it does not include operators created by gauge fields. Therefore we need to add the operators constructed from gauge fields by hand. Pure gauge field operators include the topological current $J^{t\mu}$, $F^{\mu\nu} F_{\mu\nu}$, and their composite operators. $J^{t\mu}$ is already added to the table by hand, whereas $F^{\mu\nu} F_{\mu\nu}$ mixes with the $SU(4)$ singlet four-fermion operator. It is of course also possible to have composite operators between $J^{t\mu}$ and operators constructed from the fermion modes, which are annotated with a $^*$. We will frequently refer to this table when introducing assumptions on the spectrum in our bootstrap equations.

\section{\texorpdfstring{$SU(4)$}{SU(4)} adjoint fermion bilinear bootstrap}
\label{sec: Fermion bilinear scalar bootstrap}

The fermion bilinear scalar $r_i^j\equiv\psb_i\psi^j-\frac{1}{N_f}\delta_i^j\psb_k\psi^k$ is one of the lowest-dimension gauge-invariant operators in QED$_3$, making it a natural candidate for bootstrap studies of IR fixed points of gauge theories coupled with fermions; see e.g. \cite{Berkooz:2014yda, Nakayama:2017vdd, Iha:2016ppj, Li:2018lyb, Li:2020bnb}. 
A main challenge in the fermion bilinear bootstrap comes from the $SO(N_f^2-1)$ symmetry enhancement in the crossing equations  \cite{Li:2020bnb, Li:2020tsm}. To bootstrap conformal QED$_3$ with a proper $SU(N_f)$ symmetry, one has to resolve the $SO(N_f^2-1)$ symmetry enhancement in the crossing equations. In this section, we will describe how the $SO(N_f^2-1)$ symmetry enhancement can be slightly broken by introducing gap assumptions inspired by the perturbative $N_f=4$ conformal QED$_3$ spectrum, and the resulting bootstrap bounds have kinks which could conjecturally be connected to $N_f=4$ conformal QED$_3$. Nevertheless, the positions of the kinks are sensitive to the gap assumptions, so even under this conjecture more input needs to be given or more constraints need to be imposed in order to extract the physical solution of \qed.

\subsection{Crossing equations with different symmetries and gap-dependence} \label{creqid}

In certain theories, there exists an $SO(N)$ symmetry enhancement of the crossing equations which affect general single correlator bootstrap bounds \cite{Poland:2011ey,Li:2020bnb, Li:2020tsm}. In particular, there is a unique map up to normalization which transforms the $SU(N_f)$ adjoint crossing equation into the $SO(N_f^2-1)$ vector crossing equations; see \cite{Li:2020tsm} and \cite{Manenti:2021elk}. Here we will follow \cite{Li:2020bnb} and provide a more detailed study of its effect on the bootstrap bounds. 

The operators that can appear in the $r\times r$ OPE are provided in (\ref{OPE}). The crossing equations of the four-point correlator $\langle r(x_1)r(x_2)r(x_3)r(x_4) \rangle$ can be written in the vector form
\be 
\label{eq:rrrrCrossingEq}
\sum_{\cO\in\ell^+} \lambda_\cO^2\Vec{V}^+_{(000)}+\sum_{\cO\in\ell^+} \lambda_\cO^2\Vec{V}^+_{(211)}+\sum_{\cO\in\ell^-} \lambda_\cO^2\Vec{V}^-_{(211)}\\+\sum_{\cO\in\ell^-} \lambda_\cO^2\Vec{V}^-_{(310)_R}+\sum_{\cO\in\ell^+} \lambda_\cO^2\Vec{V}^+_{(220)}+\sum_{\cO\in\ell^+} \lambda_\cO^2\Vec{V}^+_{(422)}=0\;,
\ee 
where the vector $\Vec{V}^\pm_{R}$ is a $6$-component vector corresponding to the $SU(4)$ representation $R$ with even/odd spin.\footnote{The vector
$\Vec{V}_{(310)_R}$ corresponds to the real combination of $\Vec{V}_{(310)}$ and $\Vec{V}_{(332)}$.
} The crossing equations can be captured by a $6\times 6$ matrix:
\be
\cM_{SU(4)\textrm{-ad}}=
\left(
\begin{array}{cccccc}
 0 & 0 & 0 & -F & F & F \\
 0 & \frac{1}{2} F& 0 & 0 & -\frac{1}{2} F& \frac{1}{6}F \\
 0 & -F & -F & \frac{1}{4} F& \frac{1}{2}F & \frac{1}{6} F\\
 F & -4 F & 0 & 0 & \frac{16}{3}F & \frac{16 }{15}F \\
 H & -H & 0 & -H & -\frac{2}{3}  H & -\frac{14}{15}  H \\
 0 & H & -H & \frac{1}{4} H& \frac{1}{2}H & -\frac{7}{6}  H\\
\end{array}
\right)\,,
\ee
where the columns of the matrix correspond to the vectors $\Vec{V}^\pm_{R}$ in the order 
\small
\be 
\cM_{SU(4)\textrm{-ad}}=\left( \Vec{V}^+_{(000)}, \Vec{V}^+_{(211)}, \Vec{V}^-_{(211)},
\Vec{V}^-_{(310)_R},  \Vec{V}^+_{(220)}, \Vec{V}^+_{(422)}\right)_{SU(4)\textrm{-ad}}\,,
\ee 
\normalsize
and the variables $F,H$ denote the symmetrized/anti-symmetrized conformal blocks
	\begin{table*}
	\centering 
	\caption{\label{tab:largeNSpectrum} 
		A summary of estimates for the low-lying spectrum appearing in our bootstrap crossing equations obtained using the large $N_f$ expansion. The $SO(2)$ irrep, $SU(4)$ irrep, spin, the lowest 2 or 3 scaling dimensions, and the OPE channels that the operators contribute to are shown for each type of operators. The dimensions correspond to the scaling dimension of operators constructed using the fermion mode creation operators, $J^t$, and their composition. Whenever subleading order corrections are available in the literature, we added them as well. The dimension is annotated with $^*$ if the corresponding operator is a composite operator involving $J^t$. Note that the $SO(2)$ irrep encodes both the U$(1)$ charge and the parity information: the $SO(2)$ irreps $S$ and $A$ have U$(1)$ charge $q=0$ and are parity even and odd, respectively, whereas the $SO(2)$ irreps $V$ and $T$ have the respective U$(1)$ charges $q=1/2$ and $q=1$ while they can have either parity. Special operators are highlighted in the table using square brackets. 
	}
	\begin{tabularx}{.72\textwidth}{c@{\hskip 2em}l@{\hskip 2em}c@{\hskip 2em}c@{\hskip 1em}c@{\hskip 2em}c@{\hskip 2em}}
		\hline\hline\\[-1em]
		$SO(2)$ \textbf{rep}& $SU(4)$ \textbf{rep} & \textbf{spin-}$j$ & $\Delta_1$ & $\Delta_2$ & \textbf{OPE} 
		\\[.1em]\hline\\[-1em]
		S & ($000$) ($\mathsf{Singlet}$)  & 0 & $4 +\frac{64(2\pm\sqrt{7})}{3\pi^2N_f}= \substack{6.510\\ 3.651}$ & $5.00^*$ & $\lambda_{rrO}$, $\lambda_{MMO}$
		\\[.1em]\hline\\[-1em]
		S & ($211$) ($\mathsf{Adj}$)  & 0 & $4 + \frac{8(25\pm\sqrt{2317})}{3\pi^2 N_f} = \substack{8.940\\ 2.437}$ & $5.00^*$ & $\lambda_{rrO}$
		\\[.1em]\hline\\[-1em]
		S & ($211$) ($\mathsf{Adj}$)  & 1 & 2.00 [$J_f$] & 4.00 & $\lambda_{rrO}$, $\lambda_{MMO}$ 
		\\[.1em]\hline\\[-1em]
		S & ($220$) ($\mathsf{A\bar A}$)  & 0 & $4 -\frac{64}{\pi^2 N_f} = 2.379$ & 6.00 & $\lambda_{rrO}$, $\lambda_{MMO}$
		\\[.1em]\hline\\[-1em]
		S & $(310)_R$ ($\mathsf{S\bar A}$) & 1 & 5.00 & 6.00 & $\lambda_{rrO}$ 
		\\[.1em]\hline\\[-1em]
		S & ($422$) ($\mathsf{S\bar S}$) & 0 & $4+\frac{64}{3\pi^2N_f} = 4.540$ & 6.00 & $\lambda_{rrO}$
		\\[.1em]\hline\\[-1em]
		A & ($000$) ($\mathsf{Singlet}$)  & 1 & 2.00 [$J^t$] & 3.00 & $\lambda_{MMO}$
		\\[.1em]\hline\\[-1em]
		A & ($211$) ($\mathsf{Adj}$)  & 0 & $2 -\frac{64}{3\pi^2 N_f} +\frac{256(28-3\pi^2)}{9\pi^4 N_f^2}= 1.43$ [$r$] & 4.00 & $\lambda_{MMO}$ 
		\\[.1em]\hline\\[-1em]
		A & ($220$) ($\mathsf{A\bar A}$)  & 1 & 4.00 & 6.00 & $\lambda_{MMO}$ 
		\\[.1em]\hline\\[-1em]
		T & ($000$) ($\mathsf{Singlet}$)  & 0 & 4.424 & 6.156 & $\lambda_{MMO}$ 
		\\[.1em]\hline\\[-1em]
		T & ($211$) ($\mathsf{Adj}$)  & 1 & 2.692 & 4.424 & $\lambda_{MMO}$ 
		\\[.1em]\hline\\[-1em]
		T & ($220$) ($\mathsf{A\bar A}$)  & 0 & $2.499$ [$M_1$] & 6.156 & $\lambda_{MMO}$ 
		\\[.1em]\hline\\[-1em]
		V & ($110$) ($\mathsf{Anti}$)  & 0 & $1.022$ [$M_{1/2}$] & 3.888 & $\lambda_{rMO}$ 
		\\[.1em]\hline\\[-1em]
		V & ($110$) ($\mathsf{Anti}$)  & 1 & 2.474 & $3.060^*$ & $\lambda_{rMO}$ 
		\\[.1em]\hline\\[-1em]
		V & ($200$) ($\mathsf{Sym}$)  & 0 & 3.888 & $4.474^*$ & $\lambda_{rMO}$ 
		\\[.1em]\hline\\[-1em]
		V & ($200$) ($\mathsf{Sym}$)  & 1 & 2.474 & 3.888 & $\lambda_{rMO}$ 
		\\[.1em]\hline\\[-1em]
		V & ($321$) ($\mathsf{AAdj}$)  & 0 & 3.888 & 5.303 & $\lambda_{rMO}$ 
		\\[.1em]\hline\\[-1em]
		V & ($321$) ($\mathsf{AAdj}$)  & 1 & 3.888 & 4.924 & $\lambda_{rMO}$ 
		\\[.1em]\hline\hline
	\end{tabularx}
\end{table*}
\bea[eq:conformal block functions]
F &= v^{\Delta_{r}}g_{\Delta,\ell}(u,v)-u^{\Delta_{r}}g_{\Delta,\ell}(v,u)\;, \label{Fblock}\\
H &= v^{\Delta_{r}}g_{\Delta,\ell}(u,v)+u^{\Delta_{r}}g_{\Delta,\ell}(v,u)\;. \label{Hblock}
\eea
A notable property of the above $SU(4)$ adjoint crossing equations is that they admit a unique (up to normalization) transformation $\cT_{SU(4)\textrm{-ad}}$
\beq
\cT_{SU(4)\textrm{-ad}}=
\left(
\begin{array}{cccccc}
 1 & \frac{226}{119} & \frac{4}{7} & 0 & 0 & 0 \\
 -1 & \frac{894}{119} & -\frac{4}{7} & 1 & 0 & 0 \\
 0 & 0 & 0 & 0 & 1 & \frac{4}{7} \\
\end{array}
\right)\,,
\eeq
which maps the $SU(4)$ adjoint crossing equations to the $SO(15)$ vector crossing equations 
\beq
 \cM_{SO(15)}= \left(\Vec{V}^+_S, \Vec{V}^+_T, \Vec{V}^-_A   \right)_{SO(15)}=
\left(
\begin{array}{ccc}
 0 & F & -F \\
 F & \frac{13}{15}F & F \\
 H & -\frac{17}{15} H & -H \\
\end{array}
\right)\,, \label{SO15ceq}
\eeq
in which $S,T,A$ represent the singlet, traceless symmetric, and antisymmetric representations of $SO(15)$, respectively.

The action of $\cT_{SU(4)\textrm{-ad}}$ is
\small 
\be 
(\cT\cdot \cM)_{SU(4)\textrm{-ad}}= 
\left(
\begin{array}{cccccc}
 0 & \frac{45 }{119}F & -\frac{4}{7} F& -\frac{6}{7} F & \frac{40 }{119}F & \frac{24 }{17}F \\
 F & \frac{39 }{119}F & \frac{4 }{7}F & \frac{6 }{7}F & \frac{104 }{357}F & \frac{104 }{85}F \\
 H & -\frac{3}{7}  H & -\frac{4}{7}  H & -\frac{6}{7}  H & -\frac{8}{21}  H & -\frac{8}{5}  H \\
\end{array}
\right)\,, \label{TMSU4}
\ee
\normalsize
which can be briefly expressed in a vector form
\small 
\be 
\cT_{SU(4)\textrm{-ad}}\cdot\left( \Vec{V}^+_{(000)},\; \Vec{V}^+_{(211)},\; \Vec{V}^-_{(211)},\;
\Vec{V}^-_{(310)_R},\;  \Vec{V}^+_{(220)},\; \Vec{V}^+_{(422)}\right)_{SU(4)\textrm{-ad}} 
\\
=\left( \Vec{V}^+_{S},\; x_1\Vec{V}^+_{T},\; x_2\Vec{V}^-_{A},\;
x_3\Vec{V}^-_{A},\; x_4\Vec{V}^+_{T},\; x_5\Vec{V}^+_{T}\right)_{SO(15)}, \label{SU4toSO15}
\ee 
\normalsize
with positive coefficients $x_i$
\beq
\Vec{x}=\left(\frac{45}{119},~ \frac{4}{7},~\frac{6}{7}, ~\frac{40}{119},~ \frac{24}{17}\right)\,.
\eeq
We will show that the positivity of these coefficients is critical in the bootstrap algorithm.

We can summarize the above by saying that the the transformation $\cT_{SU(4)\textrm{-ad}}$ maps the channels of the $SU(4)$ adjoint crossing equations $\cM_{SU(4)\textrm{-ad}}$ to the channels of the $SO(15)$ vector crossing equations $\cM_{SO(15)}$ through the branching rules
\bea[eq:branching-rules]
\mathbf{SO(15)}& \qquad\mathbf{SU(4)} \nn \\
S&\longleftrightarrow(000)^+\;,  \label{branch1}\\
T&\longleftrightarrow(211)^+\bigoplus (220)^+\bigoplus  (422)^+\;, \label{branch2} \\
A&\longleftrightarrow(211)^-\bigoplus (310)_R^-\;. \label{branch3}
\eea

The goal of the conformal bootstrap algorithm is to find linear functionals $$\Vec{\beta}\equiv (\beta_1, \, \beta_2,\, \beta_3, \, \beta_4, \, \beta_5, \, \beta_6)$$ whose action on the crossing equations $\cM_{SU(4)\textrm{-ad}}$ satisfies
\small 
\be
\hspace*{-.5em}\Vec{\beta}\cdot \cM_{SU(4)\textrm{-ad}}= (\beta_{(000)}^+, \, \beta_{(211)}^+,\, \beta_{(211)}^-, \, \beta_{(310)_R}^-, \, \beta_{(220)}^+, \, \beta_{(422)}^+)\succeq 0_{1\times 6}\;,\\ \forall \Delta\geqslant \Delta_{R_i, \ell}^*\;, \label{bteq}
\ee
\normalsize
where $\Delta_{R_i, \ell}^*$ is the assumed minimum scaling dimension of any spin $\ell$ operator in the $R_i$ representation of $SU(4)$.\footnote{$\Delta_{R_i, \ell}^*$ is either the unitary bound or a specific value above the unitary bound.} Due to the algebraic relation (\ref{SU4toSO15}), any linear functional $\Vec{\alpha}\equiv (\alpha_1,\, \alpha_2, \, \alpha_3)$ satisfying the $SO(15)$ bootstrap equations 
\small 
\be
\Vec{\alpha}\cdot \cM_{SO(15)}=\Vec{\alpha}\cdot\left(\Vec{V}^+_S, \Vec{V}^+_T, \Vec{V}^-_A   \right)_{SO(15)}= (\alpha_{S}^+, \, \alpha_{T}^+,\, \alpha_{A}^-)\succeq 0_{1\times 3}\;,\\ \forall \Delta\geqslant \Delta_{S/T/A, \ell}^*\;, 
\ee
\normalsize
can be used to construct linear functionals in the $SU(4)$ adjoint bootstrap 
\beq
\Vec{\beta}_{1\times6}=\Vec{\alpha}_{1\times3}\cdot \left(\cT_{SU(4)\textrm{-ad}}\right)_{3\times6}\;,
\eeq
which also satisfies the $SU(4)$ adjoint bootstrap equations
\footnotesize
\be\label{eq:functional-coincidence}
\Vec{\beta}\cdot \cM_{SU(4)\textrm{-ad}}&=\left(\Vec{\alpha}\cdot \cT_{SU(4)\textrm{-ad}}\right)\cdot\cM_{SU(4)\textrm{-ad}} \\
&=\Vec{\alpha}\cdot \left( \Vec{V}^+_{S}, \, x_1\Vec{V}^+_{T},\, x_2\Vec{V}^-_{A},\,
x_3\Vec{V}^-_{A}, \, x_4\Vec{V}^+_{T},\, x_5\Vec{V}^+_{T}\right)_{SO(15)}\\
&=\left( \alpha_{S}^+, \, x_1\alpha_{T}^+,\, x_2\alpha_{A}^-,\,x_3\alpha_{A}^-, \, x_4\alpha_{T}^+,\, x_5\alpha_{T}^+  \right)\succeq 0_{1\times 6}\;,\\&\hspace*{16em} \forall \Delta\geqslant \Delta_{R_i, \ell}^*\,, 
\ee
\normalsize
given that the gap assumptions $\Delta_{R_i, \ell}^*$ are consistent with those in the $SO(15)$ vector bootstrap $\Delta_{S/T/A, \ell}^*$ following the branching rules (\ref{eq:branching-rules}). Note in the second line we have employed the identity (\ref{SU4toSO15}) and the positivity condition in the third line follows from the positivity of $\alpha_{S/T/A}^\pm$ owing to the positive coefficients $x_i$. 

The relation (\ref{eq:functional-coincidence}) suggests that
the bounds from $SU(4)$-adjoint bootstrap cannot be weaker than that from the $SO(15)$ vector bootstrap, i.e.
$\Delta^*_{SO(15)-v}\geq \Delta^*_{SU(4)-ad}$,
because any linear functional that excludes some CFT data in the $SO(15)$ vector bootstrap must exclude the same data in the $SU(4)$-adjoint bootstrap.
On the other hand, because any four-point correlator of the $SO(15)$ vectors can be decomposed into four-point correlators of the $SU(4)$ adjoint scalar, the inverse is true, i.e. $\Delta^*_{SO(15)-v}\leq \Delta^*_{SU(4)-ad}$. Therefore we have exactly the same bounds from $SO(15)$ vector bootstrap and $SU(4)$-adjoint bootstrap computations, $\Delta^*_{SO(15)-v}= \Delta^*_{SU(4)-ad}$, under the condition that sectors on both sides that are related by the branching rules (\ref{eq:branching-rules}) have the same gap assumptions.

The above arguments show
that due to the algebraic relation (\ref{SU4toSO15}), the $SU(4)$ adjoint bootstrap problem with suitably related gap assumptions is equivalent to the $SO(15)$ vector bootstrap and admits the same solutions. The differences between the two bootstrap setups come from the gap assumptions $\Delta_{R_i, \ell}^*$.
To illustrate, let us consider the upper bounds on the scaling dimensions of the lowest non-identity singlet scalar $\Delta_0$, without imposing any gap assumptions besides the unitary bounds in other sectors; i.e., our assumptions are
\beq
 \Delta_{(000),\ell=0}\geqslant \Delta_0\;,\quad\Delta_{\textrm{other sectors}}\geqslant \textrm{unitary bounds}  
\eeq
in the $SU(4)$ adjoint bootstrap and 
\beq
\Delta_{S,\ell=0}\geqslant \Delta_0\;,\quad\Delta_{\textrm{other sectors}}\geqslant \textrm{unitary bounds}  
\eeq
in the $SO(15)$ vector bootstrap. The two sets of assumptions are consistent with the $SO(15)\rightarrow SU(4)$ branching rules (\ref{eq:branching-rules}). In consequence the singlet bounds obtained from the $SU(4)$ adjoint bootstrap and $SO(15)$ vector bootstrap are exactly the same. 

Another interesting example is given by the upper bound on the scaling dimension of the lowest $SO(15)$ traceless symmetric scalar $\Delta_{1}$ obtained from the $SO(15)$ vector bootstrap. Without imposing any extra gap assumptions, the assumptions are
\beq
\Delta_{T,\ell=0}\geqslant \Delta_1\;,\quad\Delta_{\textrm{other sectors}}\geqslant \textrm{unitary bounds}\,. \label{SO15Tbd}
\eeq 
In the $SU(4)$ adjoint bootstrap, if we want to get the upper bound on the scaling dimension of the lowest scalar in a sector like the $(422)$ representation without imposing extra gap assumptions, the assumptions are
\beq
\Delta_{(422),\ell=0}\geqslant \Delta_1\;,\quad\Delta_{\textrm{other sectors}}\geqslant \textrm{unitary bounds}\,. \label{SU4SSbbd}
\eeq
According to the branching rule (\ref{branch2}), the \(SO(15)\) assumptions in (\ref{SO15Tbd}) are actually equivalent to 
\be
\Delta_{T,\ell=0}\rightarrow  \left\{ \begin{array}{c}
 \Delta_{(422),\ell=0} \geqslant \Delta_1\,,\\
 \Delta_{(220),\ell=0} \geqslant \Delta_1\,, \\
 \Delta_{(211),\ell=0} \geqslant \Delta_1\,, \\
\end{array} \right.
 \\\Delta_{\textrm{other sectors}}\geqslant \textrm{unitary bounds}\,,  \label{gaps2}
\ee
which is stronger than the assumptions (\ref{SU4SSbbd}) in the $SU(4)$ adjoint bootstrap.  Consequently, the upper bound on the scaling dimensions of the lowest $(422)$ scalar in the $SU(4)$ adjoint bootstrap is weaker than the bound on the lowest traceless symmetric scalar in the $SO(15)$ vector bootstrap.\footnote{In principle, it is possible that there could be no solution to the crossing equations between the two gap sets (\ref{SO15Tbd}) and (\ref{gaps2}). In this case, the two bootstrap boundary conditions (\ref{SO15Tbd}) and (\ref{gaps2}) can actually generate the same bootstrap bound. In this work, we find the bootstrap bounds with such different boundary conditions are indeed different at finite derivative order $\Lambda$.} Nevertheless, the two bounds coincide with each other if we impose the assumption that the scalars in the three sectors $(422)$, $(220)$, and $(211)$ all have the same minimum scaling dimension $\Delta_1$.

The symmetry enhancement (\ref{SU4toSO15}) thus leads to a surprising fact, that in the single correlator bootstrap, although the crossing equations admit $SU(4)$ symmetry, it
cannot be distinguished from an $SO(15)$ symmetry at the crossing equation level. The constraints specific to $SU(4)$ symmetric theories can only be obtained from the gap assumptions that break the $SO(15)$ symmetry explicitly. 
This suggests that the gap assumptions in the bootstrap conditions are the only ingredients that we may resort to to carve out the parameter space of non-$SO(N)$ symmetric CFTs, while the role of the non-$SO(N)$ symmetric crossing equations is to provide access to individual sectors branched from the $SO(N)$ representations. Our bootstrap bounds for non-$SO(N)$ symmetric theories are obtained based on non-$SO(N)$ symmetric gap assumptions, and the bounds directly rely on the magnitudes of gaps in certain sectors, i.e., they are gap-dependent.

For the ambitious bootstrap dream which aims to completely solve the IR fixed points of gauge theories, this gap-dependence could be a serious problem. 
One hopes that the bootstrap bounds can provide numerical solutions of targeted theories with only few reliable and general assumptions. 
On the other hand, the gap-dependence of the bootstrap bounds indicates that the physical solutions may not saturate the bootstrap bounds unless there are sufficiently precise gaps input to the bootstrap equations. Below we will show several examples of the the gap-dependence of the bootstrap bounds and study possible approaches to partially resolve this problem.

\subsection{\texorpdfstring{$SU(4)$}{SU(4)} adjoint bootstrap results}

In this section we study the constraints on the CFT data of $N_f=4$ conformal QED$_3$ by bootstrapping the $SU(4)$ adjoint fermion bilinear scalars. 
The main results are that the bootstrap approach indeed can provide nontrivial constraints on the putative CFT data of the theory, and after imposing certain gaps inspired by the QED$_3$ spectrum, there are prominent kinks in the bootstrap bounds on scaling dimensions of operators in different $SU(4)$ representations, indicating the existence of a special solution to the crossing equations containing an $SU(4)$ adjoint scalar. Notably, the dimension of this scalar is near the perturbative and lattice results of $N_f=4$ QED$_3$. However, as discussed above, the precise locations of these kinks are gap-dependent, and consequently we need more information or constraints to pin down the underlying theories of these kinks using the conformal bootstrap and to firmly establish their connection to \qed.

The fermion bilinear scalar $r$ is parity odd in QED$_3$ and the operators appearing in the $r\times r$ OPE are parity even. The lowest scalars on the RHS of (\ref{eq:branching-rules}) are parity even four-fermion operators, which have scaling dimensions $4\pm O(1/N_f)$ and break $SO(N_f^2-1)$ symmetry by their $1/N_f$ corrections, see Table \ref{tab:largeNSpectrum} for details on the subleading corrections of the scaling dimensions of these four-fermion operators. Another notable factor breaking the $SO(15)$ symmetry appears on the RHS of (\ref{branch3}): in the $(211)^-$ sector, the lowest operator is the spin-$1$ conserved current corresponding to the $SU(4)$ symmetry, while in the $(310)_R^-$ sector, the lowest spin-$1$ operator has scaling dimension $5\pm O(1/N_f)$. Its subleading correction is not known yet, while the scaling dimension of this operator is expected to be notably higher than the unitary bound.

\begin{figure}
\centering
\hspace*{-1em}
\includegraphics[width=1\linewidth]{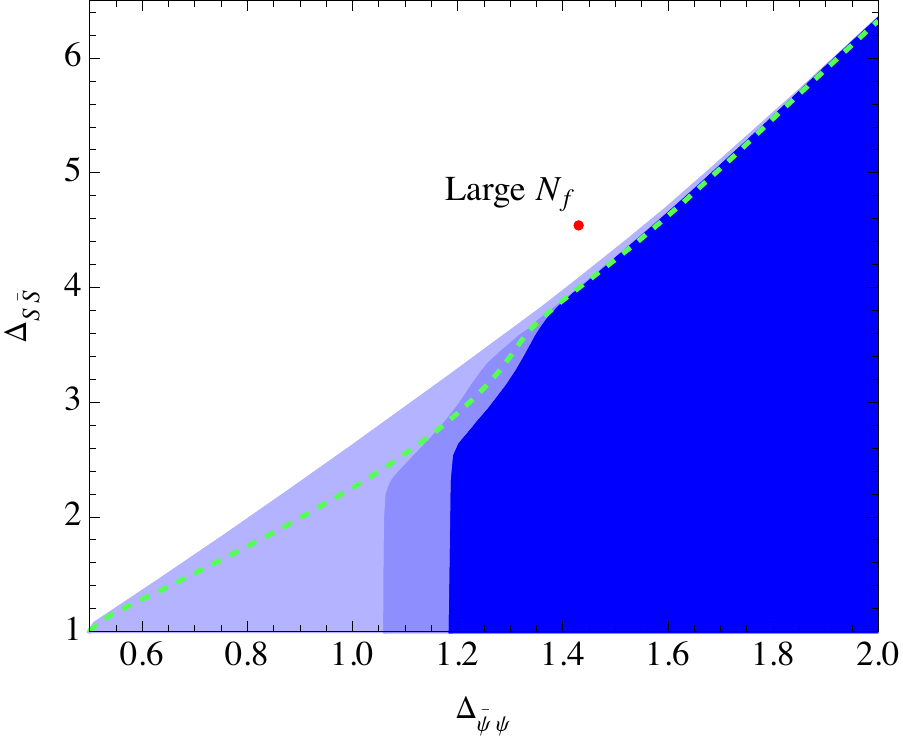}
\rlap{\hspace{-3in}\raisebox{1.5in}{%
\includegraphics[width=0.33\linewidth]{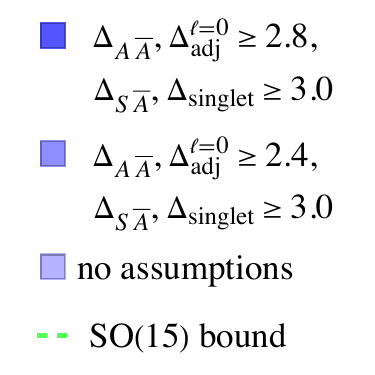}}}
\caption{
Upper bounds ($\Lambda=31$) on the scaling dimension of the lowest scalar in the $(422)$ representation under various conditions: no gaps (lightest blue region), gaps $2.4$ in the $(211)^+$ and $(220)^+$ sectors and $3.0$ in the $(422)^+$ and  $(310)_R^-$ sectors (light blue),  gaps $2.8$ in the $(211)^+$ and $(220)^+$ sectors and $3.0$ in the $(422)^+$ and  $(310)_R^-$ sectors (dark blue). The green line denotes the upper bound on the scaling dimension of the lowest $SO(15)$ traceless symmetric scalar obtained from the $SO(15)$ vector bootstrap, which is identical to the $SU(4)$ adjoint bootstrap bound on the scaling dimensions of the four-fermion scalars on the RHS of \ref{branch2} with the assumption that these four-fermion scalars have the same scaling dimension. In the physical spectrum of $N_f=4$ QED$_3$, this assumption is violated by subleading $1/N_f$ corrections. The kink near $(\Delta_{\bar{\psi}\psi}\simeq 1.35, ~ \Delta_{S\bar{S}}\simeq 3.7)$ in the green dashed line  remains in the $SU(4)$ adjoint bootstrap bound after introducing different gaps inspired by the $1/N_f$ perturbative results. Nevertheless, the position of this kink depends on the gaps. The red dot denotes the $1/N_f$ perturbative results on the scaling dimensions of the $SU(4)$ adjoint fermion bilinear and the lowest scalar in the $(422)$ representation.
} \label{ssbbd}
\end{figure}

In Fig.~\ref{ssbbd} we show the bootstrap bounds on the scaling dimension of the lowest scalar in the $S\bar{S}$  sector. The lightest blue shadowed region denotes the bootstrap bound obtained from the $SU(4)$ adjoint bootstrap without imposing any gap assumptions. The bootstrap bound is smooth without any significant structure, nevertheless, it is already quite interesting even without any extra input information specific to QED$_3$. The red dot represents the $1/N_f$ perturbative results for the scaling dimensions of the fermion bilinear $r$ (at order $1/N_f^2$) and the leading scalar in the $S\bar{S}$ sector (at order $1/N_f$). The perturbative data is notably above the bootstrap bounds and cannot belong to a unitary CFT, which suggests that at least one of the operators will receive significant higher order corrections.

The green dashed line gives the bootstrap bound on the lowest traceless symmetric scalar from the $SO(15)$ vector bootstrap. 
The same bound appears in the $SU(4)$ adjoint bootstrap if the sectors on the RHS of (\ref{branch2}) have the same gap assumptions, due to the bound coincidence explained previously.
The bootstrap bound shows a sharp kink near $\Delta_{r}\sim 1.35$, close to the expected scaling dimension of the $SU(4)$ adjoint fermion bilinear scalar in $N_f=4$ QED$_3$. Comparing with the lightest blue shadowed region, the gap assumption helps to rule out the regions on the left of the kink, while the bootstrap bound to the right of the kink is only mildly modified. 
This shows heuristically how the gap assumptions help in forming the kink structure in the $SU(4)$ adjoint bootstrap bound, and it indicates that a special solution stands out under the constraints posed by the gap assumptions.

The $SO(15)$ vector bootstrap bounds can be obtained in the $SU(4)$ adjoint bootstrap with the $SO(15)$ symmetric gap assumptions given in (\ref{gaps2}). In $N_f=4$ QED$_3$, this is only satisfied  by the tree level scaling dimensions of four-fermion operators on the RHS of (\ref{branch2}).
In the physical spectrum of $N_f=4$ QED$_3$, these four-fermion scalars have different higher order corrections, which are summarized in Table \ref{tab:largeNSpectrum}. After taking this difference into account, the gap assumptions in (\ref{gaps2}) need to be sightly modified and the bootstrap bound, especially the position of the kink will be shifted. 

According to the $1/N_f$ perturbative results in Table \ref{tab:largeNSpectrum}, at order $O(1/N_f)$ the lowest scalars in the $(211)$ and $(220)$ representations have scaling dimensions $\Delta\sim2.4$, while the higher order corrections are expected to be significant, as shown in Fig.~\ref{ssbbd} for the leading scalar in the $(422)$ representation. In Fig.~\ref{ssbbd} we tested the gaps $\Delta>2.4$ (light blue region) and $\Delta>2.8$ (dark blue region) in both the $(211)$ and $(220)$ sectors.\footnote{A natural choice of the gaps in these sectors is the irrelevance condition $\Delta>3$, which can affect whether \qed can be realized in lattice models~\cite{He:2021sto}. However, for reasons that will be explained in our monopole bootstrap study, we chose to a make a slightly more conservative gap assumption $\Delta>2.8$ instead. The bounds with gaps $\Delta>3$ in the $(211)$ and $(220)$ sectors have slightly stronger but similar shapes as the bounds shown in this work.} In addition, we also imposed the gaps $\Delta>3$ for the lowest operators in the parity even singlet and $S\bar{A}$ sectors. In the new bootstrap bounds with these gaps there are vertical left cuts caused by the gaps $\Delta>2.4$ or $\Delta>2.8$ in the $(211)$ and $(220)$ sectors. The prominent kinks remain in the new bootstrap bounds, while their positions are slightly shifted in comparison with the kink in the $SO(15)$ vector bootstrap bound.

\begin{figure*}
	\centering
	\hspace*{-1em}
	$
	\begin{aligned}
		\includegraphics[width=0.4\linewidth]{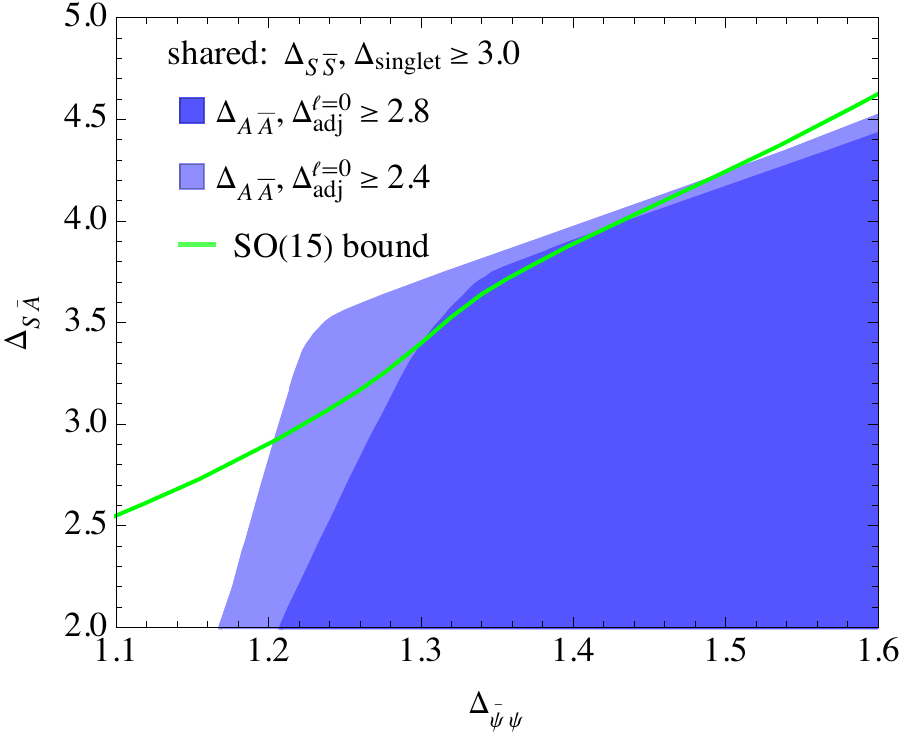}
		\includegraphics[width=0.4\linewidth]{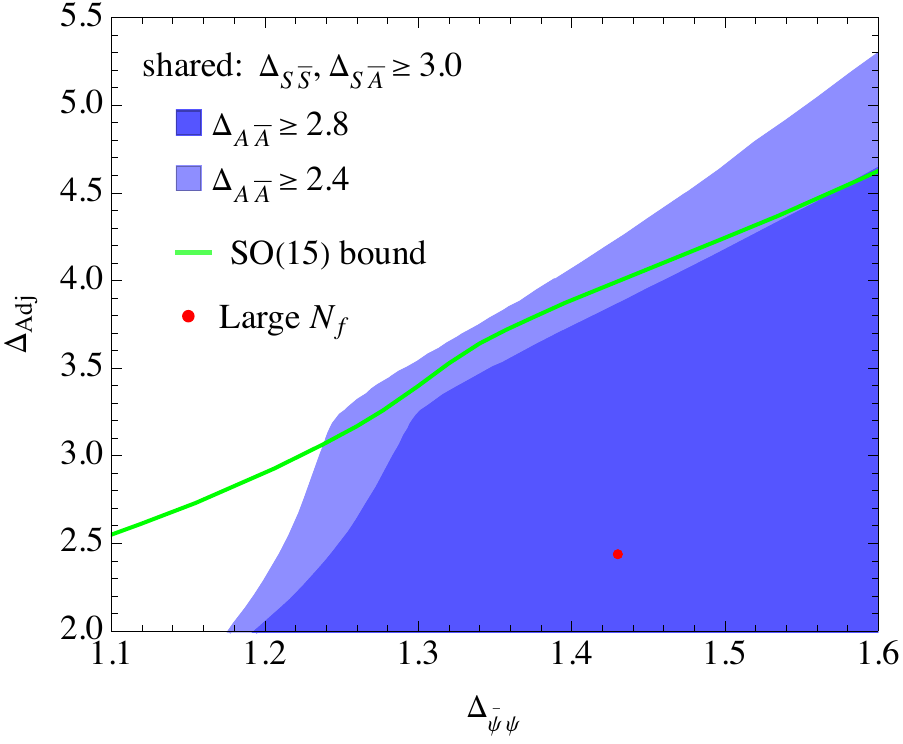}\\
		\includegraphics[width=0.4\linewidth]{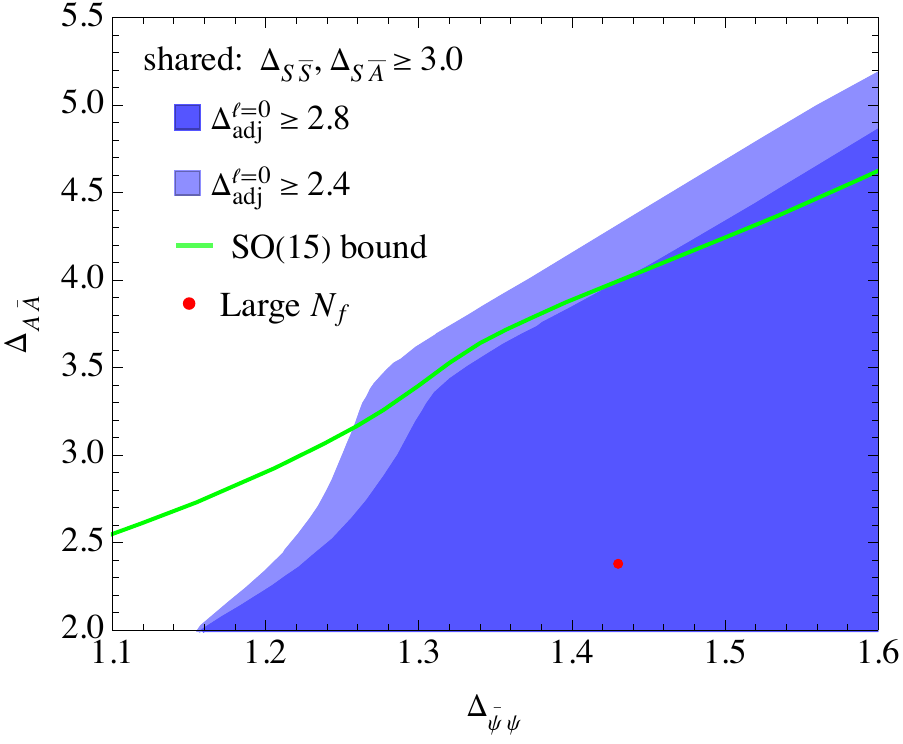}
		\includegraphics[width=0.4\linewidth]{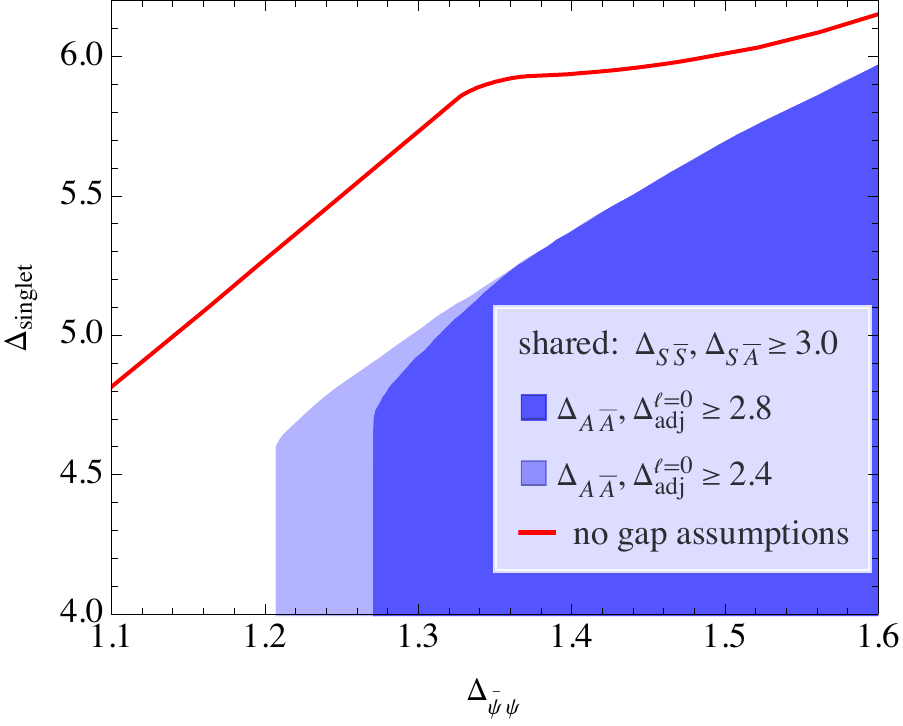}
	\end{aligned}
	$
	\caption{ 
		The (light) blue regions give bootstrap bounds ($\Lambda=31$) on the scaling dimensions of lowest operators in the $S\bar{A}$ (upper left), $Adj_{\ell=0}$ (upper right), $A\bar{A}$ (lower left), and singlet (lower right) representations of $SU(4)$ with certain gap assumptions.  The green lines denote the upper bound on the scaling dimension of the lowest $SO(15)$ traceless symmetric scalar obtained from the $SO(15)$ vector bootstrap. The red line in the lower right plot represents the $SU(4)$ singlet upper bound without any gap assumptions, which coincides with the singlet upper bound from the $SO(15)$ vector bootstrap. The red dots denote the large $N_f$ perturbative results. In the gap assumptions, we require the lowest operators in the $R = \{S\bar{A}$, $S\bar{S}, \textrm{singlet}\}$ sectors are all irrelevant $\Delta_R > 3$; while the lowest scalars in the $R = \{A\bar{A}, \textrm{Adj}_{\ell=0}\}$ sectors are above $\Delta_R>2.4$ (light blue) or $\Delta_R>2.8$ (blue). 
	} \label{multifbbd}
\end{figure*}

In Fig.~\ref{multifbbd} we show more bootstrap bounds on the scaling dimensions of operators in different representations of $SU(4)$. Generally the bootstrap bounds of non-singlet operators show prominent kinks near the kink of the $SO(15)$ vector bootstrap bound, and the positions of the kinks depend on the gaps. Note the upper-left plot of Fig.~\ref{multifbbd} gives an upper bound on the lowest spin 1 operator in the $S\bar{A}$ sector. Its branching rule is given in (\ref{branch3}), which is part of the spin 1 operator in the anti-symmetric representation of $SO(15)$ symmetry. So its bound is independent of the bound of the $SO(15)$ traceless-symmetric scalar given by the green line. Interestingly, it still shows a sharp kink with $\Delta_r$ close to the kink in the green line. 

The kink in the singlet sector (right bottom) is less significant in comparison with the non-singlet sectors. Nevertheless, a mild kink-like structure appears in the dark blue shadowed region, obtained after imposing gaps $\Delta>2.8$ in the $Adj$ and $A\bar{A}$ sectors. An interesting fact here is that after imposing gaps $\Delta>3$ in the $S\bar{A}$ and $S\bar{S}$ sectors, the singlet upper bound decreases significantly in comparison to the singlet upper bound without any gap assumptions (red line). The singlet upper bound has been observed to be significantly weaker than the expected value $\Delta_{\text{singlet}} \in (3, 4)$ in interesting physical theories. By introducing gaps inspired by the QED$_3$ spectrum which break the $SU(4)\rightarrow SO(15)$ symmetry enhancement (\ref{eq:branching-rules}), the singlet bound can be notably improved. The gap in the $S\bar{A}$ sector is especially helpful to resolve the $SO(15)$ symmetry enhancement in that its dimension is much higher than the unitary bound of spin 1 currents which forbids a conserved current for $SO(15)$ symmetry. According to the large $N_f$ spectrum in Table \ref{tab:largeNSpectrum} and the bootstrap bounds in Fig.~\ref{multifbbd}, we expect a stronger gap in this sector is also allowed and that the singlet upper bound can potentially be improved further. 

We emphasize that gap assumptions, even those such in Fig.~\ref{ssbbd} and \ref{multifbbd} which only slightly break the $SO(15)\rightarrow SU(4)$ relations (\ref{eq:branching-rules}), play a critical role in bootstrapping a specific theory such as conformal $N_f=4$ QED$_3$. With insufficient gap assumptions, many undesired potential solutions to the $SU(4)$ or $SO(15)$ crossing equations may still be around, obscuring a physical solution (which may relate to a kink structure). Recently the authors of \cite{He:2021sto} observed that the kink in the $SU(4)$ adjoint scalar bootstrap singlet bound smooths out and perhaps disappears when one imposes a gap in only the spin-1 $S\bar{A}$ sector. We do not view this as a major surprise since it is not clear that a single $S\bar{A}$ gap is sufficient to pick out the conformal $N_f=4$ QED$_3$ solution. For several sectors shown
in Fig.~\ref{ssbbd} and \ref{multifbbd}, when we use gaps inspired by the perturbative expectations for $N_f=4$ QED$_3$, the kinks remain and some become even sharper compared with those first found in \cite{Li:2018lyb}.

\begin{figure*}
\centering
\hspace*{-2em}
$
\begin{aligned}
\Delta_{S_{(000)}}^{\ell=0}\geq3.0 ,~
\Delta_{S_{(310)}}^{\ell=0}\geq3.0 ,~ 
\Delta_{S_{(220)}}^{\ell=0} &\geq 2.8 ,~ 
\Delta_{S_{(211)}}^{\ell=0}\geq2.8 ,~
\Delta_{S\bar{S}'}\geq4.5  
\\
\includegraphics[width=0.45\linewidth]{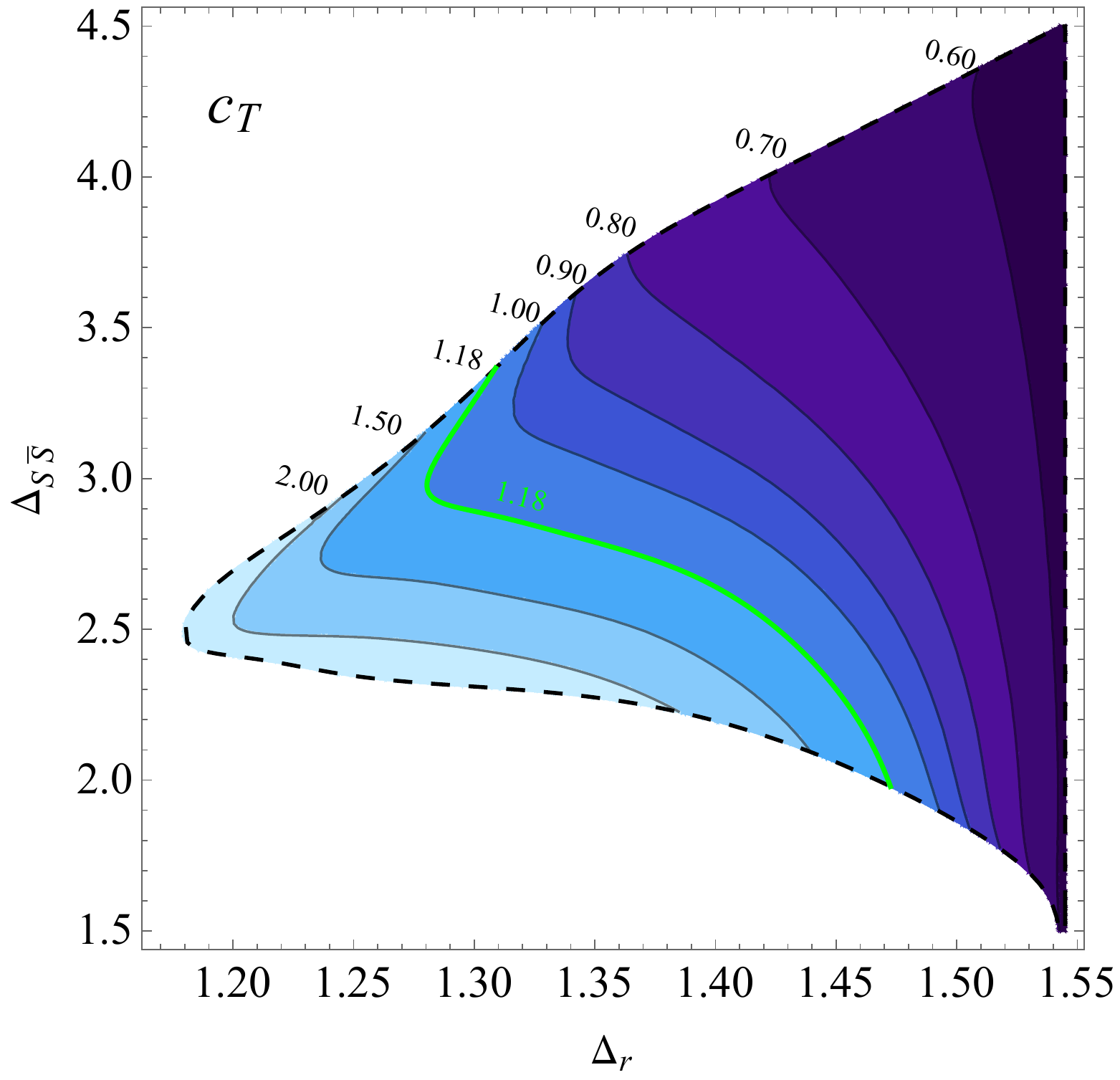} &
\includegraphics[width=0.45\linewidth]{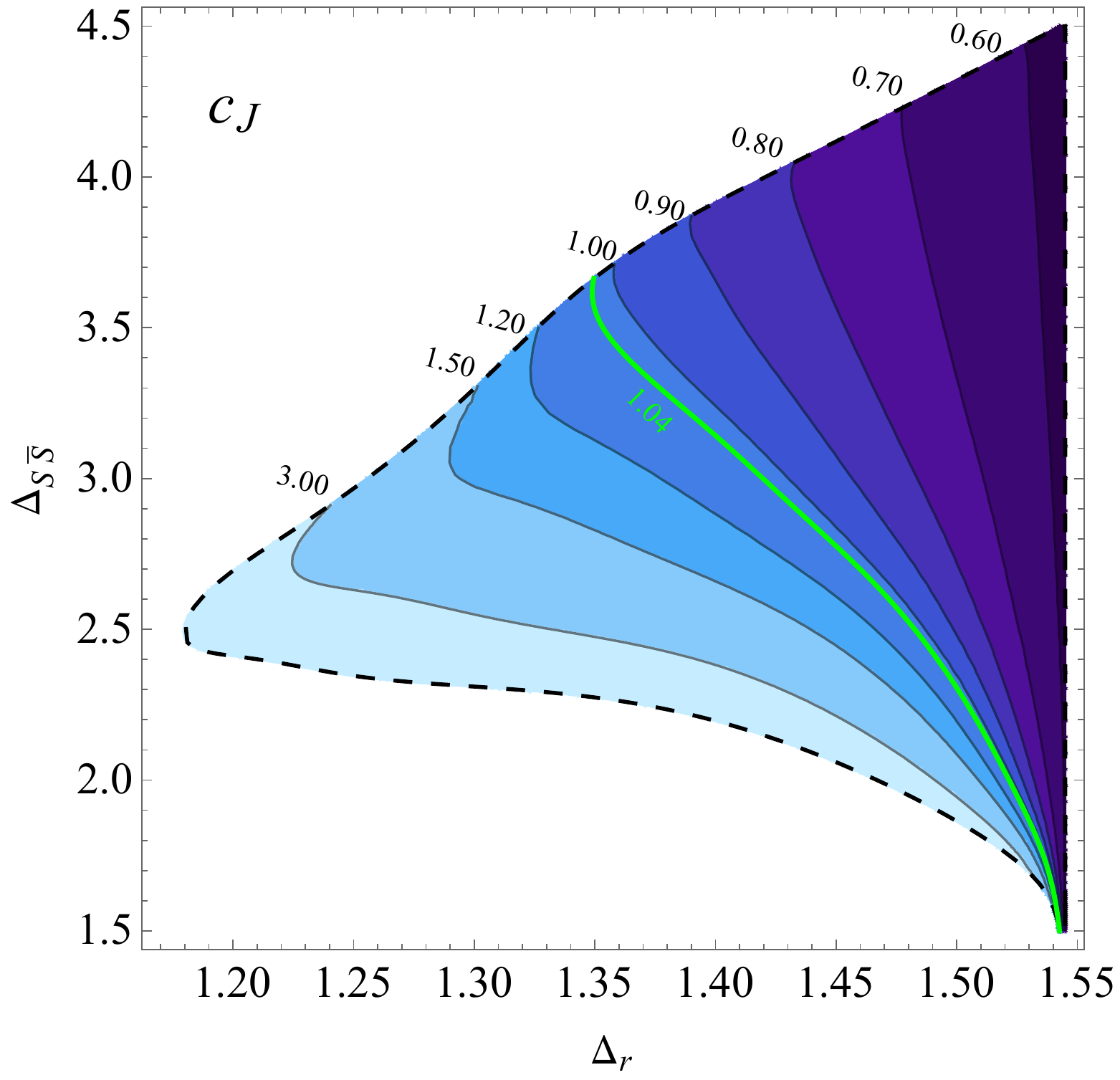}
\end{aligned}
$
\caption{Contour plots of the stress tensor central charge $c_T$ (left panel) and the $SU(4)$ conserved current central charge $c_J$ (right panel). The bounds are obtained at $\Lambda=21$ with the gap assumptions: $\Delta_R>3$ for the lowest operators in the $SU(4)$ representations $R = \{\text{singlet},S\bar{A}\}$, $\Delta_R >2.8$ for the lowest scalars in the $R = \{A\bar{A}, \text{Adj}\}$ representations, and the second lowest scalar in the $S\bar{S}$ sector is above $4.5$. The green contours denote the values of $c_T$ and $c_J$ in $N_f=4$ conformal QED$_3$ obtained from the $1/N_f$ expansion.
} \label{T2J1}
\end{figure*}

Fig.~\ref{T2J1} shows the bootstrap bounds on another two important physical quantities in CFTs, the stress tensor central charge $c_T$ and the $SU(4)$ conserved current central charge $c_J$. In the plot we have imposed the gap assumptions $\Delta>2.8$ in the $Adj$ and $A\bar{A}$ sectors and $\Delta>3$ in the singlet and $S\bar{A}$ sectors. Besides, we also assume the second $S\bar{S}$ scalar satisfies $\Delta_{S\bar{S}'}>4.5$, which leads to a lower cut in the bootstrap bound. The second lowest $S\bar{S}$ scalar has scaling dimension 6 in the large $N_f$ limit, see Table \ref{tab:largeNSpectrum}. The gap $\Delta_{S\bar{S}}'>4.5$ for the second $S\bar{S}$ scalar is slightly above the scaling dimension of the lowest $S\bar{S}$ scalar near the kink at $\Delta_{\bar{\psi}\psi}\sim1.35$.  This gap introduced a lower cut in the bound on the scaling dimension of the lowest scalar in the $S\bar{S}$ sector.
Contours denoting the $1/N_f$ perturbative results on $c_T$ and $c_J$ given in Eqs.~\ref{eq:centralcharges} are highlighted in Fig.~\ref{T2J1}, which are remarkably close to the bootstrap lower bounds on $c_T$ and $c_J$ near the kink.

The bounds on $c_T$ and $c_J$ shown in Fig.~\ref{T2J1} are especially interesting for bootstrap studies of conformal QED$_3$. A widely recognized difficulty in bootstrapping conformal gauge theories is how to distinguish theories with different gauge groups and matter representations. The conformal bootstrap focuses on gauge invariant operators, in which information about the gauge group has been obscured and the low-lying operators can be similar in different gauge theories. For instance, the $SU(N_f)$ adjoint fermion bilinear operators and four-fermion operators also exist in $SU(N_c)$ gauge theories coupled to $N_f$ fundamental fermions. Without extra constraints on the gauge interactions, it is difficult for the bootstrap algorithm to distinguish the scaling dimensions of operators in conformal QED$_3$ from those in other gauge theories. In this sense, it is not surprising that by introducing a gap on the second $S\bar{S}$ scalar, the lower region is not carved out significantly. Information about the gauge group actually appears in the central charges $c_T$ and $c_J$. The central charges can be viewed as rough measures of the number of degrees of freedom, which are significantly affected by the rank of gauge groups and their representations. 

In \cite{Giombi:2016fct} the central charges $c_T$ and $c_J$ in $SU(N_c)$ gauge theories coupled to $N_f$ fundamental fermions were computed perturbatively, which gives at leading order $N_c$ times of the central charges of $N_f$ flavor QED$_3$.  Therefore the central charges $c_T$ and $c_J$  provide critical parameters to distinguish QED$_3$ from 3d Yang-Mills theories. Going back to the bootstrap bounds on $c_T$ and $c_J$ in Fig.~\ref{T2J1}, the lower bounds on $c_T$ and $c_J$ near the kink are close to the perturbative results of QED$_3$, while significantly lower than the central charges of QCD$_3$, giving evidence that the underlying theory of the kink could be QED$_3$ or a similar $U(1)$ gauge theory. Moreover, near the lower cut caused by the gap for the second $S\bar{S}$ scalar, $c_T$ and $c_J$ have much stronger lower bounds. This region may be excluded at higher $\Lambda$ and does not clearly appear to correspond to any physical theories. Alternatively this region may contain solutions of certain Yang-Mills gauge theories with scaling dimensions $(\Delta_r, \Delta_{S\bar{S}})$ in between the kink and lower cut, which cannot be excluded by the gap $\Delta_{S\bar{S}'}>4.5$ for the second $S\bar{S}$ scalar and have central charges $c_T$ and $c_J$ significantly larger than those of QED$_3$. By inputting upper bounds on the central charges, the bootstrap solutions near the lower cut of the $\Delta_{S\bar{S}}$ allowed region can be excluded. The central charges may thus play a more efficient role in excluding Yang-Mills gauge theory solutions compared with imposing gap assumptions in the spectrum.

 In consideration of the special role that the central charges play in the bootstrap bounds, it would be very interesting to bootstrap mixed correlators between $SU(4)$ adjoint fermion bilinears and $SU(4)$ conserved currents. The roles of conserved currents in the 3d numerical bootstrap have been studied in \cite{Dymarsky:2017xzb,Dymarsky:2017yzx,Reehorst:2019pzi}. Another motivation to study mixed correlators involving the $SU(4)$ conserved currents is that they may play an interesting role in resolving the bootstrap bound coincidence caused by the algebraic relation between the crossing equations of $SU(4)$ adjoint scalars and the $SO(15)$ vector scalars (see \cite{Li:2020bnb} for more discussions). We leave this direction for future study.

\section{Monopole single correlator bootstrap revisited}
\label{sec:monopole}
As noted in the introduction, monopole operators are particularly interesting in studying conformal \qed, since the topological $U(1)_t$ symmetry provides an opportunity to distinguish conformal QED$_3$ from QCD$_3$ with $SU(N_c)$ gauge interactions, which are otherwise difficult for the bootstrap to distinguish just based on their flavor symmetries and the perturbative gauge invariant spectrum.\footnote{QCD$_3$ theories which contain $U(1)$ factors, e.g.~$U(N_c)$ QCD$_3$, also contain monopole operators charged under the topological $U(1)^t$ generated by the current $J_\mu^t=\epsilon_{\mu\nu\rho}\mathrm{tr}(F^{\nu\rho})$. Both the monopole spectrum and central charges have a strong dependence on $N_c$ \cite{Dyer:2013fja}, which can be useful for distinguishing these theories.
} 
Bootstrap studies of the monopole four-point correlator in this theory were performed previously in \cite{Chester:2016wrc, Chester:2017vdh}. The key results were that after imposing certain gaps, the bootstrap bounds show kink-like structures. Nevertheless, the kinks are gap dependent, meaning it may be hard to pin down the conformal QED$_3$ solution with the monopole bootstrap and just a few reliable and general assumptions. We will focus our attention on a less ambitious but still nontrivial task, which is to test the perturbative and lattice results of conformal QED$_3$ using the monopole bootstrap. 

Along the way, we will show an algebraic relation between the crossing equations of the four-point functions of the monopole operator $\cM_{1/2}$ and the crossing equations of the $SO(12)$ vector scalar, which in turn leads to a coincidence of bootstrap bounds between the monopole bootstrap and the $SO(12)$ vector bootstrap. We find that gaps inspired by the perturbative spectrum which take advantage of parity symmetry can play an important role in resolving this $SO(12)$ symmetry enhancement in the bootstrap bounds and in carving out peninsula structures. Based on these, we will then introduce {\it interval positivity} constraints in the bootstrap setup, with which the allowed parameter space can be further isolated into a closed island.

\subsection{Single correlator crossing equations of the monopole operator \texorpdfstring{$\cM_{1/2}$}{M1/2}}
The crossing equations for the monopole four-point correlator were computed in  \cite{Chester:2016wrc}.
The monopole $\cM_{1/2}$ with lowest $U(1)_t$ charge $q=\frac{1}{2}$ forms the $(110)$ representation of $SU(4)$. This monopole operator is not parity-definite: parity flips the sign of the $U(1)$ gauge flux and maps the monopole operator $\cM_{1/2}$ to the anti-monopole $\cM_{-1/2}$. It is convenient to rewrite the $U(1)_t$ charged monopole ($\cM_{1/2}$) and anti-monopole ($\cM_{-1/2}$) operators in an $SO(2)$ vector form $\cM_{1/2}^a$ with
\small 
\be
\cM_{1/2}^{a=1}=(\cM_{1/2}+\cM_{-1/2})/2\;,\quad \cM_{1/2}^{a=2}=-i(\cM_{1/2}-\cM_{-1/2})/2\;, 	
\ee 
\normalsize
where the $SU(4)$ indices have been assumed implicitly. Note that these are now parity-definite. Our crossing equations are of the monopole four-point correlator
\beq
\langle \cM_{1/2}^a (x_1) \cM_{1/2}^b (x_2)\cM_{1/2}^c (x_3)\cM_{1/2}^d (x_4)  \rangle\;. \label{4Ms}
\eeq
There are $9$ sectors with different $SU(4)\times SO(2)$ representations or parity charge which appear in the OPE of $\cM_{1/2}^a\times \cM_{1/2}^b$. 
We can understand the algebraic structure of the crossing equations from (\ref{4Ms}) with the tensor product of the monopole's $SU(4)$ and $SO(2)$ representations:
\begin{equation}\label{eq:repsMM}
\begin{aligned}
SU(4):\;& (110) \bigotimes (110)= (000) \bigoplus (211) \bigoplus (220)\;, \\
SO(2):\;& V \bigotimes V= S \bigoplus T \bigoplus A\;,
\end{aligned}
\end{equation}
where $V,S, T, A$ denote vector, singlet, traceless-symmetric tensor, and antisymmetric tensor representations of $SO(2)$. The $S$ and $A$ sectors are isomorphic for $SO(2)$, but they have different spin selection rules and parity charges; see Table \ref{slrule}.

\begin{table}
\centering
\caption{Spin selection rules $(\ell^\pm)$ and parity charges $(P^\pm)$ for the monopole crossing equations. There are no definite parity charges in the $T$ sectors. Sectors of the same colors correspond to the same sectors in the $SO(12)$ vector crossing equations.  \label{slrule}
}
		\begin{tabular}{cccc}
\hline\hline\\[-1em]
			& ~~\textbf{(000)}~~ & ~~\textbf{(211) }~~& ~~\textbf{(220) }~~
 \\[.1em]\hline\\[-1em]
			~$S$ & $l^+, P^+$                  & \color{red}$\ell^-,P^+ $        &  \color{blue}$\ell^+, P^+$ 
 \\[.1em]\hline\\[-1em]
		~	$A$ & \color{red}$\ell^-, P^-$   & \color{blue}$\ell^+,P^-$         &  \color{red}$\ell^-, P^-$  
 \\[.1em]\hline\\[-1em]
		~	$T$ &  \color{blue}$\ell^+$                & \color{red}$\ell^-$           &        \color{blue}$\ell^+$ 
 \\[.1em]\hline\hline 
		\end{tabular}
\end{table}

The crossing equations can be summarized by the vector equation
\beq\label{eq:MMMMCrossingEq}
\sum_{\cO,i} \lambda_\cO^2\Vec{V}_{S_i}^\pm+\sum_{\cO,i} \lambda_\cO^2\Vec{V}_{A_i}^\pm+\sum_{\cO,i} \lambda_\cO^2\Vec{V}_{T_i}^\pm=0\;,
\eeq
in which the vector $\Vec{V}_{R_i}^\pm$ has an even/odd spin selection rule and its subscript $R_i$ denotes a sector with $SO(2)$ representation $R=S/A/T$ and $SU(4)$ representation $i=(000), (211)$, or $(220)$. The vectors $\Vec{V}$ have $9$ components and the crossing equations 
\be 
\cM_{\text{monopole}}\equiv \Big( \Vec{V}^+_{S_{(000)}}, \Vec{V}^-_{S_{(211)}}, \Vec{V}^+_{S_{(220)}},\Vec{V}^-_{A_{(000)}}, \Vec{V}^+_{A_{(211)}},\\ \Vec{A}^-_{A_{(220)}},\Vec{V}^+_{T_{(000)}}, \Vec{V}^-_{T_{(211)}}, \Vec{V}^+_{T_{(220)}}\Big)_{\text{monopole}}
\ee 
can be written in a $9\times9$ square matrix form, as expected in the single correlator bootstrap with general global symmetries \cite{Rattazzi:2010yc}:
\footnotesize
\be
\cM_{\text{monopole}}=
\left(
\begin{array}{ccccccccc}
 0 & 0 & 0 & 0 & F & -F & 0 & -F & F \\
 0 & 0 & 0 & -F & -F & -\frac{2}{3}F & F & F & \frac{2}{3}F \\
 0 & -F & F & 0 & -F & F & 0 & 0 & 0 \\
 F & F & \frac{2 }{3}F & F & F & \frac{2}{3} F & 0 & 0 & 0 \\
 F & -F & -\frac{4}{3}F & -F & F & \frac{4 }{3} F& -2 F & 2 F & \frac{8 }{3} F\\
 0 & 0 & 0 & -H & H & \frac{4 }{3}H & H & -H &- \frac{4}{3}H \\
 H & -H & -\frac{4}{3} H & H & -H & -\frac{4}{3}H & 0 & 0 & 0 \\
 0 & -H & H & 0 & H & -H & 0 & 2 H & -2 H \\
 H & H & \frac{2 }{3} H& -H & -H & -\frac{2}{3} H & -2 H & -2 H & -\frac{4}{3} H \\
\end{array}
\right)\,, \label{matrix}
\ee
\normalsize
where $F/H$ are the symmetrized/anti-symmetrized conformal block functions (\ref{eq:conformal block functions}).

It turns out that there is a relation which maps the above crossing equations (\ref{matrix} onto the much simpler \(SO(12)\) vector crossing equations which was not noted in previous monopole bootstrap works \cite{Chester:2016wrc, Chester:2017vdh}.  Following the procedure discovered in \cite{Li:2020bnb}, there is a $3\times9$ matrix
\footnotesize
\be
\cT_{\text{monopole}}=\left(
\begin{array}{ccccccccc}
 1 & \frac{19}{154} & \frac{75}{154} & \frac{5}{308} & -\frac{5}{308} & 0 & 0 & 0 & 0 \\
 0 & \frac{40}{77} & \frac{12}{77} & \frac{62}{77} & \frac{15}{77} & 0 & 0 & 0 & 0 \\
 0 & 0 & 0 & 0 & 0 & \frac{5}{11} & \frac{15}{22} & \frac{1}{11} & \frac{7}{22} \\
\end{array}
\right)\,,
\ee 
\normalsize
which can transform the monopole crossing equations into the $SO(12)$ vector four-point crossing equations 
\be
 \cM_{SO(12)}= \left(\Vec{V}^+_S, \Vec{V}^+_T, \Vec{V}^-_A   \right)_{SO(12)}=
\left(
\begin{array}{ccc}
 0 & F & -F \\
 F & \frac{5 }{6}F & F \\
 H & -\frac{7}{6} H & -H \\
\end{array}
\right)\,.  \label{SO12}
\ee
Its action on the monopole crossing equations gives
\footnotesize
\begin{multline}
(\cT\cdot \cM)_{\text{monopole}}=
\\\left(
\begin{array}{ccccccccc}
	0 & -\frac{5}{11} F& \frac{40 }{77} F & -\frac{1}{11}F & \frac{30  }{77}F & -\frac{20}{33}  F & \frac{12 }{77}F & -\frac{10}{11}  F & \frac{80 }{77}F \\
	F & \frac{5 }{11}F & \frac{100 }{231}F & \frac{1}{11}F & \frac{25 }{77}F & \frac{20 }{33}F & \frac{10 }{77}F & \frac{10 }{11}F & \frac{200 }{231}F \\
	H & -\frac{5}{11}  H & -\frac{20}{33}  H & -\frac{1}{11}H & -\frac{5}{11}  H & -\frac{20}{33}  H & -\frac{2}{11}  H & -\frac{10}{11}  H & -\frac{40}{33}  H \\
\end{array}
\right)\,,  
\end{multline}
\normalsize
which can be briefly written in the vector form
\footnotesize
\begin{multline}
\hspace*{-2em}\left[\cT\cdot\left( \Vec{V}^+_{S_{(000)}}, \Vec{V}^-_{S_{(211)}}, \Vec{V}^+_{S_{(220)}},\Vec{V}^-_{A_{(000)}}, \Vec{V}^+_{A_{(211)}}, \Vec{A}^-_{A_{(220)}},\Vec{V}^+_{T_{(000)}}, \Vec{V}^-_{T_{(211)}}, \Vec{V}^+_{T_{(220)}}\right)\right]_{\text{monopole}}\\
=\left( \Vec{V}^+_{S}, ~ x_1\Vec{V}^-_{A},~ x_2\Vec{V}^+_{T},~
x_3\Vec{V}^-_{A}, ~ x_4\Vec{V}^+_{T},~ x_5\Vec{V}^-_{A},~ x_6\Vec{V}^+_{T}, ~ x_7\Vec{V}^-_{A},~ x_8\Vec{V}^+_{T}\right)_{SO(12)}\,, \label{mp-SO}
\end{multline}
\normalsize
with positive coefficients $x_i$
\beq
\Vec{x}=\left\{\frac{5}{11},~\frac{40}{77},~\frac{1}{11},~\frac{30}{77},~\frac{20}{33},~\frac{12}{77},~\frac{10}{11},~\frac{80}{77}\right\}\,.
\eeq
Therefore the transformation $\cT_{\text{monopole}}$ maps the monopole crossing equations into the $SO(12)$ vector crossing equations,
combined with the $SO(12)\rightarrow SU(4)\times SO(2)$ branching rules
\bea[eq:s012 branching rules]
\mathbf{SO(12)}& \qquad\mathbf{SU(4)\times SO(2)} \nn \\
S&\longleftrightarrow S_{(000)}\;,  \label{branching1}\\
T&\longleftrightarrow S_{(220)}\bigoplus A_{(211)}\bigoplus T_{(000)}\bigoplus T_{(220)}\;, \label{branching2} \\
A&\longleftrightarrow  S_{(211)}\bigoplus A_{(000)}\bigoplus A_{(220)}\bigoplus T_{(211)}\;. \label{branching3}
\eea
Note that only even (odd) spins appear in the $S, T$ ($A$) sectors of $SO(12)$, consistent with the spin selection rules of the different $SU(4)\times SO(2)$ representations shown in Table \ref{slrule}.

Positivity of $x_i$ implies that the coefficients in the $N_f=4, q=1/2$ monopole crossing equations have the same positivity properties as in the $SO(12)$ vector crossing equations. 
This agrees with the results in \cite{Li:2020bnb, Li:2020tsm}, that in general for a scalar in a representation with $N^*$ degrees of freedom, its four-point crossing equations relate to the $SO(N^*)$ vector crossing equations through a unique linear transformation. 
As proved in \cite{Li:2020bnb} and the section \ref{creqid} of this paper, this relation combined with suitable boundary conditions can lead to coincidences between the monopole and $SO(12)$ vector bootstrap bounds. 
Indeed one can show that the bootstrap bound on the lowest non-unit scalar in the $S_{(000)}$ sector coincides with the singlet bound in $SO(12)$ vector bootstrap. 
Such a bound coincidence can be broken using non-$SO(N^*)$ symmetric boundary conditions in the bootstrap implementation. 

It is very interesting to compare the branching rules in the monopole crossing equations (\ref{eq:s012 branching rules}) with those in the $SU(4)$ adjoint fermion bilinear crossing equations (\ref{eq:branching-rules}). A major difference is that in (\ref{eq:branching-rules}) all the operators on the RHS are parity even, while in (\ref{eq:s012 branching rules}), the $SU(4)\times SO(2)$ representations branched from $SO(12)$ $A$ or $T$ sectors contain both parity even and parity odd representations, as well as $T_{\Vec{x}}$ monopole sectors without a definite parity charge. Specifically, the lowest scalar in the $S_{(220)}$ sector is a parity even four-fermion operator while the lowest scalar in the $A_{(211)}$ sector is just the parity odd fermion bilinear $r$, which have quite different scaling dimensions. The lowest scalars in the $T_{(000)}$ and $T_{(220)}$ sectors also have rather different scaling dimensions at leading order, see Table \ref{tab:largeNSpectrum}. Similar differences appear in the branching rule of the $SO(12)$ $A$ sector (\ref{branching3}). This is different from the fermion bilinear $r$ crossing equations (\ref{eq:branching-rules}), in which the $SO(15)$ symmetry enhancement is broken only at the subleading order $O(1/N_f)$. Therefore, the monopole crossing equations perhaps provide the strongest way to break the $SO(N)$ symmetry enhancement appearing in bootstrap studies for gauge theories with smaller symmetry.

Based on the above facts, it is possible to introduce highly restrictive gap assumptions in the QED$_3$ monopole bootstrap. Perturbative calculations can provide valuable guidance on the possible gaps in different sectors. However, one needs to use this information carefully as the CFT data may receive notable higher order corrections. On the other hand, the monopole bootstrap can provide a nonperturbative check on whether the perturbative (or lattice) results can be consistent with conformality and unitarity.

\subsection{Monopole bootstrap bounds with gaps inspired by \texorpdfstring{QED$_3$}{QED3} spectrum}
In this section we explore bootstrap constraints from the crossing equations of the four-point correlator  $\langle \cM_{1/2}\cM_{1/2}\cM_{1/2}\cM_{1/2}\rangle$. The symmetry enhancement of the crossing equations (\ref{mp-SO}) strongly affects the monopole bootstrap bounds. Both singlet and non-singlet bounds coincide with the $SO(12)$ vector bootstrap results unless the symmetry is strongly broken by gap assumptions. However, interesting bootstrap results can be obtained after introducing gap assumptions inspired by the perturbative spectrum of QED$_3$, shown in Table \ref{tab:largeNSpectrum}.

\begin{figure}
\centering
\hspace*{-1.5em}
$\begin{aligned}
\includegraphics[width=0.53\linewidth]{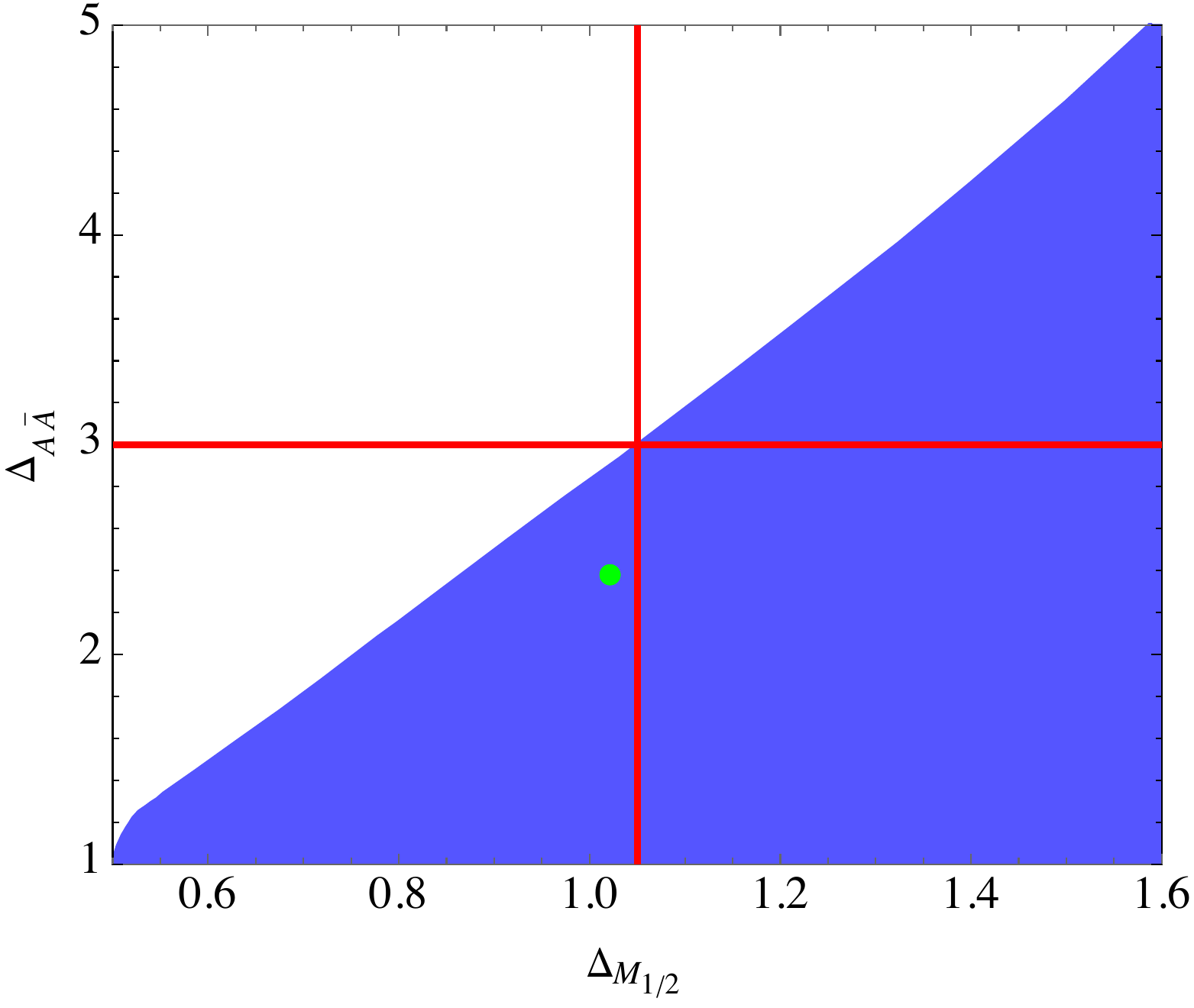}
\includegraphics[width=0.53\linewidth]{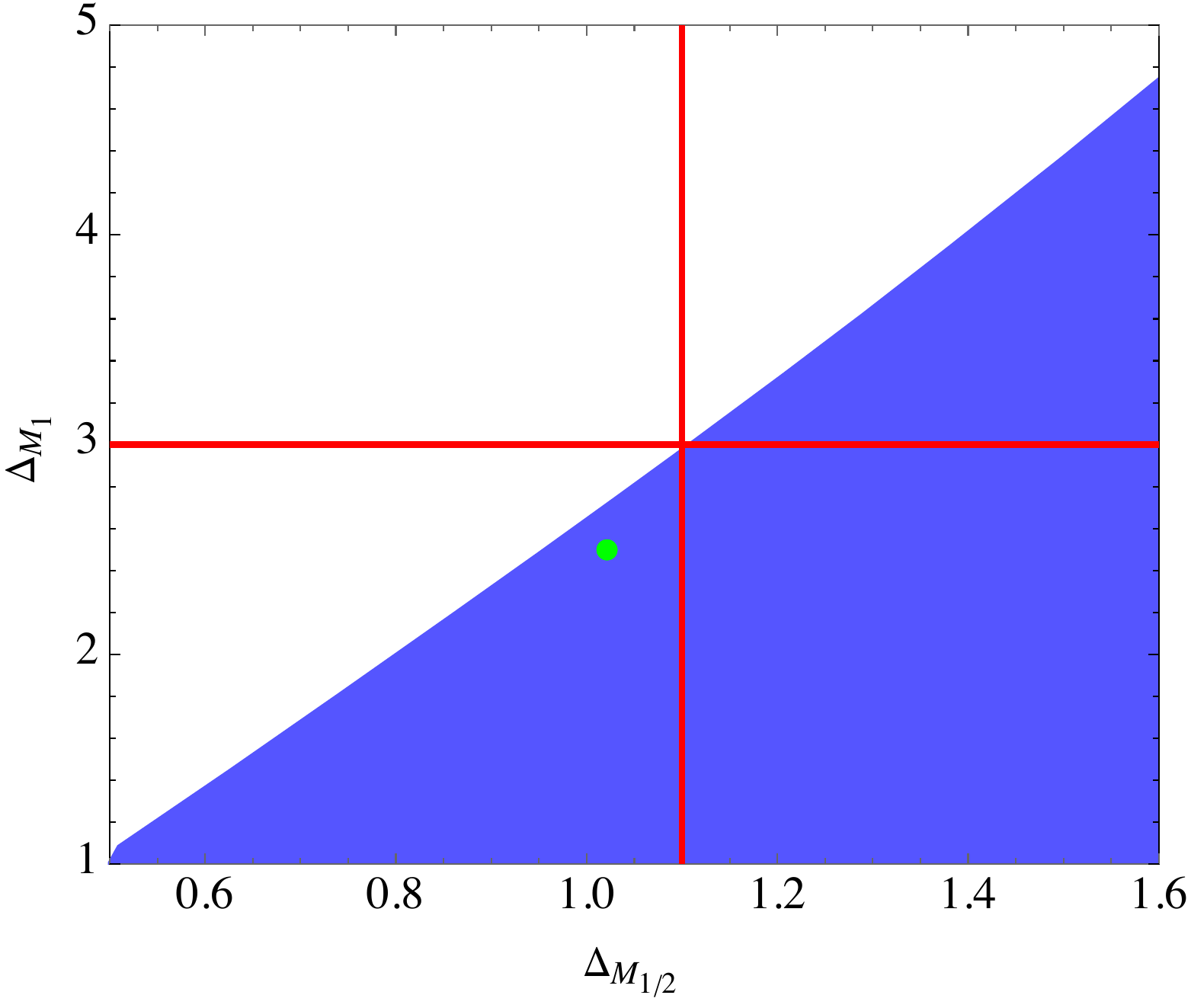}
\end{aligned}$
\caption{
Bounds on the scaling dimensions of the lowest scalar in the $S_{(220)}$ sector (left) and the charge 1 monopole in the $T_{(220)}$ sector (right) at $\Lambda=31$.
The green dots denote the $1/N_f$ perturbative predictions.
The axes highlighted with red color are positioned at the intersection of the bounds and $\Delta_{A\bar{A}},\Delta_{M_1} = 3$. Note that the perturbative results are ruled out if we assume $\Delta\geq 3$ in either sector.
} \label{mp:aab-T2}
\end{figure}

In Fig.~\ref{mp:aab-T2}, we show the bootstrap bounds on the scaling dimensions of the lowest scalars in the $S_{(220)}$ (left panel) and $T_{(220)}$ (right panel) sectors  without imposing any gap assumptions.
The bootstrap bounds are close to straight lines in the regions away from the unitary bound $\Delta=1/2$. 
A direct consequence is that by imposing a gap $\Delta^*$ for the lowest   scalar in $S_{(220)}$ sector: $\Delta_{S_{(220)}}\geqslant \Delta^*$, there will be a minimal $\Delta_{\cM_{1/2}}$  in  the bootstrap allowed region proportional to the gap $\Delta^*$.
This explains the $S_{(220)}$-gap dependent bootstrap bounds observed in \cite{Chester:2016wrc}.
The red dots in Fig.~\ref{mp:aab-T2} denote the $1/N_f$ perturbative results, which locate in the physically allowed regions and are well consistent with the bootstrap bounds without imposing any gap assumptions.

The lowest scalar in the $S_{(220)}$ sector is the four-fermion operator with scaling dimension $\Delta\simeq2.4$ at subleading order in the $1/N_f$ expansion, see Table \ref{tab:largeNSpectrum}. Its scaling dimension is expected to receive notable corrections from higher order terms. 
An interesting question is whether this operator is relevant or not. Assuming the lowest scalar in the $S_{(220)}$ sector is irrelevant, the bootstrap bound in Fig.~\ref{mp:aab-T2} introduces a lower cut on the scaling dimension of the monopole $\cM_{1/2}$: $\Delta_{\cM_{1/2}}>1.05$ at $\Lambda=31$. This is consistent with the lattice result \cite{Karthik:2019mrr} but excludes the perturbative prediction at subleading order $\Delta_{\cM_{1/2}}\simeq 1.022$. We do not have solid evidence on the relevance of the lowest scalar in the $S_{(220)}$ sector and we will adopt a weaker gap assumption in the $S_{(220)}$ sector with which the perturbative result on $\Delta_{\cM_{1/2}}$ is still in the allowed region of the bootstrap bound.  

Due to the gap dependence of the bootstrap bound, it is likely too much to hope that our current bootstrap setup can {\it solve} the $N_f=4$ conformal QED$_3$ as a special solution saturating the bootstrap bound. However, it is still interesting to know whether by imposing gaps inspired by the perturbative monopole spectrum, will the bootstrap bounds converge to the region near perturbative CFT data of $N_f=4$ conformal QED$_3$ or completely exclude it? 
In the monopole spectrum,
the subleading order corrections on the scaling dimensions of the low-lying monopole operators have been shown to be small: only $3.6\%$  ($7.2\%$ ) of the leading term for $\cM_{1/2}$ ($\cM_{1}$). If this is also true for higher order corrections, i.e.~the large $N_f$ expansion is still converging, then the current perturbative results should be close to the physical spectrum. In contrast, subleading order corrections of the four-fermion operators are more significant and the perturbative results have been shown in Fig.~\ref{ssbbd} to be not reliable. The readers should be reminded that our assumptions on the gap $2.8$ in the $S_{(220)}$ sector and the monopole spectrum have not been strictly established yet and the bootstrap computations should be considered as numerical experiments before more solid evidence on these assumptions can be obtained.

\begin{figure}
\centering
\hspace*{-1em}			\includegraphics[width=1\linewidth]{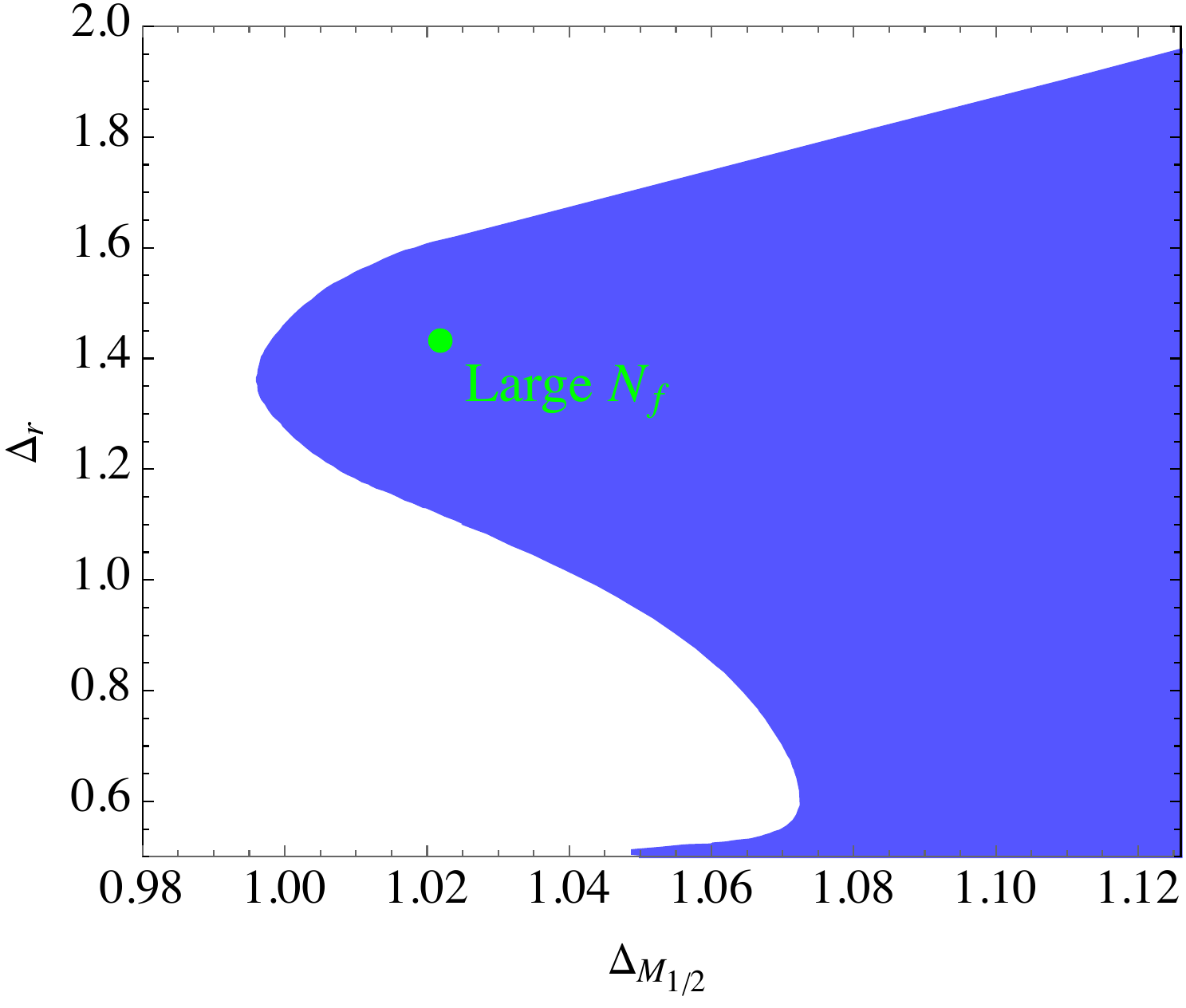}
    \rlap{\hspace{-2.9in}\raisebox{0.9in}{ 
        $
		\begin{aligned}
			\Delta_{S_{(000)}}^{\ell=0} &\geq 3.0 \\
			\Delta_{S_{(220)}}^{\ell=0} &\geq 2.8 \\
			\Delta_{r'} &\geq 3.0 \\
			\Delta_{T_{(000)}}^{\ell=0} &\geq 4.0
		\end{aligned}
		$}}
	\caption{
		Bootstrap bound on the scaling dimensions of the monopole $\cM_{1/2}$ and adjoint fermion bilinear operator $r$ at $\Lambda=31$. To get this bound, we assumed the scaling dimensions of the lowest parity even singlet scalar and second parity odd $SU(4)$ adjoint scalar are irrelevant, the lowest scalar in the $A_{(220)}$ sector is above $\Delta>2.8$, and the lowest scalar in the $T_{(000)}$ sector has scaling dimension $\Delta>4.0$. The green dot denotes the large $N_f$ expansion estimate. 
	} \label{mp:r-gaps}
\end{figure}

In Fig.~\ref{mp:r-gaps} we show the bootstrap bound on the scaling dimension of the lowest parity odd $SU(4)$ adjoint scalar $r$ in the $A_{(211)}$ sector. To obtain the result, we have imposed gaps $\Delta_{S_{(000)}}\geq3.0$, $\Delta_{S_{(220)}}\geq2.8$, $\Delta_{T_{(000)}}\geq4.0$, and $\Delta_{A'_{(211)}}\geq3.0$ for the second lowest scalar in the $A_{(211)}$ sector. In the conformal phase of QED$_3$, the lowest parity even singlet scalar is expected to be irrelevant. The gap $\Delta_{S_{(220)}}\geq2.8$ is  weaker than the marginality condition and it can generate a lower cut on $\Delta_{\cM_{1/2}}$ below the perturbative result $1.022$. The gap in $\Delta_{T_{(000)}}$ can affect the upper bound in Fig.~\ref{mp:r-gaps}.  A weaker gap in this sector gives a higher upper bound on $\Delta_r$. According to the $1/N_f$ expansion results in Table \ref{tab:largeNSpectrum}, the leading order result gives $\Delta_{T_{(000)}}\simeq 4.42$, so the gap $\Delta_{T_{(000)}}\geqslant 4.0$ actually assumes the higher order corrections will not reduce the scaling dimension drastically. The next scalar in the parity odd $A_{(211)}$ sector can be constructed by contracting the spin indices of the $SU(4)$ conserved current and the topological $U(1)_t$ conserved current $J^{f}_{\mu}J^{t\mu}$, which has scaling dimension 4 in the large $N_f$ limit. We assume this operator remains irrelevant at $N_f=4$. The gap in this sector can affect the lower bound on $\Delta_r$. 

The three sectors with gaps, $S_{(220)}$, $A_{(211)}$, and $T_{(000)}$, together with the isolated operator $r$, appear in the $SO(12)\rightarrow SU(4)\times SO(2)$ branching rule (\ref{branching2}). In the physical spectrum of $N_f=4$ QED$_3$, the lowest scalars in these four sectors have rather different scaling dimensions, as they carry different charges under the parity symmetry. Therefore, the  spectrum in the monopole bootstrap strongly breaks the enhanced $SO(12)$ symmetry in the algebraic relation (\ref{branching2})!   
In contrast, in the $SU(4)$ adjoint crossing equations,  all the operators appearing in the $SO(15)\rightarrow SU(4)$ branching rule (\ref{branch2}) are parity even and the lowest operators in these sectors have the same scaling dimensions at the leading order. The $SO(15)$ symmetry is only broken mildly by the higher order $1/N_f$ corrections. 

\begin{figure*}
	\centering
	$\begin{aligned}
		\includegraphics[width=0.4\linewidth]{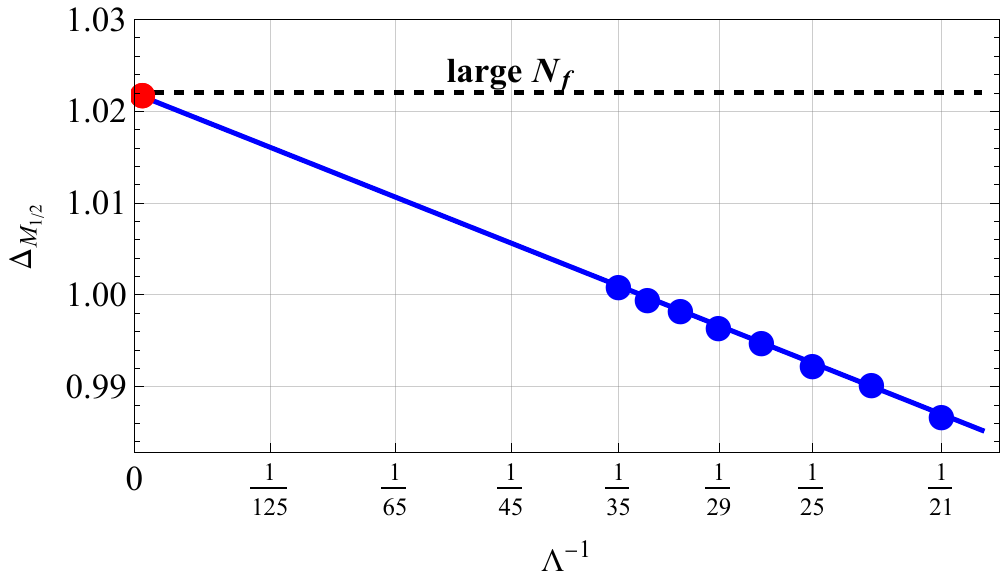} \quad
		\includegraphics[width=0.4\linewidth]{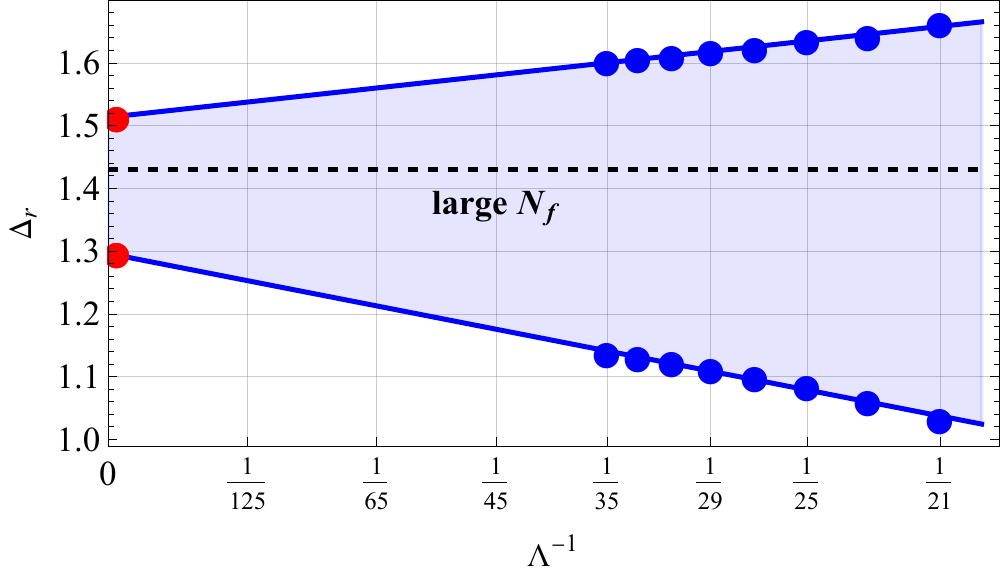} \\
		\includegraphics[width=0.4\linewidth]{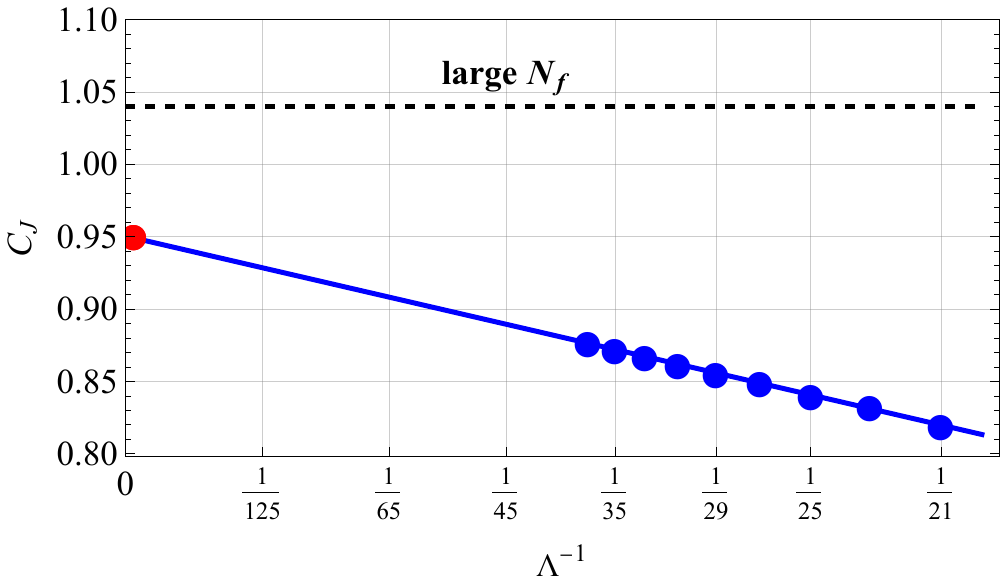} \quad
		\includegraphics[width=0.4\linewidth]{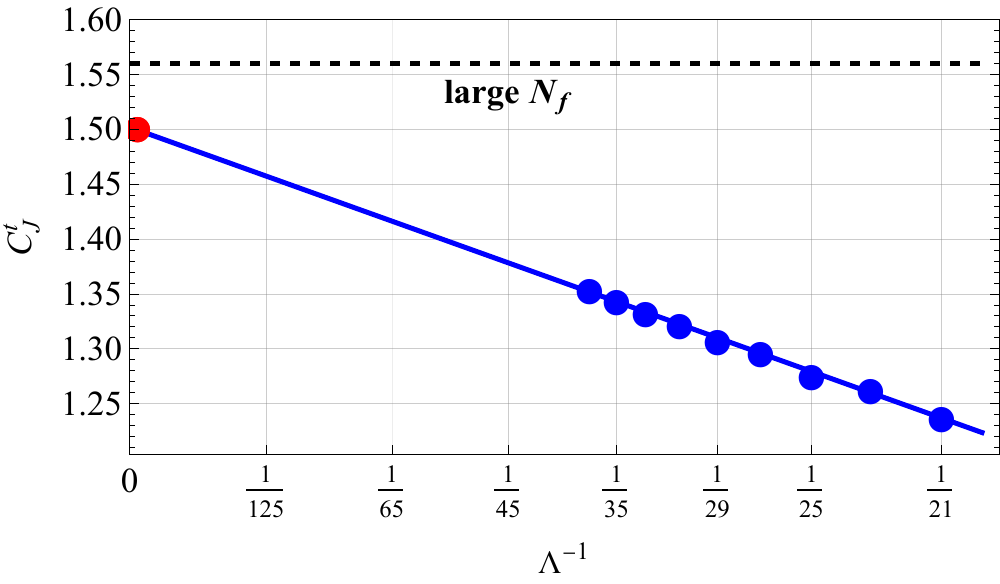}
	\end{aligned}$
	\caption{
		\textbf{Upper left: }Extrapolations of the bootstrap bounds on scaling dimensions of $\Delta_{M_{1/2}}$ with fixed $\Delta_r=1.43$.
		\textbf{Upper right: }Extrapolations of the bootstrap bounds on scaling dimensions of $\Delta_r$ with fixed $\Delta_{\cM_{1/2}}=1.022$.
		\textbf{Lower left: }Extrapolations of the central charge of the $SU(4)$  conserved current at fixed $\Delta_{\cM_{1/2}}=1.022$ and $\Delta_r=1.43$.
		\textbf{Lower right: }Extrapolations of the central charge of topological $U(1)_t$  conserved current at fixed $\Delta_{\cM_{1/2}}=1.022$ and $\Delta_r=1.43$.
		The red dots denote the extrapolation's prediction at $\Lambda=\infty$, and black dashed lines denote the large-$N_f$ perturbation theory prediction.
	} \label{extraplt1}
\end{figure*}

With the above gap assumptions, the bootstrap bound on the scaling dimension of the lowest parity odd $SU(4)$ adjoint scalar forms an interesting peninsula structure and the $1/N_f$ expansion results locate near the tip of the peninsula. Due to the special role of parity symmetry, the monopole bootstrap combined with gap assumptions inspired by the large $N_f$ QED$_3$ spectrum is more effective at carving out the CFT parameter space as compared with the $SU(4)$ fermion bilinear bootstrap shown in Fig.~\ref{T2J1}. 

We would like to make two remarks about the results in Fig.~\ref{mp:r-gaps}. Firstly the bound has a clear gap-dependence. The boundary in different directions is determined by the gaps in certain sectors. Due to this fact, our current bootstrap setup cannot be used to {\it solve} the target theory without extra specific input. We think this is a general problem for the bootstrap studies of non-$SO(N)$ vector scalars with scaling dimensions notably above the unitary bound. Another fact that the readers should keep in mind is that though we have pushed the bootstrap numerical precision to $\Lambda=31$, the bound is far from converged. This fact can be seen in Fig.~\ref{extraplt1}. In the top two panels of Fig.~\ref{extraplt1}, we show the extrapolations of the bootstrap bounds at different $\Lambda$ with fixed $\Delta_r=1.43$ or $\Delta_{\cM_{1/2}} = 1.022$. 
It requires much higher $\Lambda$ to have the lower or upper bounds close to the optimal bounds in the linear extrapolations.  In the lower two panels of Fig.~\ref{extraplt1}, we show the extrapolations of the lower bounds on the central charges of the $SU(4)$ conserved current and the topological $U(1)_t$ conserved current with fixed $\{\Delta_{\cM_{1/2}},\Delta_r\}=\{1.022, 1.43\}$.
 Interestingly, the $SU(4)$ conserved current central charge bound has a large $\Lambda$ extrapolation at $c_J\simeq 0.95$, not far from the $1/N_f$ perturbative prediction $c_J\simeq 1.04$. Similarly, the $U(1)_t$ conserved current central charge has a large $\Lambda$ extrapolation at $c_J^t\simeq 1.50$, and  the $1/N_f$ expansion at subleading order predicts $c_J\simeq 1.56$. Extrapolation of the stress tensor central charge goes to $c_T\simeq 0.89$ at large $\Lambda$, which is somewhat lower than the $1/N_f$ expansion result $c_T\simeq 1.18$. This is consistent with the observation that the bootstrap bounds in the singlet sectors are relatively weaker than those in the non-singlet sectors.

\subsection{Closed islands from monopole single correlator bootstrap with interval positivity assumptions} \label{islands}
In the last section we have shown that non-trivial peninsula structures show up if we break the $SO(12)\rightarrow SU(4)\times U(1)_t$ symmetry enhancement by physically inspired gap assumptions. There we did not impose any assumptions on the spectrum in the $T_{(220)}$ sector. 
As a part of the branching rule (\ref{branching2}), the spectrum in this sector also plays an important role in the monopole bootstrap.

The $N_f=4$ QED$_3$ spectrum in the $T_{(220)}$  sector is shown in Table \ref{tab:largeNSpectrum}. According to the $1/N_f$ expansion at sub-leading order, the lowest charge 1 monopole operator in the $T_{(220)}$ 
sector has scaling dimension $\Delta_{\cM_1}\simeq 2.5$. An interesting fact is 
that the second scalar in this sector has a significantly higher scaling dimension at leading order $\Delta_{\cM'_1}^0\simeq 6.16$. To take advantage of this big gap while still allowing uncertainty about the precise value of $\Delta_{\cM_1}$, we employ an {\it interval positivity} assumption, namely, we assume an upper bound on the dimension of the lowest charge 1 monopole operator, $\Delta_{\cM_1}\leqslant\Delta_{\cM_1}^{\rm max}$, together with a lower bound on the dimension of the next operator in the same channel $\Delta_{\cM'_1}\geqslant\Delta_{\cM_1'}^{\rm min}>\Delta_{\cM_1}^{\rm max}$. Assumptions of this type can be efficiently studied with a modification to the bootstrap algorithm, see Appendix \ref{sec:software} for more details.

We refer to the perturbative results given in Table \ref{tab:largeNSpectrum} when making assumptions on $\Delta_{\cM_1}^{\rm max}$ and $\Delta_{\cM_1'}^{\rm min}$. Specifically, we take $\Delta_{\cM_1'}^{\rm min}=5.0<\Delta_{\cM'_1}^0\simeq 6.16$, and will test gaps $\Delta_{\cM_1}^{\rm max}=2.5, 2.6$, which are inspired by the $1/N_f$ expansion result $\Delta_{\cM_1}\simeq 2.5$.
The  {\it interval positivity} assumptions can provide surprisingly strong constraints on the CFT data. We will then compare the bootstrap results with the perturbative and lattice CFT data of $N_f=4$ QED$_3$.

\begin{figure}
\centering
\hspace*{-1em}	
\includegraphics[width=0.89\linewidth]{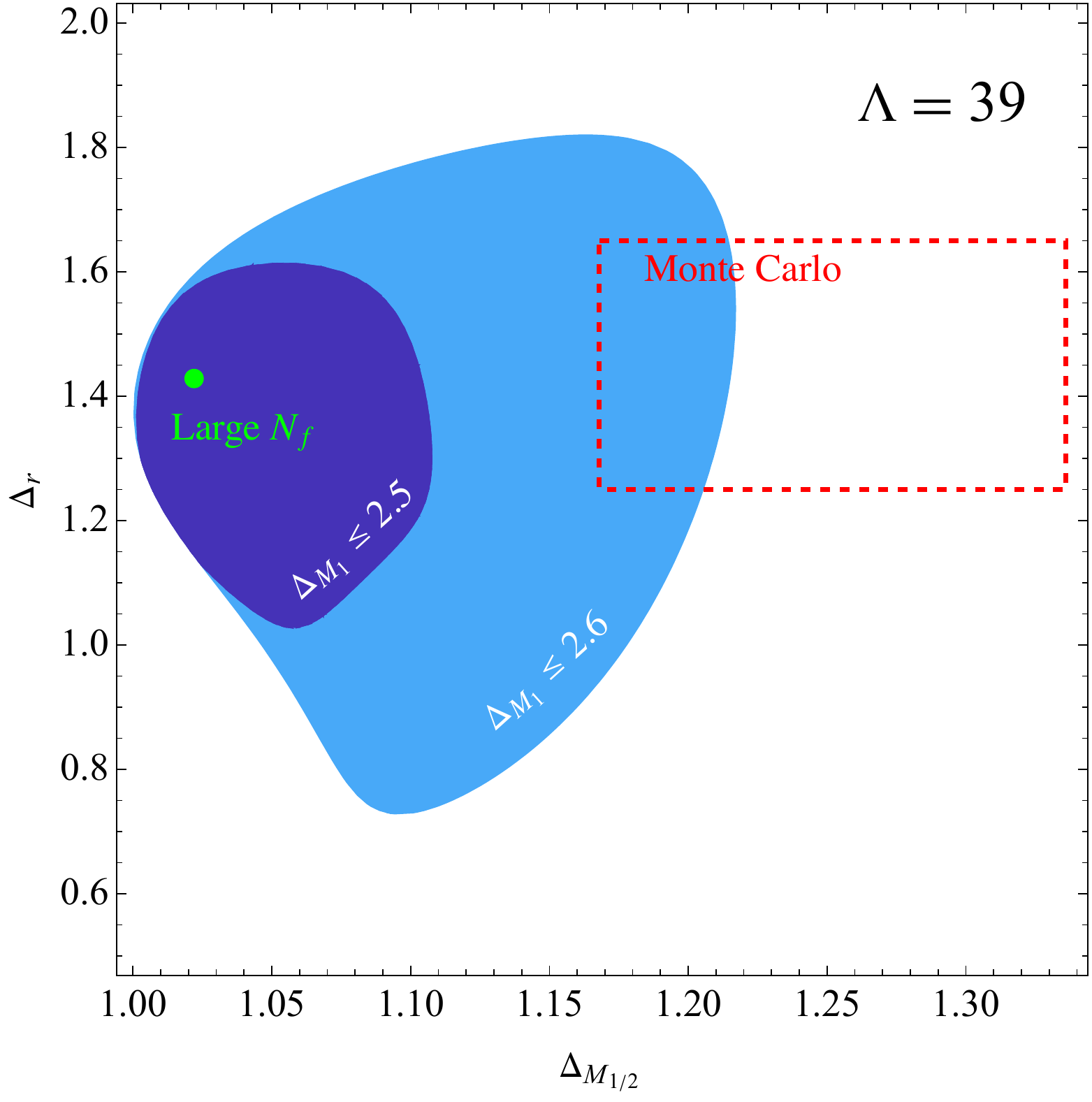}
\rlap{\hspace{-0.9in}\raisebox{1.0in}{ $ 
\begin{aligned}
	\Delta_{S_{(000)}}^{\ell=0} &\geq 3.0 \\
	\Delta_{S_{(220)}}^{\ell=0} &\geq 2.8 \\
	\Delta_{r'} &\geq 3.0 \\
	\Delta_{T_{(000)}}^{\ell=0} &\geq 4.0 \\
	\Delta_{\cM_1'} &\geq 5.0
\end{aligned}$ }}

\caption{Bounds on the scaling dimensions of $(\Delta_{\cM_{1/2}}, \Delta_r)$ with the same gaps as in Fig.~\ref{mp:r-gaps} along with the interval positivity assumptions: $\Delta_{\cM'_1}\geqslant 5.0$ and $\Delta_{\cM_1}\leqslant 2.5, 2.6$. We used $\Lambda=39$ in the bootstrap computations.
} \label{islands-interval}
\end{figure}

\begin{figure}
\centering
$\begin{aligned}
\includegraphics[width=0.75\linewidth]{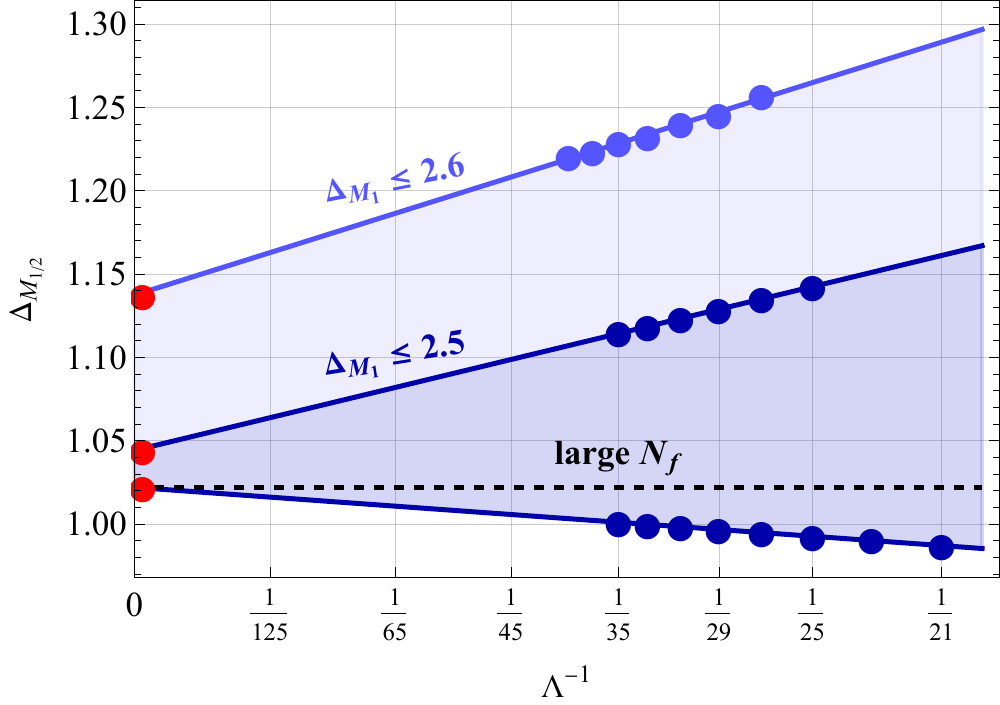}\\
\includegraphics[width=0.75\linewidth]{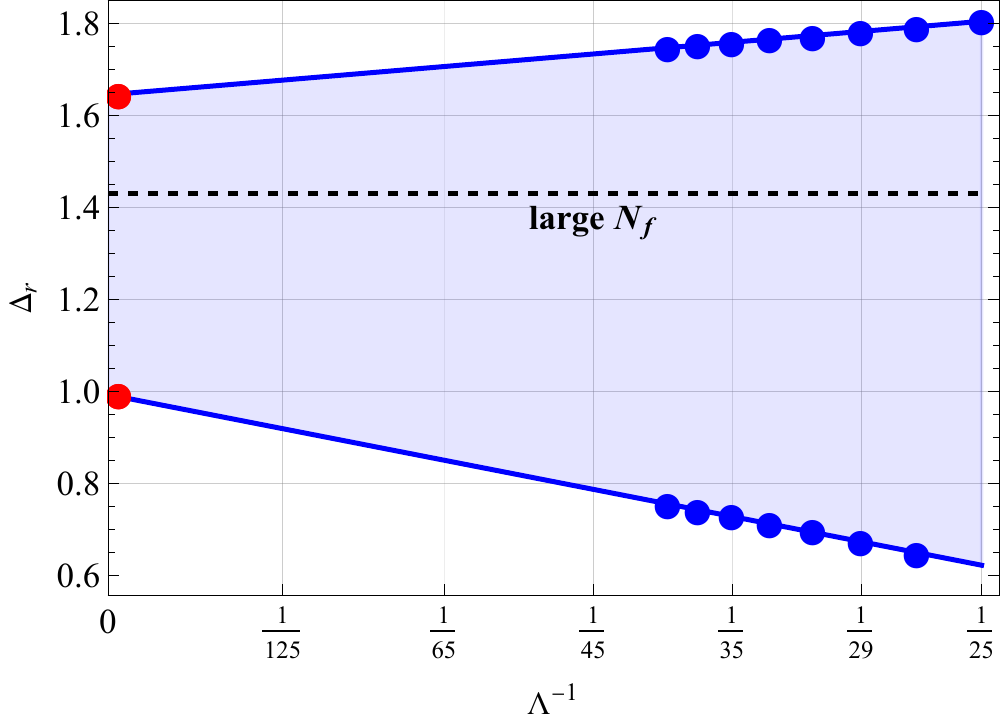}
\end{aligned}$
\caption{\textbf{Top panel:} extrapolations of the maximum $\Delta_{\cM_{1/2}}$ at fixed $\Delta_r=1.43$ in the islands with gaps $\Delta_{\cM_1}\leqslant 2.5, 2.6$. \textbf{Bottom panel:} extrapolations of the upper and lower bounds on the scaling dimension $\Delta_r$ in the island with $\Delta_{\cM_1}^{\rm max}=2.6$, at fixed $\Delta_{\cM_{1/2}}=1.08$, which is the center of the range $\Delta_{\cM_{1/2}}\in (1.02, 1.14)$ obtained from the large $\Lambda$ extrapolation. The red dots denote the $1/N_f$ perturbative results for $\Delta_{\cM_{1/2}}$ and $\Delta_r$.
} \label{lsand-r-cut-ext}
\end{figure}

Bootstrap results with these different interval positivity assumptions are shown in Fig.~\ref{islands-interval}. Remarkably, with these gap assumptions inspired by the perturbative $N_f=4$ QED$_3$ spectrum, the CFT data $(\Delta_{\cM_{1/2}}, \Delta_r)$ can be restricted into closed islands! 
The shapes of the islands are gap-dependent, and become very small if we take $\Delta_{\cM_1}^{\rm max}=2.4$ and disappear with smaller $\Delta_{\cM_1}^{\rm max}$. The island is still closed at $\Delta_{\cM_1}^{\rm max}=2.65 \,(\Lambda=31)$, extending to a maximum $\Delta_{\cM_{1/2}}\simeq 1.4$.
Note that the bounds shown in the plot are computed with relatively high numerical precision ($\Lambda=31$), however, they are not well converged yet and are actually affected by the issue of slow convergence. This can be qualitatively seen through the linear extrapolation of the bound to the large $\Lambda$ limit. In Fig.~\ref{lsand-r-cut-ext}, we show the maximum values of $\Delta_{\cM_{1/2}}$ at fixed $\Delta_r=1.43$ in the islands computed at different values of $\Lambda$, and their linear extrapolation to $\Lambda=\infty$. Surprisingly, if we set the gap $\Delta_{\cM_1}^{\rm max}$ at the perturbative estimate $\Delta_{\cM_1}^{\rm max}=2.5~(\simeq \Delta_{\cM_1})$, the upper bound on $\Delta_{\cM_{1/2}}$ extrapolates to $\Delta_{\cM_{1/2}}\simeq1.04$, close to the perturbative result $\Delta_{\cM_{1/2}}\simeq1.02$. The left part of the island coincides with the tip of the peninsula structure in Fig.~\ref{mp:r-gaps}, in which the minimum $\Delta_{\cM_{1/2}}$ with fixed $\Delta_r=1.43$ extrapolates to $\Delta_{\cM_{1/2}}\simeq1.02$, as shown in Fig.~\ref{extraplt1}.
Therefore with the interval positivity assumptions $\Delta_{\cM_{1}}\leqslant 2.5$, $\Delta_{\cM'_{1}}\geqslant 5.0$, our bootstrap implementation gives a closed island in $(\Delta_{\cM_{1/2}}, \Delta_r)$, which shrinks to a rather small region consistent with the perturbative predictions. 

Here we would like to remind the readers that the gap $\Delta_{A_{(220)}} \geqslant 2.8$, which effectively determines the minimum $\Delta_{\cM_{1/2}}$, is chosen by hand (but without tuning), and the agreement between the linear extrapolation of the left edge of the bootstrap result and the perturbative result could be considered accidental. On the other hand, the assumed maximum value $2.5$ for $\Delta_{\cM_1}$, which affects the maximum $\Delta_{\cM_{1/2}}$ in the island, is coming from the perturbative result at subleading order. These gaps together conspiratorially restrict the CFT data close to the perturbative QED$_3$ spectrum. If we relax the maximum of $\Delta_{\cM_1}$ to $2.6$, the left part of the closed island remains the same, while its right side increases to $\Delta_{\cM_{1/2}}\simeq 1.25$ ($\Lambda=31$), which overlaps with the lattice results enclosed by the red dashed rectangle in Fig.~\ref{islands-interval}. However, the right part of the island shrinks a lot at higher $\Lambda$. The linear extrapolation of the maximum $\Delta_{\cM_{1/2}}$ at fixed $\Delta_r=1.43$ in the islands gives the estimate $\Delta_{\cM_{1/2}}\simeq 1.14$ at $\Lambda=\infty$, which marginally excludes the lattice results.

\begin{figure*}
	\centering
$
	\begin{aligned}
		\Delta_{S_{(000)}}^{\ell=0} \geq 3.0, ~
		\Delta_{S_{(220)}}^{\ell=0} \geq 2.8, ~
		\Delta_{r'}\geq 3.0,
		\Delta_{T_{(000)}}^{\ell=0} \geq 4.0, ~
		\Delta_{\cM_1} \leq 2.6, ~
		\Delta_{\cM_1'} \geq 5.0 \hspace{4cm}
 \\
		\includegraphics[width=0.33\linewidth]{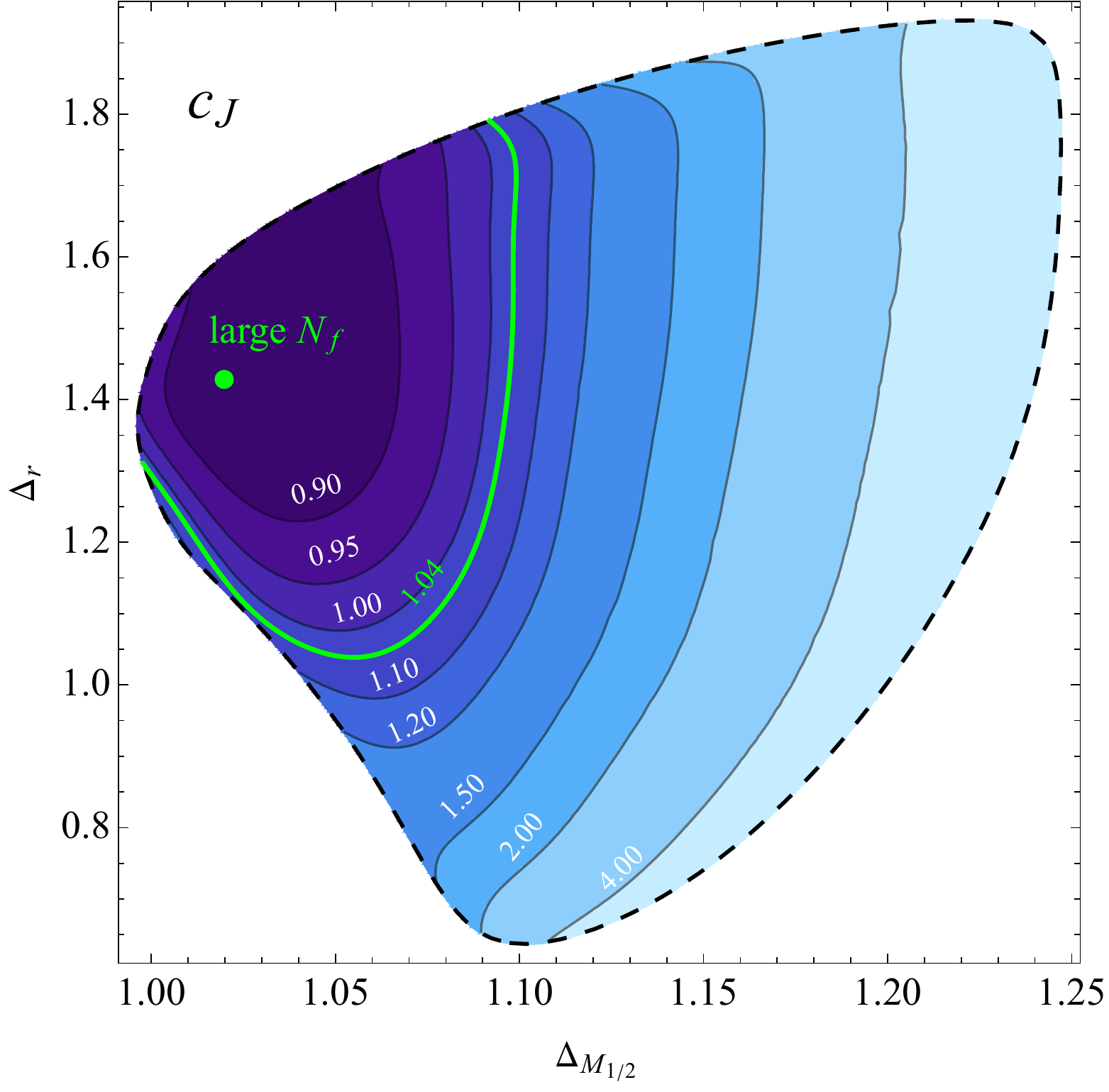}
		\includegraphics[width=0.33\linewidth]{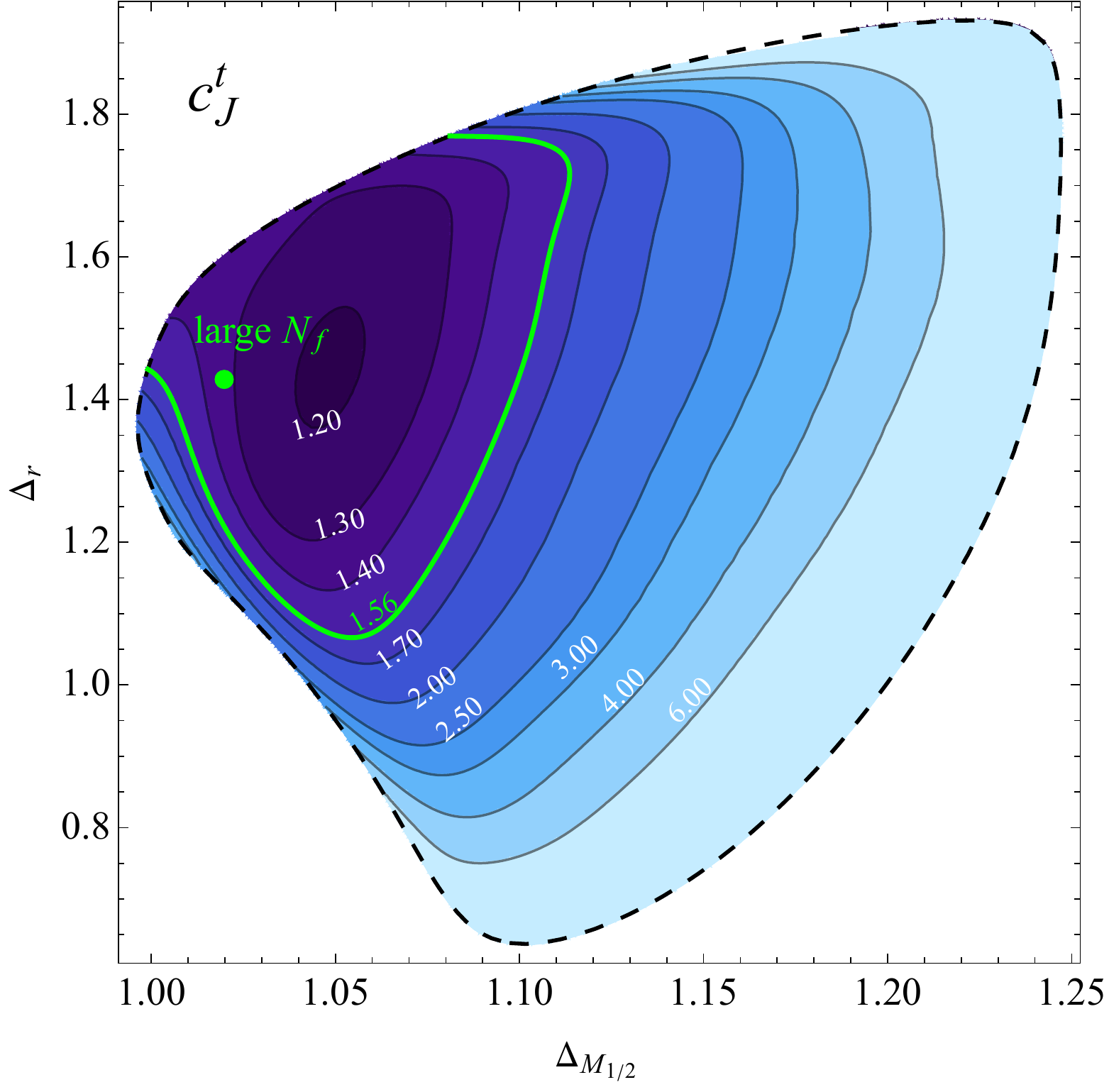}
		\includegraphics[width=0.33\linewidth]{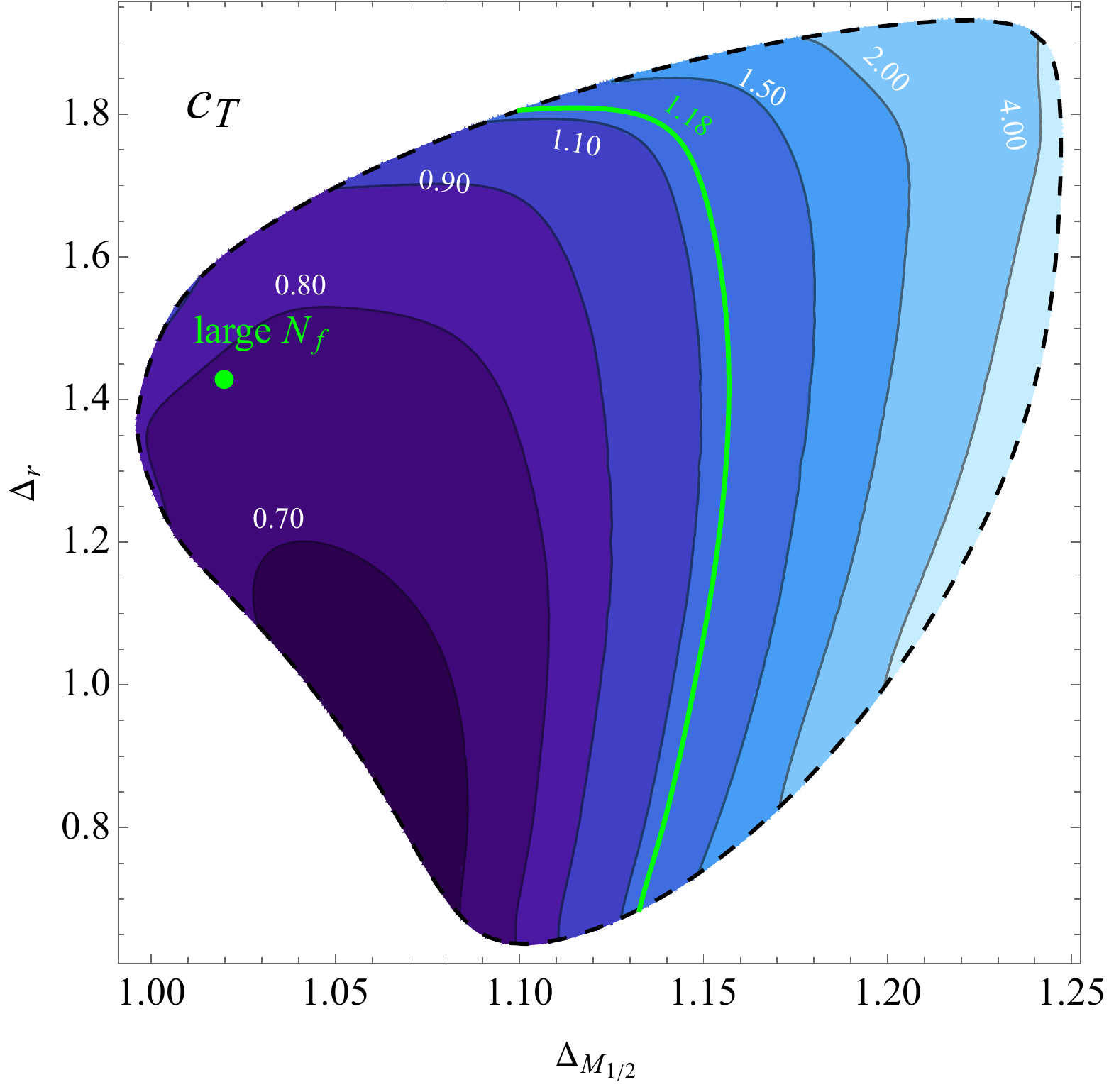}
	\end{aligned}
	$
	\caption{Lower bounds ($\Lambda=31$) on the $SU(4)$ conserved current central charge $c_J$ (left), topological $U(1)_t$ conserved current central charge $c_J^t$ (middle), and the stress tensor central charge $c_T$ (right), inside the island of Fig.~\ref{islands-interval} with the interval positivity assumption $\Delta_{\cM_1}^{\rm max}=2.60$ (other gap assumptions are the same as in Fig.~\ref{islands-interval}). The green contours denote the $1/N_f$ perturbative results at subleading order: $c_J=1.04$, $c_J^t=1.56$, and $c_T=1.18$. For the $(\Delta_{\cM_{1/2}},\Delta_{r})$ inside the green contours, the bounds on the central charges are consistent with the perturbative results, while in the right part of the island, they are significantly higher than the perturbative results. 
	} \label{islands-cJ}
\end{figure*}

More restrictive constraints come from the lower bounds on the central charges $c_J$, $c_J^t$, and $c_T$, which are shown in Fig.~\ref{islands-cJ}. 
The large $N_f$ perturbative results on the central charges are given by the green contours. Inside the contours the central charges have lower bounds below the perturbative results.
In the right part of the island with $\Delta_{\cM_{1/2}}>1.15$, the lower bounds on conserved current central charges quickly increase to the range $c_J>1.5$ and $c_J^t>2.5$, significantly above the $1/N_f$ perturbative results at subleading order $c_J\simeq 1.04$ and $c_J^t\simeq 1.56$. Such big discrepancies are unlikely to be explained by the higher order corrections, which indicate the bootstrap bounds in Fig.~\ref{islands-cJ} are inconsistent with the lattice results on $N_f=4$ QED$_3$.
Nevertheless, this contradiction  should not be simply interpreted to exclude the lattice results, as our bootstrap bounds are gap-dependent. 
By relaxing the gap assumptions, e.g., using an interval positivity assumption with $\Delta_{\cM_1}^{\rm max} > 2.6$ in the bootstrap implementation, one can obtain weaker bootstrap bounds in which the lattice results locate in the allowed region. In the next subsection we will study additional bootstrap bounds with different gap assumptions which provide some necessary conditions for the lattice results to be physical. Here the roles of central charges are quite reminiscent of their roles in Fig.~\ref{T2J1}, where in comparison with the allowed parameter space of the operator scaling dimensions, bounds on the central charges provide more restrictive constraints for conformal QED$_3$.

The above numerical experiment is surprising to us in two aspects. From the bootstrap point of view, it is a welcome surprise that the bootstrap algorithm, though affected by the gap-dependence problem, can effectively capture a special solution which is rather close to the perturbative estimates of $N_f=4$ QED$_3$. Note that due to the parity symmetry, operators in different sectors have diversified scaling dimensions; the conserved current central charges also have notable differences both in their physical meanings and  magnitudes.  Therefore it is highly nontrivial that several of these properties can be simultaneously satisfied by the bootstrap constraints. From the QED$_3$ side, we do not have solid evidence on the gap $2.8$ in the $A_{(220)}$ sector, and the current perturbative results on the monopole spectrum and central charges may still receive notable higher order corrections. In this sense, it is surprising that the perturbative CFT data taken at face value can seemingly provide a consistent solution to the bootstrap equations.

Since our bootstrap results are gap-dependent, their physical relevance relies on the validity of the gap assumptions in our bootstrap implementation. 
Given our gap assumptions are consistent or close to the physical spectrum, then our bootstrap results are closely relevant to the physical solution of $N_f=4$ conformal QED$_3$, which have significant meanings both for understanding the IR phases of QED$_3$ and its applications in condensed matter systems.
On the other hand, we cannot exclude the possibility, although less likely, that few of our gap assumptions  strongly violate the physical spectrum, and the coincidences between our bootstrap results and perturbative CFT data are purely accidental. 
To verify the two possibilities, we suggest to compute the CFT data using other nonperturbative approaches, e.g., the lattice simulations. The scaling dimensions of the lowest scalar in $A_{(220)}$ and the charge 1 monopole operators are especially important in our bootstrap setup. Reliable estimations of these operators can verify whether our assumptions are consistent with the physical spectrum.

\subsection{Bound on the charge 1 monopole operator \texorpdfstring{$\cM_1$}{M1} and the lattice results}
In this subsection we study the bootstrap bounds on the scaling dimension of charge 1 monopole operator $\cM_1$ and the bounds on the central charges in the resulting allowed region. The results will explain why the interval positivity assumptions can generate closed islands. We will additionally provide more comparisons between the bootstrap bounds and the lattice results. Since our bootstrap results are gap-dependent and the gap assumptions are not strictly established yet, our results cannot verify or exclude the lattice results by themselves. Nevertheless, they can provide strong necessary conditions for the lattice results to be physical.

\begin{figure}
    \centering
    \includegraphics[width=0.8\linewidth]{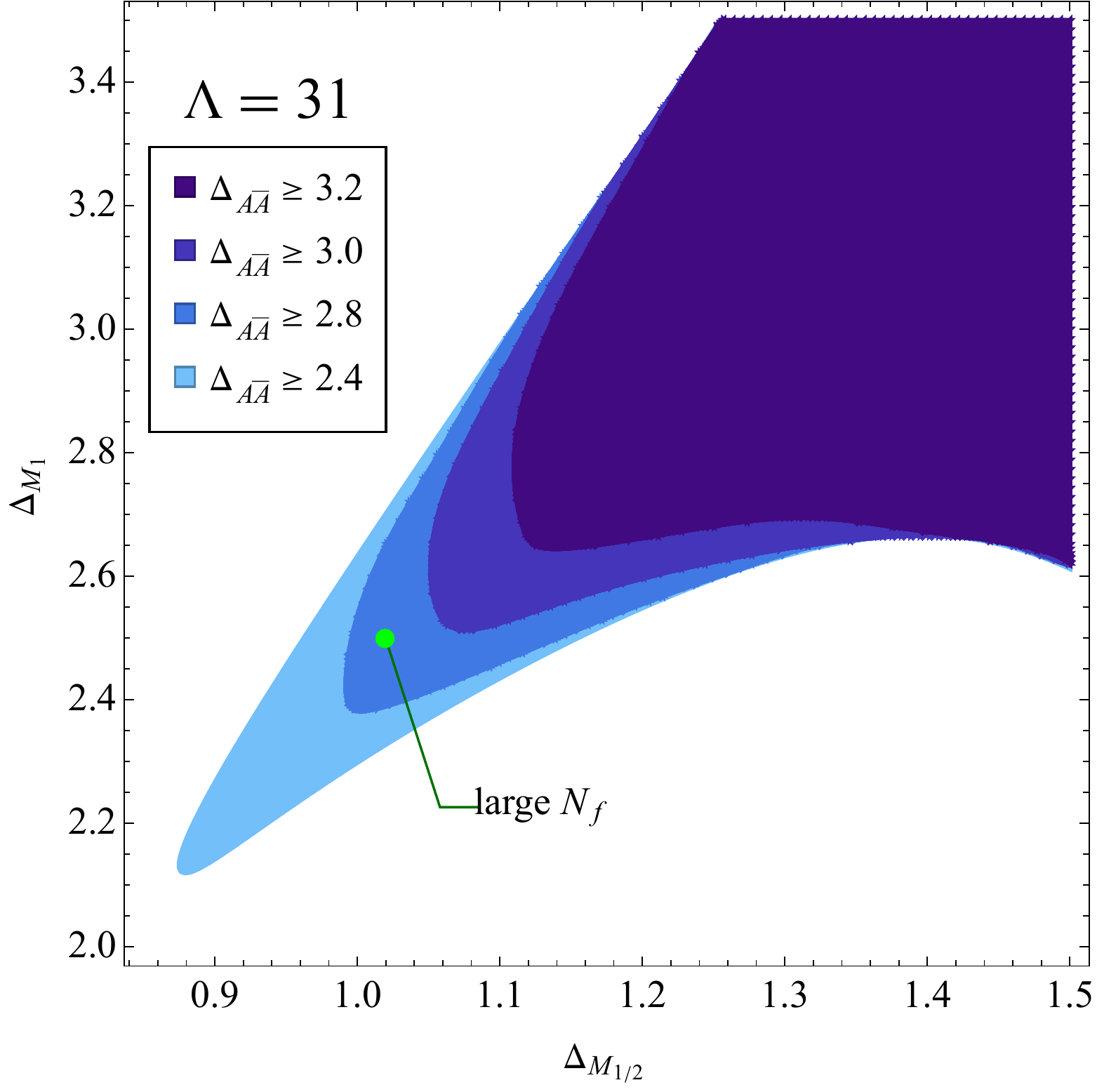}
\caption{
Bootstrap bounds on the scaling dimension of the lowest charge 1 monopole $\cM_{1}$ at $\Lambda=31$. 
To get the bounds, we have employed gap assumptions on the lowest parity even singlet scalar $\Delta_{S_{(000)}}\geqslant 3.0$, the second lowest charge 1 monopole operator $\cM_{1}'$ in the $T_{(220)}$ sector $\Delta_{\cM_1'}\geqslant 5.0$, and the lowest scalar in the $S_{(220)}$ sector $\Delta_{A\bar{A}}\geqslant 2.4, 2.8, 3.0,3.2$.
}  \label{fig:M1wave}
\end{figure}

First let us consider the bootstrap bounds on the scaling dimension of the lowest charge 1 monopole operator $\cM_{1}$. The results are shown in Fig.~\ref{fig:M1wave}. 
To get the bounds we used the gap assumptions $\Delta_{S_{(000)}}\geqslant3.0$ and $\Delta_{\cM_1'}\geqslant 5.0$ for the second lowest charge 1 monopole $\cM_1'$. The bootstrap bounds change notably with different gaps $\Delta_{A\bar{A}}\geqslant \Delta^*$ for the lowest scalar in the $S_{(220)}$ sector.
The most interesting point in  Fig.~\ref{fig:M1wave} is that the bootstrap allowed region forms a wave structure when the $S_{(220)}$ gap is in the range $\Delta^*\leqslant 3$. 
The $1/N_f$ perturbative results locate near the tip of the bootstrap bound associated with the gap $\Delta^*=2.8$. For larger gaps $\Delta^*\geqslant 3.2$ the wave structure disappears.
The wave structures in the $\cM_1$ bootstrap bounds are reminiscent of the bootstrap bound on the 3d critical Ising model with a gap on the second $\mathbb{Z}_2$ even scalar \cite{El-Showk:2012cjh},  
while the gaps imposed here are not fully justified as our knowledge of $N_f=4$ QED$_3$ is limited. Due to the wave structure in Fig.~\ref{fig:M1wave}, the interval positivity condition $\Delta_{\cM_1}\leqslant \Delta_{\cM_1}^{\rm max}$ truncates the tip of the wave structure below $\Delta_{\cM_1}^{\rm max}$, which disconnects from the right part of the bulk region and forms a closed island. Bootstrap bounds on the scaling dimension of the monopole $\cM_1$ were first presented in \cite{Chester:2016wrc}, in which the authors introduced a weaker gap assumption on the scaling dimension of the second lowest monopole $\cM'_1$ and the bootstrap bound shows a weaker peninsula structure. The sharp wave structure we see here can not appear unless the stronger gap assumption on  $\Delta_{\cM'_{1}}$ is imposed.

\begin{figure*}
	\centering
\hspace*{-2em}
$
\begin{aligned}
\Delta_{S_{(000)}}^{\ell=0}\geq3.0 ,
\Delta_{S_{(220)}}^{\ell=0}\geq \mathbf{2.8} ,
\Delta_{\cM_1'}\geq5.0  \hspace{1.2cm}
\\
\includegraphics[width=0.45\linewidth]{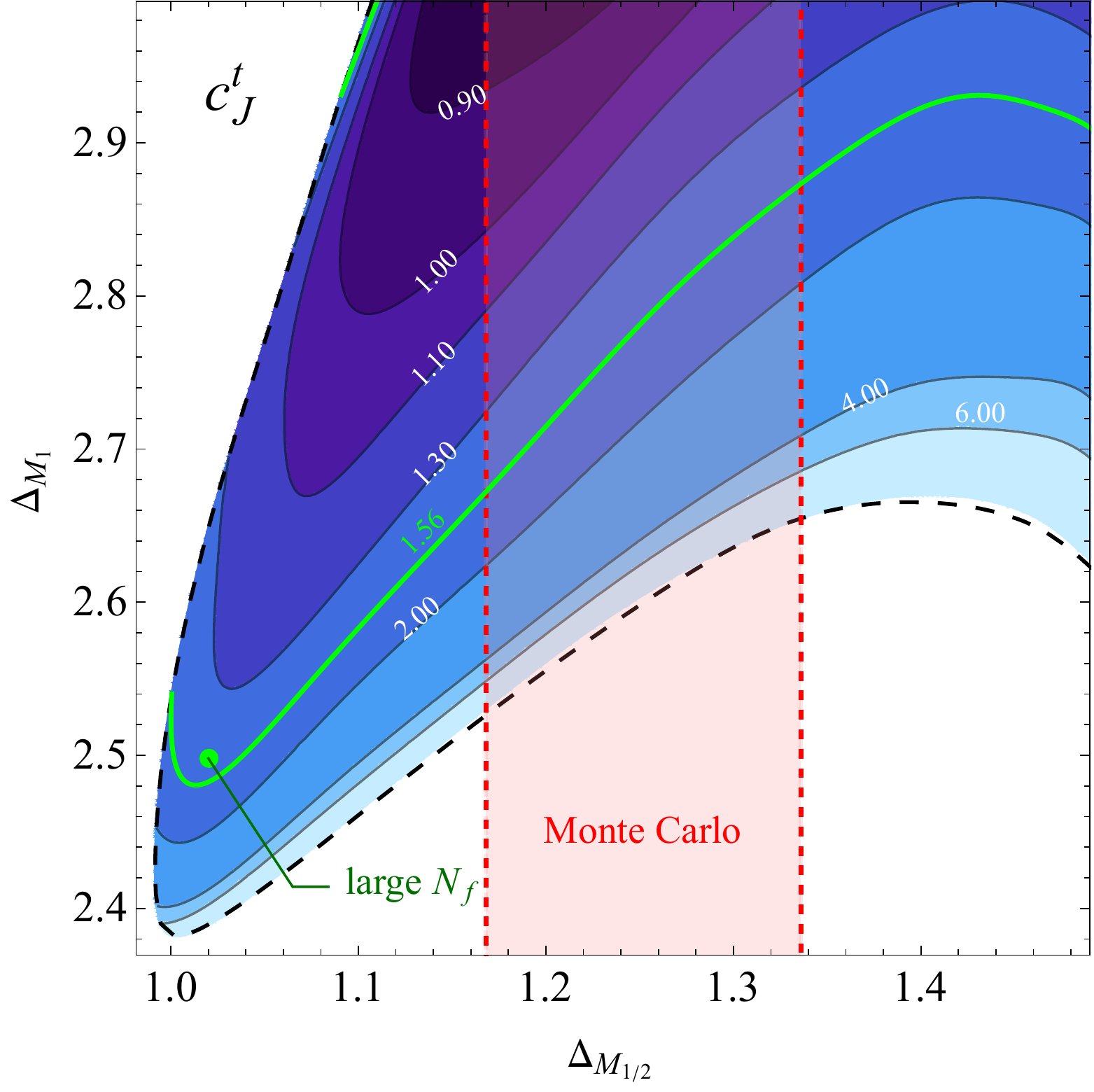}
\end{aligned}
\;\;
\begin{aligned}
\Delta_{S_{(000)}}^{\ell=0}\geq3.0 ,
\Delta_{S_{(220)}}^{\ell=0}\geq\mathbf{3.0} ,
\Delta_{\cM_1'}\geq5.0  \hspace{1.2cm}
\\
\includegraphics[width=0.45\linewidth]{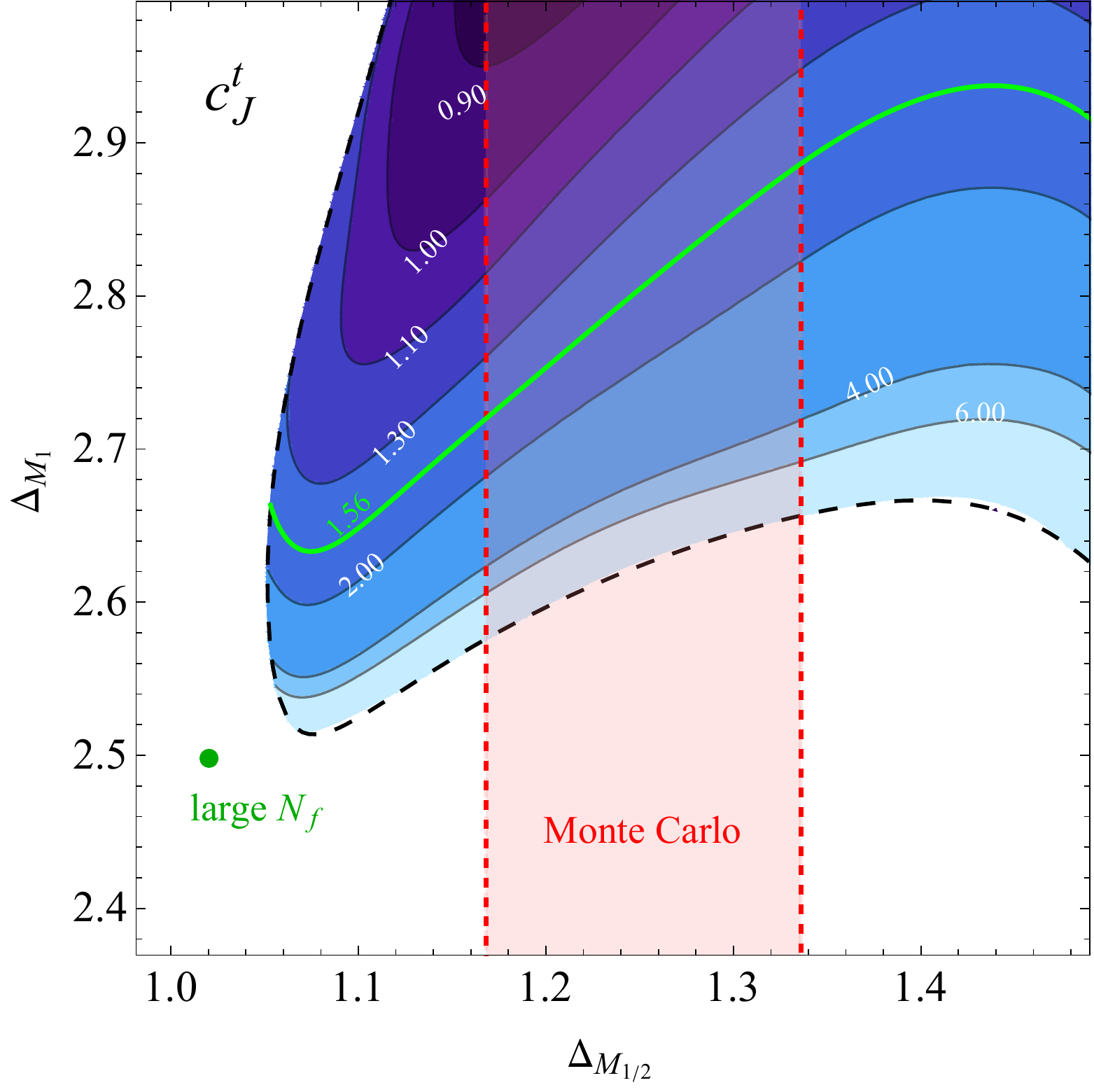}
\end{aligned}
$	
	\caption{
		Lower bounds on the  topological $U(1)_t$ central charge $c_J^t$ with different gap assumptions at $\Lambda=31$. 
		The green dot denotes the perturbative results of the monopole scaling dimensions $(\Delta_{\cM_{1/2}},\Delta_{\cM_1})\simeq(1.022, 2.499)$, and the green line shows the contour $c_J^t=1.56$ predicted by the $1/N_f$ expansion. The lattice value $\Delta_{\cM_{1/2}}=1.252(84)$ \cite{Karthik:2019mrr} is given by the pink shaded region. 
	} \label{mp-mp1}
\end{figure*}

In Fig.~\ref{mp-mp1}, we also present the lower bound on the topological conserved current central charge $c_J^t$ inside the wave structure, where its $1/N_f$ perturbative prediction is given by the green contour. Similar to the results in Fig.~\ref{islands-cJ}, the central charge $c_J^t$ has a much higher lower bound in the right part of the allowed region. According to the $c_J^t$ lower bound, it requires the scaling dimension of the monopole $\cM_1$ to be above $\Delta_{\cM_1}\geqslant 2.67$, or even higher values for the $\Delta_{\cM_{1/2}}$ in the range predicted by the lattice results \cite{Karthik:2019mrr}.  

\section{Bootstrapping mixed correlators with \texorpdfstring{$\cM_{1/2}$}{M1/2} and \texorpdfstring{$r$}{r} }
\label{sec:mixed}

We have shown that the single correlator bootstrap results can provide strong constraints on the conformal $N_f=4$ QED$_3$. 
To improve the the bootstrap results, the key is to find a more restrictive bootstrap implementation. A straightforward generalization of our work is to bootstrap mixed correlators with multiple operators in $N_f=4$ QED$_3$. In this section, we perform a mixed correlator bootstrap study of conformal $N_f=4$ QED$_3$ with emphasis on the two low-lying scalars $r$ and $\cM_{1/2}$. We will show that this bootstrap setup indeed can significantly improve the lower cuts of the closed islands obtained in \secref{\ref{islands}}.\footnote{
A similar study of the same mixed correlators was also performed in \cite{He:2021sto}, which obtained general constraints on the possible stable critical phases of Dirac spin liquids on triangular and kagome lattices.}

Bounds on the scaling dimensions of the operators $\cM_{1/2}$ and $r$ obtained from the $r-\cM_{1/2}$ mixed correlator bootstrap are shown in Fig.~\ref{islands-interval-mixed}. Details on the mixed correlator bootstrap implementation are presented in Appendix \ref{rMmixing}. In the mixed correlator bootstrap, we used the same gap assumptions as in the monopole single correlator bootstrap, including the {\it interval positivity} assumption. In addition, we also required that the lowest scalar in the $S\bar{S}$, i.e. $(422)$, representation of the $SU(4)$ flavor symmetry is irrelevant. Compared with the single correlator bootstrap bound, the mixed correlator bootstrap significantly improves the lower bound on the scaling dimension of $\Delta_r$ in the closed island: $\Delta_r\geqslant 1.12$ at $\Lambda=27$. 

\begin{figure}
	\centering
	\hspace*{-2em}
$
\begin{aligned}
	\includegraphics[width=0.8\linewidth]{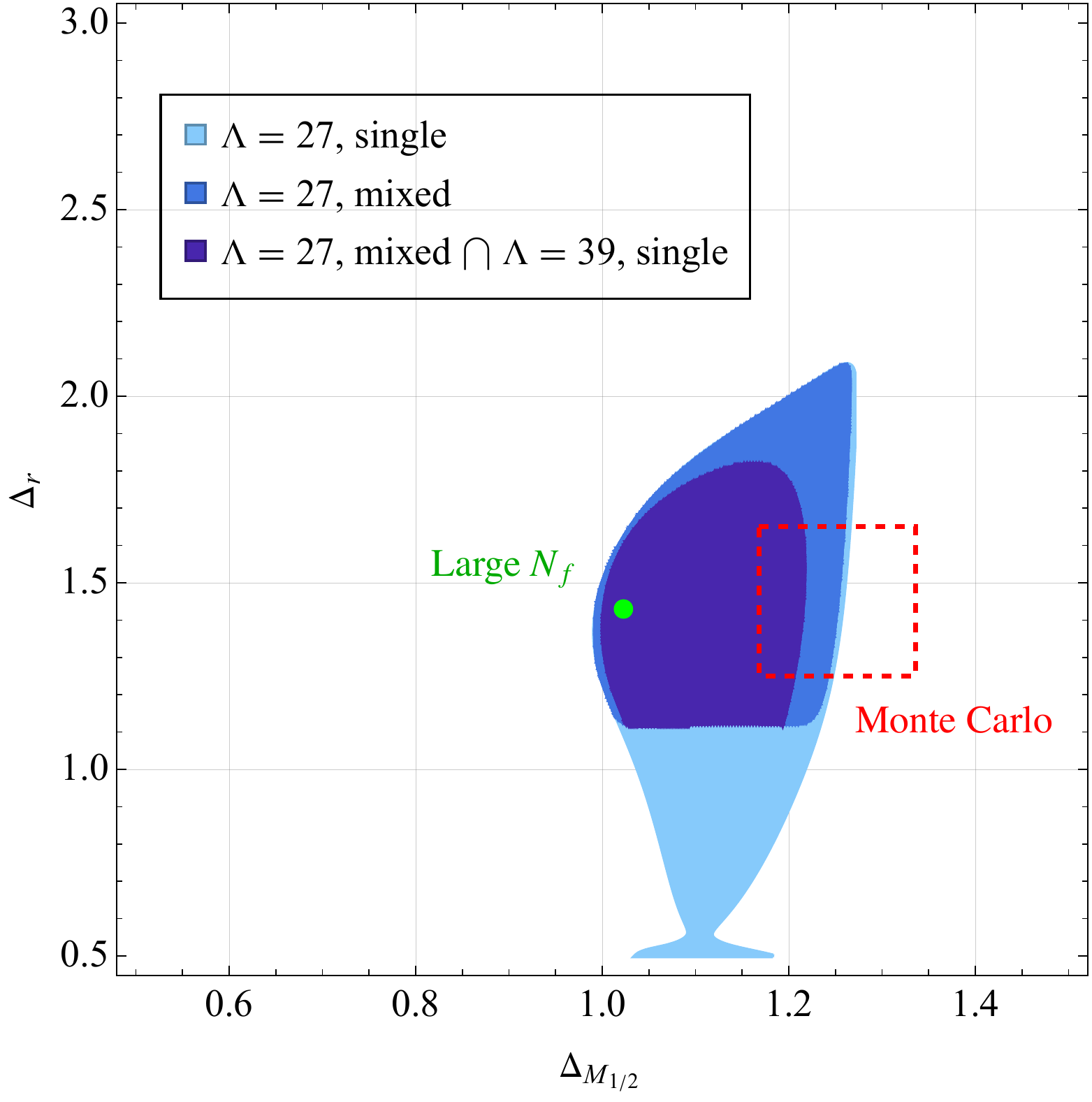}
\end{aligned}
\;\;
\begin{aligned}
	\Delta_{S_{(000)}}^{\ell=0} &\geq 3.0 \\
	\Delta_{S_{(220)}}^{\ell=0} &\geq 2.8 \\
	\Delta_{r'} &\geq 3.0 \\
	\Delta_{T_{(000)}}^{\ell=0} &\geq 4.0 \\
	\Delta_{\cM_1} &\leqslant 2.6 \\
	\Delta_{\cM_1'} &\geq 5.0  \\ \\
	\textrm{mixed:} \\
	\Delta_{S\bar{S}}^{\ell=0} &\geq 3.0
\end{aligned}
$
	\caption{Bounds on the scaling dimensions of $(\Delta_{\cM_{1/2}}, \Delta_r)$ with the interval positivity assumption $\Delta_{\cM_1}\leqslant 2.60$,
		comparing monopole single correlator bootstrap results with the monopole-adjoint mixed correlator bootstrap results at $\Lambda=27$.
	} \label{islands-interval-mixed}
\end{figure}

The large $N_f$ prediction and fermion bilinear bootstrap bounds on the scaling dimension of the lowest scalar in the $S\bar{S}$ sector were shown in Fig.~\ref{ssbbd}, from which we expect the gap $3.0$ is a reliable assumption. Moreover, the fermion bilinear bootstrap bound on $\Delta_{S\bar{S}}$ in Fig.~\ref{ssbbd} explains why the we can obtain a stronger minimum on $\Delta_{r}$ after introducing the gap $\Delta_{S\bar{S}}\geqslant3.0$: the upper bound on $\Delta_{S\bar{S}}$ cannot be higher than 3 for $\Delta_r<1.12$. This provides a nice example which illustrates how the mixed correlator bootstrap can help to get stronger bounds with reliable assumptions. The bootstrap bounds in certain sectors are more restrictive and the mixed correlator bootstrap implementation can help to exploit the constraints in these sectors. We expect there are extra sectors, especially in the $\cM_{1/2}$ and $\cM_1$ mixed correlator setup, which can provide strong constraints on the CFT data with reliable gap assumptions. We hope to give a more systematic study of these constraints in our next work.

\begin{figure}
	\centering
\hspace*{-2em}
$
\begin{aligned}
	\includegraphics[width=0.8\linewidth]{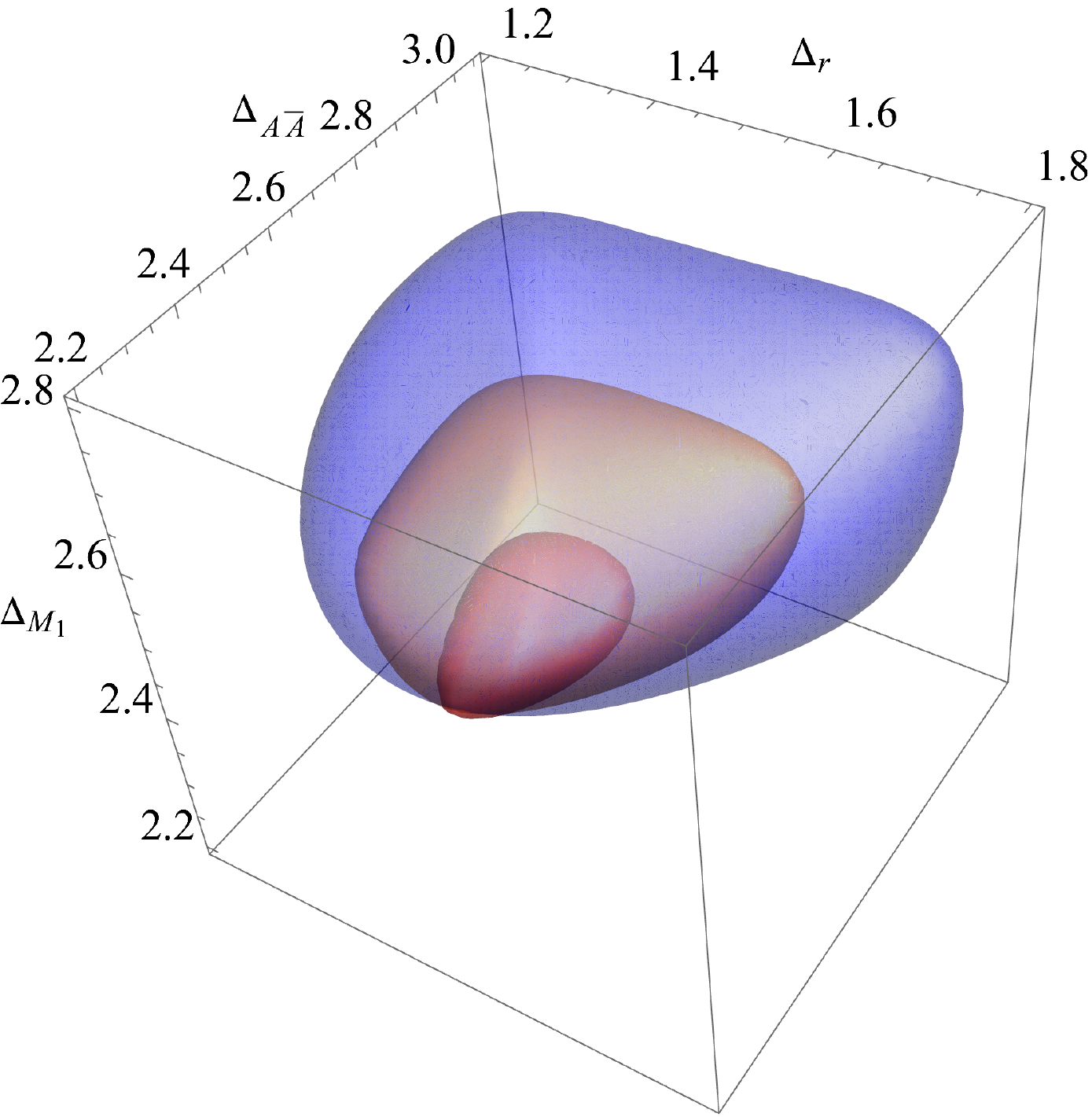}
\end{aligned}
\;\;
\begin{aligned}
	\Delta_{\cM_{1/2}'} &\geqslant 3.0 \\
	\Delta_{r'} &\geqslant 3.0 \\
	\Delta_{\cM_1'} &\geqslant 4.5 \\
	\Delta_{A\bar{A}'}^{\ell=0} &\geqslant 4.5 \\
	\Delta_{S_{(000)}}^{\ell=0} &\geqslant 3.0 \\
	\Delta_{S\bar{S}}^{\ell=0} &\geqslant 3.5 \\
	\Delta_{V_{(200)}}^{\ell=0} &\geqslant 3.0 \\
	\Delta_{V_{(321)}}^{\ell=0} &\geqslant 3.0 \\
	\Delta_{V_{(321)}}^{\ell=1} &\geqslant 3.0 \\
	\Delta_{T_{(000)}}^{\ell=0} &\geqslant 4.0 \\
\end{aligned}
$
	\caption{Bounds on the scaling dimension of $(\Delta_r, \Delta_{A\bar{A}}, \Delta_{\cM_{1}})$ from the monopole-adjoint mixed correlator bootstrap ($\Lambda=19$) at fixed $\Delta_{\cM_{1/2}}$. The islands in the plot, from the largest to the smallest, correspond to $\Delta_{\cM_{1/2}} = 1.06, 1.02,$ and $0.98$, respectively. The full dynamical version of this 3-dimension plot is included in the attached {\texttt Mathematica} notebook.
		\label{fig:islands-3d} }
\end{figure}

In the bootstrap studies of $N_f=4$ conformal QED$_3$, several operators in different sectors play important roles. Their scaling dimensions relate to higher dimensional structures in the parameter space of CFT data. In Fig.~\ref{fig:islands-3d} we make a first attempt to map out such a higher dimensional structure at \(\Lambda=19\). Specifically we show the closed 3d allowed region in the space $(\Delta_r,\Delta_{\cM_1},\Delta_{A\bar{A}})$ with different fixed scaling dimensions of $\cM_{1/2}$, $\Delta_{\cM_{1/2}}=0.98, 1.02, 1.06$, making a set of plausible gap assumptions. The bootstrap allowed regions are 3d slices of a more complicated higher dimensional geometric structure and so have interesting shapes. The ranges of the islands with different fixed $\Delta_{\cM_1/2}$ have been summarized in Table \ref{tab:range-mixed}. In particular, by taking the large $N_f$ result $\Delta_{\cM_1/2}\simeq 1.02$, the perturbative predictions of $\Delta_r\simeq1.43$ and $\Delta_{\cM_1}\simeq2.50$ are located inside the 3d island. The large $N_f$ prediction of $\Delta_{A\bar{A}}\simeq2.38$ is slightly outside of the island, while the gap $\Delta_{A\bar{A}}\geqslant 2.8$ used in other sections is consistent with the range of $\Delta_{A\bar{A}}$ in the 3d island.

\section{Conclusions and discussions}
\label{sec:conclusion}
The broad goal of the conformal bootstrap project is to find and classify CFTs. 
On the other hand, non-supersymmetric gauge theories have so far shown resistance to being solved numerically using bootstrap methods.\footnote{However, remarkable progress towards numerically solving conformal gauge theories with extended supersymmetry has been made in the recent work \cite{Agmon:2019imm, Chester:2021aun}.}
In this work we have attempted to study the presumed IR fixed point of $N_f=4$ QED$_3$ using the conformal bootstrap. Most notably, we found that after imposing some assumptions inspired by perturbative computations for \(N_f=4\) \qed we can obtain a closed island in parameter space. The ranges of the islands under different bootstrap setups are summarized in Table \ref{tab:range-main}.
Promisingly, bounds in this island on the scaling dimensions $\Delta_{\cM_{1/2}}$, $\Delta_{\cM_1}$, $\Delta_r$ as well as on the central charges $c_J$, $c_J^t$ and $c_T$ are consistent with their $1/N_f$ perturbative results, which in turn are close to saturating the bootstrap bounds. It is important to be clear that the physical relevance of these results relies on the validity of the gap assumptions used in our bootstrap computations, but nevertheless we believe our work has progressed our understanding of $N_f=4$ conformal QED$_3$.

A major challenge in getting precise results from the \qed bootstrap is the notable sensitivity of the bounds to assumed gaps in the spectrum, closely connected to the symmetry-enhancement phenomena discussed in \cite{Li:2020bnb,Li:2020tsm}. The crossing equations of the single four-point correlators $\langle rrrr\rangle$ and $\langle  \cM_{1/2}\cM_{1/2}\cM_{1/2}\cM_{1/2}\rangle$ have positivity properties that can be mapped to the crossing equations of $SO(15)$ and $SO(12)$ vectors, respectively. To bootstrap non-$SO(N)$ symmetric theories, one has to impose gap assumptions which explicitly break the $SO(N)$ symmetries, and intriguing kinks and peninsulas which appear in the bootstrap bounds show a clear dependence on these gap assumptions. Despite this gap sensitivity, we believe that these discontinuities could still be of physical relevance to our understanding of \qed, in the sense that they could be directly connected to the physical \qed solution through larger geometrical structures in scaling-dimension space. We have gained some confidence in this interpretation by inputting a set of gap assumptions inspired by the perturbative spectrum, and seeing that lower bounds on the stress tensor and current central charges near these kinks are nicely compatible with their estimated values from $1/N_f$ perturbation theory. In particular, this makes it seem unlikely that the kinks are related to non-Abelian gauge theories, which have significantly larger values of the central charges.

\begin{table}
	\centering
\caption{The ranges of scaling dimensions of the adjoint fermion bilinear $r$, lowest charge 1 monopole $\CM_1$, and the lowest $A\bar{A}$-rep scalar, in the 3-dimensional islands of fixed $\Delta_{\CM_{1/2}}$ values. The islands and the gap assumptions are shown in Figure~\ref{fig:islands-3d}. The island corresponding to the large-$N_f$ value of $\Delta_{\CM_{1/2}}$ is highlighted in bold font. See the body of this paper for further discussion of these values. }
	$\begin{array}{llll}
		\hline\hline\\[-1em]
		\Delta_{\CM_{1/2}}  \hspace*{2em}& \Delta_{r} \hspace*{5em}& \Delta_{\CM_1} \hspace*{4.5em}& \Delta_{A\bar{A}} \\
		[.1em]\hline\\[-1em]
		0.98 & (1.36, 1.52)~~ & (2.27, 2.46)~~ & (2.43, 2.76)~~ \\
		[.1em]\hline\\[-1em]
		\bf 1.02 & \bf(1.30, 1.66)~~ & \bf(2.28, 2.60)~~ & \bf(2.39, 2.91)~~\\
		[.1em]\hline\\[-1em]
		1.06 & (1.26, 1.79)~~ & (2.29, 2.75)~~ & (2.33, 3.06)~~\\
		[.1em]\hline\hline 
	\end{array}$
	\label{tab:range-mixed}
\end{table}

The parity symmetry of $N_f=4$ QED$_3$ makes the monopole bootstrap particularly effective in separating \qed from other solutions to crossing equations. Operators appearing in the $\cM_{\pm1/2}\times\cM_{\pm1/2}$  OPE carry different parity charges depending on their representations of $SU(4)\times U(1)_t$, and their scaling dimensions strongly break the $SU(4)\times U(1)_t\rightarrow SO(12)$ relation between the crossing equations. In contrast, in the fermion bilinear bootstrap, operators in different sectors branched from $SO(N)$ representations have the same parity charges and their scaling dimensions only differ by loop corrections in the $1/N_f$ expansion. By inputting gap assumptions inspired by the $N_f=4$ QED$_3$ perturbative spectrum, in particular an expected large gap until the second charge $1$ monopole, we are able to find a sharp peninsula structure in \((\Delta_{\cM_{1/2}},\Delta_{\cM_1})\) whose narrow tip coincides neatly with the perturbative estimates of the theory. The peninsula structure remains gap dependent, and the gap assumption in the $S_{(2,2,0)}$ sector is particularly important as it determines the minimum value of $\Delta_{\cM_{1/2}}$. As emphasized recently in \cite{He:2021sto}, the leading operator in this sector is also physically important because its relevance or irrelevance determines whether \qed can be reached in lattice systems. We found that a gap $\Delta_{S_{(2,2,0)}}\geqslant 2.8$ allows for a nice consistency with $1/N_f$ perturbation theory, while irrelevance of this operator implies that uncomputed $1/N_f$ corrections to \((\Delta_{\cM_{1/2}},\Delta_{\cM_1})\) should be of the same order as computed ones. It will be important in future work to determine which of these scenarios is correct.

\begin{table}
	\centering
	\caption{A summary of the ranges of scaling dimensions of the lowest charge-1/2 monopole $\cM_{1/2}$ and the $SU(4)$ adjoint fermion bilinear $r$ in different setups. The shared gap assumptions are shown in Figure~\ref{islands-interval}, and the assumptions specific to the different setups are presented in the first row. }
	$\begin{array}{lll}
		\hline\hline\\[-1em]
		\textbf{setup} \hspace*{7em}&\Delta_{\cM_{1/2}}\textbf{ range} \hspace*{3em}& \Delta_{r}\textbf{ range}
		\\\hline\\[-1em]
		\begin{aligned}
			\Lambda=39~\text{single} \\ \Delta_{\cM_1} \leq 2.50
		\end{aligned}
		&(1.00, 1.10)&(1.03, 1.61)
		\\\hline\\[-1em]
		\begin{aligned}
			\Lambda=39~\text{single} \\ \Delta_{\cM_1} \leq 2.60
		\end{aligned}
		&(1.00, 1.22)&(0.73, 1.81)
		\\\hline\\[-1em]
		\begin{aligned}
			\bigcap \makecell{\Lambda=39~\text{single}\\\Lambda=27~\text{mixed}} \\
			\left\{\makecell{\Delta_{\cM_1} \leq 2.60\\\Delta_{S\bar{S}} \geq 3.50 } \right.
		\end{aligned}
		&(1.00, 1.22)&(1.12, 1.81)
		\\\hline\\[-1em]
		\begin{aligned}
			\Lambda=31~\text{single} \\ 
			\left\{\makecell{\Delta_{\cM_1} \leq 2.60\\c_J^t\leq 2 } \right.
		\end{aligned}
		&(1.00, 1.14)&(0.98, 1.77)
		\\\hline\hline
	\end{array}$
	\label{tab:range-main}
\end{table}

Adopting the assumption that we should take perturbation theory at least somewhat seriously, our most notable results are obtained by imposing an {\it interval positivity} assumption $\Delta_{\cM_1} \leqslant 2.6$, that $\Delta_{\cM_1}$ required to be near or below its subleading perturbative estimate $\Delta_{\cM_1} \approx 2.5$, which in turn restricts the peninsula structure to a closed island. Notably, this gives a closed region for the fermion bilinear dimension $\Delta_{r}$ as well as for $\Delta_{\cM_{1/2}}$. Our bootstrap island at $\Lambda=31$ overlaps with previous lattice estimates for $\Delta_r$ and $\Delta_{\cM_{1/2}}$ \cite{Karthik:2015sgq,Xu:2018wyg,Karthik:2019mrr}. However, by computing lower bounds on the central charges $c_J$, $c_J^t$ and $c_T$ inside this island, we see that the lattice estimates of $(\Delta_{\cM_{1/2}},\Delta_r)$ require $c_J$ and $c_J^t$ to be significantly higher than their $1/N_f$ perturbative estimates, suggesting that this region is likely unphysical. In contrast, the lower bounds on the central charges agree with their $1/N_f$ perturbative results in the region with $\Delta_{\cM_{1/2}}\in (1.0,1.1)$, compatible with the $1/N_f$ estimate $\Delta_{\cM_{1/2}}\simeq 1.022$. In fact, if we adopt the more restrictive assumption $\Delta_{\cM_1}\leqslant 2.5$, then a linear extrapolation of the bootstrap island suggests that it shrinks to a small range with $\Delta_{\cM_{1/2}}\in (1.02, 1.04)$, beautifully compatible with the perturbative results for both the scaling dimension and central charge data. 

The results we have laid out so far give a potentially bright outlook for the future of bootstrapping \qed, and we see several concrete directions for future work. The islands we obtained in this work rely on inputting an assumption which places either $\Delta_{\cM_1}$ or $\Delta_{\cM_{1/2}}$ near its perturbative value. It is important to find ways to get rid of this condition. Moreover, in this work some of the gap assumptions we made are not fully justified, and we hope that bootstrap results for \qed can ultimately be established using a set of sufficiently general assumptions that are more firmly established. A key point to improving this situation is to find an even more restrictive bootstrap setup, and there are a number of concrete mixed-correlator setups that could be pursued. 

The bounds on the scaling dimension of the lowest charge 1 monopole operator have an interesting wave structure, which explains why islands can be formed with interval positivity assumptions and generally provides strong constraints if one assumes that the lattice results \cite{Karthik:2019mrr} are reliable. This wave structure is reminiscent of a similar structure appearing in the bootstrap of the 3d Ising CFT, leading to the conjecture that the solution at the tip of the wave might be further isolated by bootstrapping mixed correlators of the monopoles $\cM_{1/2}$ and $\cM_1$. In this mixed correlator setup, we can get access to more representations of $SU(4)\times U(1)_t$ and further exploit the constraints from parity symmetry and gaps in the monopole spectrum (which reflect the underlying equations of motion), which have played crucial roles in generating the current bootstrap results. Besides the gaps explored in this work, there are likely to be other sectors which can also introduce strong constraints on the CFT data, especially the spin $1$ sectors appearing in the $SO(12)\rightarrow SU(4)\times U(1)_t$ branching rules. We hope to provide a more systematic exploration of these directions in future work.

\section*{Acknowledgements}
We thank Thomas Appelquist, Meng Cheng, Shai Chester, George Fleming, Liam Fitzpatrick, Luca Iliesiu, Yin-Chen He, Ami Katz, Petr Kravchuk, Walter Landry, Silviu Pufu, Slava Rychkov, William Witczak-Krempa, David Simmons-Duffin, and Ning Su for discussions. The work of SA, RE, ZL, and DP is supported by Simons Foundation grant 488651 (Simons Collaboration on the Nonperturbative Bootstrap) and DOE grant no.\ DE-SC0020318. SA is also supported by a VIDI grant of the Netherlands Organisation for Scientific Research~(NWO) that is funded by the Dutch Ministry of Education, Culture and Science~(OCW). YX is supported by a Yale Mossman Prize Fellowship in Physics. Our bootstrap computations were carried out on the Yale Grace computing cluster, supported by the facilities and staff of the Yale University Faculty of Sciences High Performance Computing Center.

\appendix
 
\section{Further details on the fermion bilinear bootstrap}
\label{sec:bilinear-details}

\subsection{Conventions}
\label{sec: group theory conventions}
In this appendix, we set the conventions to describe various operators in QED$_3$. We use the Minkowski metric in mostly plus signature $\eta_{\mu\nu}=\textrm{diag}(-1,1,1)$. We will denote a fermion in the fundamental representation and its dual with uppercase and lowercase indices respectively, i.e. $\psi^\a$ and $\psi_\a$ --- and similarly for the complex conjugate representation $\psi_{\da}$ and its dual $\psi^{\da}$. Between these representations, we have the relation $(\psi^\a)^\dagger=(\psi^{\dagger})^{\da}$ and we have the intertwining operator $\gamma^0_{\da\a}$, so that $(\psi^{\a})^\dagger\gamma^0_{\da\b}$ transforms as the dual fermion $\psi_\b$. We will then define $\widebar{\psi}_\a\coloneqq(\psi^\b)^\dagger\gamma^0_{\db\a}$, with which we can construct the invariant scalar $\widebar{\psi}_\a\psi^\a$.\footnote{
For Majorana fermions, we can convert all dotted indices to undotted ones, with which $\gamma^0_{\a\b}$ and $\psi^a$ can be interpreted as the symplectic tensor and fundamental representation of $\Sp(2,\R)$ group, as was done in \cite{Iliesiu:2015qra,Iliesiu:2017nrv,Albayrak:2019gnz,Albayrak:2020rxh}.}
Whenever there is no room for confusion, we will suppress spinor indices.\footnote{In our conventions, (un)dotted indices are contracted from north(south)-west to south(north)-east.}

We use an explicit real representation of the $\gamma^\mu$ matrices, i.e.
\be 
\gamma^0=\begin{pmatrix}
	0&1\\-1&0
\end{pmatrix}\;,\quad 
\gamma^1=\begin{pmatrix}
	0&1\\1&0
\end{pmatrix}\;,\quad 
\gamma^2=\begin{pmatrix}
	1&0\\0&-1
\end{pmatrix}\,,
\ee 
with which the Lorentz generators acting on the Dirac spinors can be written as $\gamma_{\mu\nu}=\frac{i}{4}\left(\gamma_\mu\gamma_\nu-\gamma_\nu\gamma_\mu\right)$. We take the space parity transformation as the reflection $x^2\rightarrow -x^2$, and we choose its action on the fermions as $\psi\rightarrow\gamma^2\psi$.\footnote{As the double cover of the rotation group ($\mathrm{Pin}$ group) acts on $\psi$, both $\pm\gamma^2\psi$ are valid choices. Since we will always consider operators containing an even number of fermions, this does not pose any ambiguity.} This means $\widebar{\psi}\rightarrow -\widebar{\psi}\gamma^2$, indicating that $\widebar{\psi}\psi$ transforms as a parity-odd scalar.

We can also work out how space parity transformations act on the standard $|l,m\>$ basis from the theory of angular momentum. Its coordinate representation, the spherical harmonics $Y_{lm}(\theta,\phi)$, pick up a sign under our parity transformation: $Y_{lm}(\theta,\phi)\rightarrow(-1)^{l+m}Y_{lm}(\theta,\phi)$.\footnote{We note that this is different from the standard formula $\cP: Y_{lm}(\theta,\phi)\rightarrow(-1)^{l}Y_{lm}(\theta,\phi)$ because we took our parity transformation as the reflection $x^2\rightarrow -x^2$ instead of the inversion $x^i\rightarrow-x^i$.} In the presence of a magnetic flux $q$ with $l-\abs{q}\ge0$, we will instead resort to the \emph{scalar monopole spherical harmonics} introduced in \cite{WU1976365,PhysRevD.16.1018}:
\small 
\be 
\label{eq: scalar monopole spherical harmonics}
		Y_{q,lm}(\theta,\phi) &\equiv \Theta_{q,lm} (\cos\theta) e^{i(m+\ka q)\phi}\,, \\
		\Theta_{q,lm}(x) &\equiv 2^{m-1} 
		\sqrt{\frac{(2\ell+1)(\ell-m)!(\ell+m)!}{\pi(\ell-q)!(\ell+q)!}}
		\sqrt{\frac{(1+x)^{q-m}}{(1-x)^{q+m}}} P_{\ell+m}^{-q-m,q-m}(x)\,,
\ee 
\normalsize
where $P_{n}^{\alpha,\beta}(x)$ is the Jacobi polynomial and the parameter $\ka=\pm 1$.\footnote{The value of $\ka$ depends on which coordinate chart we are using to describe $Y_{q,lm}$: if we choose the chart that includes the whole sphere minus the south (north) pole, then $\ka$ is 1 (-1). In the rest of the paper, we stick to $\ka=1$.} In our conventions,
\small 
\be 
\label{eq: transformation of scalar monopole spherical harmonics}
\text{Space Parity: }\quad Y_{q,lm}(\theta,\phi)\rightarrow(-1)^{l+m}e^{2iq\phi}Y_{-q,lm}(\theta,\phi)\;.
\ee 
\normalsize

Our conventions for the global $SU(N)$ symmetry is analogous: we write indices in the (anti)fundamental representations (downstairs)upstairs, i.e. $\cO^i$ vs $\cO_i$. However, unlike the spacetime representations, these ones are actually conjugate, hence we have $(\cO^i)^\dagger=\cO_i$. Similar to $\gamma^0$ for the Clifford algebra, we have the Levi-Civita tensor $\epsilon$ which acts an intertwining operator between these conjugate representations,\footnote{
Let an operator $\cO^{i_1\dots i_n}$ transform under the $SU(N)$ action as \mbox{$\cO^{i_1\dots i_n}\rightarrow U^{i_1}_{\;\;j_1}\dots U^{i_n}_{\;\;j_n}\cO^{j_1\dots j_n}$}, or \mbox{$\cO^I\rightarrow U^I_{\;\;J}\cO^J$} as a shorthand notation. We similarly have $\cO_I\rightarrow\cO_J(U^\dagger)^J_{\;\;I}$ for the conjugate operator. Due to the identity \mbox{$\e_{IK}U^K_{\;\;J}=(U^\dagger)^L_{\;\;I}\e_{LJ}$}, $(\e_{IJ}\cO^J)$ transforms as an operator in the conjugate representation, i.e. $(\e_{KJ}\cO^J)(U^\dagger)^K_{\;\;I}$ --- similarly, the relation \mbox{$\e^{IK}U^J_{\;\;K}=(U^\dagger)^I_{\;\;L}\e^{LJ}$} implies the inverse, i.e. \mbox{$\left(\cO_K\e^{KI}\right)\rightarrow U^I_{\;\;J}\left(\cO_K\e^{KJ}\right)$}. Note that the identity \mbox{$\e_{i_1\dots i_n k_1\dots k_{\bar{n}}}U^{k_1}_{j_1}\cdots U^{k_{\bar{n}}}_{j_{\bar{n}}}=(U^\dagger)^{l_1}_{i_1}\cdots (U^\dagger)^{l_n}_{i_n}\e_{l_1\dots l_n j_1\dots j_{\bar{n}}}$} follows from $\det U =1$ condition and hence is valid only for the special unitary group.
} hence we choose\footnote{Our choice of normalization follows from the useful identity $\epsilon_{i_1\dots i_{n}j_1\dots j_{m}}\epsilon^{k_1\dots k_{n}j_1\dots j_{m}}=n! m! \delta^{k_1}_{[i_1}\cdots \delta^{k_n}_{i_n]}$.}
\be 
\label{intertwining conjugate representations}
A_{i_1\dots i_{n}}=&\frac{1}{\sqrt{n!\bar{n}!}}\epsilon_{i_1\dots i_{n}j_1\dots j_{\bar{n}}}B^{j_1\dots j_{\bar{n}}}\;,\\
C^{i_1\dots i_m}=&\frac{1}{\sqrt{m!\widebar{m}!}}D_{j_1\dots j_{\widebar{m}}}\epsilon^{j_1\dots j_{\widebar{m}}i_1\dots i_{m}}\,,
\ee 
for operators $A,B,C,$ and $D$ in representations $\mathfrak{m}_1,\mathfrak{m}_1^*,\mathfrak{m}_2,$ and $\mathfrak{m}_2^*$ respectively. Here, we defined the shorthand notation
\be 
\bar{n}\equiv N-n
\ee 
and similarly for $m$. If $m=\widebar{m}=n=\bar{n}=\frac{N}{2}$, we can choose $B=C=\cO$ and $A=D=\cO^\dagger$, which gives the reality conditions
\be 
\label{eq: transformation of real operators}
\left(\cO^{i_1\dots i_n}\right)^\dagger=&\frac{1}{n!}\epsilon_{i_1\dots i_{n}j_1\dots j_{n}}\cO^{j_1\dots j_{n}}\;,\\
\cO^{i_1\dots i_n}=&\frac{1}{n!}\left(\cO^{j_1\dots j_n}\right)^\dagger\epsilon^{j_1\dots j_{n}i_1\dots i_{n}}\,.
\ee 
 
The generalization of our notation to mixed tensors of $SU(N)$ is straightforward. For notational brevity, we'll take consecutive indices antisymmetrized, whereas groups of indices separated by lines are symmetrized; for instance, the Young diagram for the representation of $\cO^{il|jm|k}$ reads as {\scriptsize \begin{ytableau}i& j&k\cr l& m\cr \end{ytableau}}.\footnote{We note that for the representation to correspond to a valid Young diagram, we have the constraint \mbox{$l_1\ge l_2\ge \dots \ge l_n$} for $\cO^{k_{11}\dots k_{1l_1}|k_{21}\dots k_{2l_2}|\cdots|k_{n1}\dots k_{nl_n}}$.}  The \equref{intertwining conjugate representations} then generalizes as
\small 
\be 
A_{k_{11}\dots k_{1n_1}|
k_{21}\dots k_{2n_2}|
\cdots |
k_{c1}\dots k_{cn_c}
}=
\left(
\prod\limits_{a=1}^c
\frac{\epsilon_{k_{a1}\dots k_{an_a}l_{a1}\dots l_{a\bar{n}_a}}}{\sqrt{n_a!\bar{n}_a!}}
\right)
\\\x
B^{l_{c1}\dots l_{c\bar{n}_c}|
\cdots|
l_{21}\dots l_{2\bar{n}_2}|
l_{11}\dots l_{1\bar{n}_1}
}\,,
\ee 
\normalsize
where $A$ transforms in the Young diagram of $c$ columns, each column having $n_c$ boxes (and $B$ transforms as its dual). For instance, two operators $A$ and $B$ in the conjugate representations  $\mathsf{(A\bar{S})}$ and $\mathsf{(S\bar{A})}$ of $SU(4)$ would be related as 
\small 
\be 
\left(A^{k_{11}k_{12}k_{13}|k_{21}k_{22}k_{23}|k_{31}k_{32}}\right)^\dagger=\frac{\e_{k_{11}k_{12}k_{13}l_{11}}\e_{k_{21}k_{22}k_{23}l_{21}}\e_{k_{31}k_{32}l_{31}l_{32}}}{\sqrt{3!3!2!2!}}B^{l_{31}l_{32}|l_{21}|l_{11}}\;.\label{key}
\ee 
\normalsize
 
The final group that we should set our conventions for is the $SO(2)$ group under which the monopole operators transform in the fundamental representation, i.e. $M^{I;a}$ for $a=1,2$ ($I$ denotes the collective indices for $SU(N)$). We are interested in cases where the monopole operators are real, hence \equref{eq: transformation of real operators} generalizes as\footnote{As monopole operators have the Dynkin labels $[\bar{0},2\abs{q},\bar{0}]$ ($\bar{0}$ denoting the sequence of $\frac{N-2}{2}$ many $0$'s), they are pseudoreal if $2\abs{q}+1\in2\N^+$ and $N=4n-2\text{ for }n\in\N^+$ \cite{onishchik2012lie}. For such cases, one uses $\e_{ab}$ instead of $\de_{ab}$.}
\be 
\left(\cO^{i_1\dots i_n;b}\right)^\dagger=&\frac{1}{n!}\delta_{ab}\epsilon_{i_1\dots i_{n}j_1\dots j_{n}}\cO^{j_1\dots j_{n};a}\;,\\
\cO^{i_1\dots i_n;b}=&\frac{1}{n!}\delta_{ab}\left(\cO^{j_1\dots j_n;a}\right)^\dagger\epsilon^{j_1\dots j_{n}i_1\dots i_{n}}\;.
\ee 

\subsection{Index free notation for \texorpdfstring{$SU(N_f)$}{SU(N)} tensors}
\label{sec: index-free notation}
One can represent arbitrary mixed representations of the $SU(N)$ group as polynomials of a set of commuting and anticommuting variables by constructing explicit projector operators in the basis of fundamental indices.\footnote{Interested readers can consult \cite{Costa:2014rya,Costa:2016hju} for examples of such projectors and the related illustrative birdtrack notation.} Instead of doing this, we will follow a similar approach to \cite{Kos:2014bka} and use a less systematic yet more practical approach by mixing fundamental and antifundamental indices.

Let us first consider operators of the form $\cO^{k_1|k_2|\dots|k_n}_{l_1|l_2\dots|l_m}$.\footnote{These correspond to representations with the Young diagram $\begin{aligned}
	\text{\scriptsize \begin{ytableau}1& 2&\none[\dots]&m&1&2&\none[\dots]&n\cr 2& &\none[\dots]&\cr\none[\vdots]&\none[]&\none[]&\none[\vdots]\cr $N-1$&&\none[\dots]& \end{ytableau}}
	\end{aligned}$,
where the adjoint operator is the special case with $m=n=1$.
} We can construct this tensor with auxiliary bosonic vectors $u^i$ and $\ub_i$ and define $\cO(u,\ub)\coloneqq \left(\prod_{i=1}^m u^{l_i}\right)\left(\prod_{i=1}^n \ub_{k_i}\right)\cO^{k_1|k_2|\dots|k_n}_{l_1|l_2\dots|l_m}$. One can reconstruct the tensor as $\cO^{k_1|k_2|\dots|k_n}_{l_1|l_2\dots|l_m}=\left[
\left(\prod_{i=1}^m \frac{\partial}{\partial u^{l_i}}\right)\left(\prod_{i=1}^n \frac{\partial}{\partial \ub_{k_i}}\right)\cO(u,\ub) - \text{traces}
\right]$. As $u\.\ub$ only contributes to the trace, we drop such terms in $\cO(u,\ub)$.\footnote{More precisely, $\cO(u,\ub)$ is only defined modulo the ideal of functions proportional to $u\.\ub$, hence we can restrict $\cO(u,\ub)$ to the locus  $u\.\ub=0$. For similar index-free techniques, see \cite{Giombi:2011rz,Costa:2011mg,SimmonsDuffin:2012uy}.}

For more general tensors, we need to consider other auxiliary vectors and further constraints on the polynomial. For an operator of the most general form $\cO^{k_1k_2\dots k_n|\dots}_{l_1l_2\dots l_m|\dots}$, we have the polynomial form $\cO(u^{(1)},\dots u^{(m)},\ub^{(1)},\dots \ub^{(n)})$. Symmetrization of the indices are already satisfied as we are multiplying with the same vectors for indices in the same row; to satisfy antisymmetrization between indices in different rows, we impose the constraints
\small 
\be 
u^{(a)}\.\frac{\partial}{\partial u^{(b)}}\cO(u^{(1)},\dots u^{(m)},\ub^{(1)},\dots \ub^{(n)})=0\;,\\ 
\ub^{(a)}\.\frac{\partial}{\partial \ub^{(b)}}\cO(u^{(1)},\dots u^{(m)},\ub^{(1)},\dots \ub^{(n)})=0\quad\text{ for }a\ne b\;.
\ee 
\normalsize
By using these constraints alongside $u^{(a)}\.\ub^{(b)}=0$ for any $a$ and $b$, we can construct correlation functions as polynomials of auxiliary vectors.\footnote{One can always can get back the explicit tensorial form by differentiating and subtracting the indices, however we actually do not need tensor forms for practical purposes.}

We can illustrate this with the trivial case of the two-point function of the adjoint operators:
\be 
\<\cO_{\mathsf{Adj}}\left(u_1^{(1)},\ub_1^{(1)}\right)
\cO_{\mathsf{Adj}}\left(u_2^{(1)},\ub_2^{(1)}\right)
\>\propto u_1^{(1)}\.\ub_2^{(1)}u_2^{(1)}\.\ub_1^{(1)}\;.
\ee 
As a more detailed example, let us consider the two-point function of $\mathsf{(A\bar{S})}$ and $\mathsf{(S\bar{A})}$ operators. These operators are dual of each other and have the Young diagrams $\begin{aligned}
	\text{\scriptsize \begin{ytableau}{}&&\cr &&\cr &\cr\none[\vdots] &\none[\vdots] \cr$N-1$ &$N-1$\end{ytableau}}
\end{aligned}$ and $\begin{aligned}
\text{\scriptsize \begin{ytableau}{}&&\cr\cr\none[\vdots]\cr$N-2$ \end{ytableau}}
\end{aligned}$ respectively. The two-point function then reads as
\small 
\be 
\<
\cO_{\mathsf{(A\bar{S})}}\left(u_1^{(1)},u_1^{(2)},\ub_1^{(1)}\right)\cO_{\mathsf{(S\bar{A})}}\left(\ub_2^{(1)},\ub_2^{(2)},u_2^{(1)}\right)
\>\\\propto \left(u_1^{(1)}\.\ub_2^{(1)}u_1^{(2)}\.\ub_2^{(2)}-u_1^{(1)}\.\ub_2^{(2)}u_1^{(2)}\.\ub_2^{(1)}\right)\left(\ub_1^{(1)}\.u_2^{(1)}\right)^{2}\;,
\ee
\normalsize
which is the only combination that
\begin{itemize}
	\item has the correct order in each term,
	\item is free of $u_i\.\ub_i$,
	\item satisfies the necessary conditions
	\be 
	\begin{aligned}
		u_i^{(a)}\.\frac{\partial}{\partial u_i^{(b)}}\<
		\cO_{\mathsf{(A\bar{S})}}\left(\ub_1^{(1)},\ub_1^{(2)}\right)\cO_{\mathsf{(S\bar{A})}}\left(u_2^{(1)},u_2^{(2)}\right)
		\>=&0\\
		\ub_i^{(a)}\.\frac{\partial}{\partial \ub_i^{(b)}}\<
		\cO_{\mathsf{(A\bar{S})}}\left(\ub_1^{(1)},\ub_1^{(2)}\right)\cO_{\mathsf{(S\bar{A})}}\left(u_2^{(1)},u_2^{(2)}\right)
		\>=&0
	\end{aligned}
	\\\text{ for }a\ne b=1,2\text{ and }i=1,2\;.
	\ee 
\end{itemize}

We can similarly write down three-point functions of two external adjoint operators as follows:
\bea[eq: three point su(n) structures]
\<\cO_{\mathsf{Adj}}
\cO_{\mathsf{Adj}}
\cO_{\mathsf{S\bar{A}}}
\>\propto& \left(U_{31}U_{32}\right)V_{1323}^{1112}
\;,\\
\<\cO_{\mathsf{Adj}}
\cO_{\mathsf{Adj}}
\cO_{\mathsf{A\bar{S}}}
\>\propto&\left(U_{31}U_{32}\right)^*\left(V_{1323}^{1112}\right)^*
\;,\\
\<\cO_{\mathsf{Adj}}
\cO_{\mathsf{Adj}}
\cO_{\mathsf{S\bar{S}}}
\>\propto&\left(U_{13}U_{23}\right)\left(U_{13}U_{23}\right)^*
\;,\\
\<\cO_{\mathsf{Adj}}
\cO_{\mathsf{Adj}}
\cO_{\mathsf{A\bar{A}}}
\>\propto&V_{1323}^{1112}\left(V_{1323}^{1112}\right)^*
\;,\\
\label{eq: adjoint three-point structure}
\<\cO_{\mathsf{Adj}}
\cO_{\mathsf{Adj}}
\cO_{\mathsf{Adj}}
\>\propto&\left(U_{12}U_{23}U_{31}\right)\pm \left(U_{12}U_{23}U_{31}\right)^*
\;,\\
\<\cO_{\mathsf{Adj}}
\cO_{\mathsf{Adj}}
\cO_{\mathsf{Singlet}}
\>\propto&U_{12}U_{12}^*\;,
\eea 
where
\be
V_{ijkl}^{abcd}\coloneqq U_{ij}^{ab}U_{kl}^{cd}-U_{il}^{ad}U_{kj}^{cb} 
\ee 
for
\be 
U_{ij}^{ab}\coloneqq u_i^{(a)}\.\ub_j^{(b)}\;,\quad \left(U_{ij}^{ab}\right)^*\coloneqq \ub_i^{(a)}\.u_j^{(b)}
\ee 
with the shorthand notation $U_{ij}\equiv U_{ij}^{11}$. We observe that there are two structures for 3 adjoint operators (self-dual and anti-self-dual) and that the structures for $\mathsf{A\bar{S}}$ and $\mathsf{S\bar{A}}$ are dual of each other. All the other structures are evidently self-dual.

Once we include monopole operators, the number of auxiliary variables depend on $N$, hence we will focus on $N=4$ below. For external $\cO_{\mathsf{M}}$ and $\cO_{\cM^\dagger}$, we have
\be 
\<\cO_{\mathsf{M}}
\cO_{\mathsf{M}^\dagger}
\cO_{\mathsf{Singlet}}
\>\propto& V_{1212}^{1122}
\;,\\
\<\cO_{\mathsf{M}}
\cO_{\mathsf{M}^\dagger}
\cO_{\mathsf{Adj}}
\>\propto& U_{32}^{11}V_{1312}^{1122}-U_{32}^{12}V_{1312}^{1121}
\;,\\
\<\cO_{\mathsf{M}}
\cO_{\mathsf{M}^\dagger}
\cO_{\mathsf{A\bar{A}}}
\>\propto& V_{1313}^{1122}\left(V_{2323}^{1122}\right)^*\;,
\ee 
which satisfy all the necessary conditions stated above. In addition, we can explicitly check that the structures are invariant under the combined action of conjugation with permutation of first two external operators, i.e. under $U_{ab}^{ij}\rightarrow\left(U_{ab}^{ij}\right)^*\evaluated_{\substack{u_1^{(a)}\leftrightarrow u_2^{(a)}\\\ub_1^{(a)}\leftrightarrow \ub_2^{(a)}}}$.\footnote{For instance, one can show that $\left(V_{1212}^{1122}\right)^*=V_{2121}^{1122}$ hence $V_{1212}^{1122}$ is indeed invariant after conjugation followed by 1$\leftrightarrow2$ in the lower-stair indices. To show the invariance of such structures, the identities $\left(V_{ijkl}^{abcd}\right)^*=V_{lkji}^{dcba}$, $V_{ijkl}^{abcd}=-V_{ilkj}^{adcb}=-V_{kjil}^{cbad}$ become useful.} 

A basis of four-point functions can also be constructed as polynomials of auxiliary vectors; for instance, for four external adjoint operators, there are 9 such structures:
\be 
T_{13}T_{24}\;,\quad T_{12}T_{34}\pm T_{14}T_{23} \;,\quad T_{1234}\pm T_{1423}\;,\\ T_{1243}\pm T_{1324}\;,\quad T_{1342}\pm T_{1432}
\ee 
for 
\be 
\label{eq: four point structures in index free notation}
T_{i_1i_2\dots i_n}\coloneqq U_{i_1i_2}U_{i_2i_3}\cdots U_{i_ni_1}
\ee 
where we choose combinations that are invariant under $1\leftrightarrow3$ exchange modulo a sign.

\subsection{Setup of the crossing equations}
Let us consider a four-point function $\<A_{1m}B_{2n}C_{3r}D_{4p}\>$ where $\cO_{im}\equiv\cO_m(x_i)$ for the collective global symmetry index $m$. We also assume in this section that an operator $A$ is in the representation $a$ of the global group.

In this notation, we have the conformal block decomposition
\small 
\be 
\label{eq: conformal block expansion}
\<A_{1m}B_{2n}C_{3p}D_{4r}\>=\frac{1}{x_{12}^{\De_A+\De_B}x_{34}^{\De_C+\De_D}}\left(\frac{x_{24}}{x_{14}}\right)^{\De_{AB}}\left(\frac{x_{14}}{x_{13}}\right)^{\De_{CD}}\\\x\sum\limits_{\substack{O\in A\times B\\O^\dagger\in C\times D\\i,j}}(-1)^{l_\cO}\lambda_{ABO}^{(i)}\lambda_{CDO^\dagger}^{(j)} \left(T_o^{abcd}\right)_{mnpr}^{(ij)}g_O^{ABCD}(u,v)
\ee 
\normalsize
where $T$ is the global symmetry four-point tensor structure. The summation $i,j$ is over the multiplicity of the representation $o,\bar{o}$.

For bosonic operators, we can go to a kinematic regime where $\<A_{1m}B_{2n}C_{3p}D_{4r}\>=\<C_{3p}B_{2n}A_{1m}D_{4r}\>$ by fixing the conformal frame as
\small 
\be 
\label{eq: conformal frame}
x_1=(0,0,\vec{0}),\; x_2=\left(\frac{z-\zb}{2i},\frac{z+\zb}{2},\vec{0}\right),\; x_3=(0,1,\vec{0}),\; x_4=(0,\infty,\vec{0})
\ee 
\normalsize
with $u=z\zb$ and $v=(1-z)(1-\zb)$;\footnote{An operator at infinity is defined as $\cO_4\equiv\cO(x_4)=\lim\limits_{L\rightarrow\infty}L^{2\De_{\cO}}\cO(0,L,\vec{0})$.} this leads to
\small 
\be 
	\sum\limits_{\substack{O\in A\times B\\O^\dagger\in C\times D\\i,j}}(-1)^{l_\cO}\lambda_{ABO}^{(i)}\lambda_{CDO^\dagger}^{(j)} \left(T_o^{abcd}\right)_{mnpr}^{(ij)}F_{\pm,O}^{ABCD}(u,v)
	\\\mp\sum\limits_{\substack{O\in C\times B\\O^\dagger\in A\times D\\i,j}}(-1)^{l_\cO}\lambda_{CBO}^{(i)}\lambda_{ADO^\dagger}^{(j)} \left(T_o^{cbad}\right)_{pnmr}^{(ij)}F_{\pm,O}^{CBAD}(u,v)=0
\ee 
\normalsize
where we added/subtracted $(u\leftrightarrow v)$ from the original equation. We also defined
\small 
\be 
\label{eq: definition of F and H}
F_{\pm,O}^{ABCD}(u,v)\equiv v^{\frac{\De_B+\De_C}{2}}g_O^{ABCD}(u,v)\pm u^{\frac{\De_B+\De_C}{2}}g_O^{ABCD}(v,u)\;.
\ee 
\normalsize
The crossing equation simplifies for certain correlators; for instance, for $\<ABAB\>$, it reads as
\be 
\sum\limits_{\substack{O,O^\dagger\in A\times B\\i,j}}(-1)^{l_\cO}\lambda_{ABO}^{(i)}\lambda_{ABO^\dagger}^{(j)} \Bigg(\left(T_o^{abab}\right)_{mnpr}^{(ij)}\\\mp \left(T_o^{abab}\right)_{pnmr}^{(ij)}\Bigg)F_{\pm,O}^{ABAB}(u,v)=0\;.
\ee 

The global symmetry tensor structure $T$ can be fixed once the three-point structures are chosen. To set this convention, we can define the OPE expansion as
\be 
\label{eq: OPE expansion}
A_{1m}B_{2n}=\sum\limits_{\substack{O\in A\times  B\\o\in a\times b\\i}}\lambda_{ABO}^{(i)}\left(t^{ab}_o\right)_{mn}^{(i)s}c_{ABO}(x_1,x_2,\partial_5)O_{5s}\;,
\ee 
where $\left(t^{ab}_o\right)_{mn}^{(i)s}$ are three-point structures of the global group and $c(x_1,x_2,\partial_5)$ is a differential operator containing the information of the descendants of $O$.\footnote{We are suppressing the contracted spacetime indices of the operator $\cO$ and the structure $c_{ABO}$.} If we apply this OPE inside a three-point function, we see that
\small 
\be 
\label{eq: three point structure}
\<A_{1m}B_{2n}O^\dagger_{3t}\>=\sum\limits_i\lambda_{ABO}^{(i)}\left(t^{ab}_o\right)_{mn}^{(i)s}\left(\delta^{o\bar{o}}\right)_{st}\<A_1B_2O_3^\dagger\>\;,
\ee 
\normalsize
where $\<A_1B_2O_3^\dagger\>\equiv c_{ABO}(x_1,x_2,\partial_5)\<O_5O^\dagger_3\>$ is the standard three-point structure of the conformal group with global symmetry dependence stripped off.\footnote{For 3d CFTs, we can write it down as $\frac{\<S_3X_1X_2S_3\>^l}{(X_1\.X_2)^{\#}(X_2\.X_3)^{\#}(X_3\.X_1)^{\#}}$ up to an overall factor in the embedding space formalism, where $X$ and $S$ are the position vector and auxiliary spinor respectively.} This structure has the symmetry $\<A_1B_2O_3^\dagger\>=(-1)^l\<B_2A_1O_3^\dagger\>$; as we also have $\<A_{1m}B_{2n}O^\dagger_{3t}\>=\<B_{2n}A_{1m}O^\dagger_{3t}\>$ for bosonic operators $A$ and $B$, we conclude
\be 
\label{eq: symmetry of  global symmetry three point structure}
\sum\limits_i\lambda_{ABO}^{(i)}\left(t^{ab}_o\right)_{mn}^{(i)s}=(-1)^l\sum\limits_i\lambda_{BAO}^{(i)}\left(t^{ba}_o\right)_{nm}^{(i)s}\;.
\ee 

By applying the OPE twice in a four-point function, we find the relations
\be 
\label{eq: conformal block convention}
\hspace*{-2.5em}
c_{ABO}(x_1,x_2,\partial_5)c_{CDO^\dagger}(x_3,x_4,\partial_5)\<O_5O_5^\dagger\>=\\
\frac{(-1)^{l_\cO}}{x_{12}^{\De_A+\De_B}x_{34}^{\De_C+\De_D}}\left(\frac{x_{24}}{x_{14}}\right)^{\De_{AB}}\left(\frac{x_{14}}{x_{13}}\right)^{\De_{CD}}g_O^{ABCD}(u,v)\;,
\ee 
and
\be 
\left(T_o^{abcd}\right)_{mnpr}^{(ij)}=\left(t^{ab}_o\right)_{mn}^{(i)s}
\left(t^{cd}_{\bar{o}}\right)_{pr}^{(j)t}
\left(\delta^{o\bar{o}}\right)_{st}\;.
\ee
With \equref{eq: symmetry of  global symmetry three point structure}, one can use the latter equation to obtain several relations.\footnote{One can immediately write down
	\bea[eq: symmetries of T general case]
	\sum\limits_i
	\lambda_{ABO}^{(i)}\left(T_o^{abcd}\right)_{mnpr}^{(ij)}=(-1)^{l_O}\sum\limits_i\lambda_{BAO}^{(i)}\left(T_o^{bacd}\right)_{nmpr}^{(ij)}
	\;,\\
	\sum\limits_j
	\lambda_{CDO}^{(j)}\left(T_o^{abcd}\right)_{mnpr}^{(ij)}=(-1)^{l_O}\sum\limits_j\lambda_{DCO}^{(j)}\left(T_o^{abdc}\right)_{mnrp}^{(ij)}\;.
	\eea 
	By using this, we can also obtain further relations; for instance,
	\be 
	\label{eq: symmetries of T}
	\sum\limits_{i,j}\lambda_{AAO}^{(i)}\lambda_{BBO^\dagger}^{(j)}\left(T_o^{aabb}\right)^{(ij)}_{mnpr}=\sum\limits_{\substack{i,j\\\text{even } l}}\lambda_{AAO}^{(i)}\lambda_{BBO^\dagger}^{(j)}\left(T_o^{aabb}\right)_{\{mn\}\{pr\}}^{(ij)}\\+\sum\limits_{\substack{i,j\\\text{odd } l}}\lambda_{AAO}^{(i)}\lambda_{BBO^\dagger}^{(j)}\left(T_o^{aabb}\right)_{[mn][pr]}^{(ij)}
	\ee 
	for $O_{\{ab\}}\equiv\frac{1}{2}\left(O_{ab}+O_{ba}\right)$ and $ O_{[ab]}\equiv\frac{1}{2}\left(O_{ab}-O_{ba}\right)$.
}

With all the conventions set up, we can finally choose our conformal block normalization. For this, we consider the normalization of the differential operator $c_{ABO}(x_1,x_2,\partial_5)$. In the OPE limit, we choose it such that\footnote{This form is schematic in that it only determines the overall scaling while suppressing the spacetime tensor structure.} 
\small 
\be 
\hspace*{-1em}
c_{ABO}(x_1,x_2,\partial_3)\sim\sqrt{\frac{2^l\G(l+\half)}{\sqrt{\pi}\G(l+1)}}x_{12}^{\De_\cO-\De_A-\De_B}\;,\quad\abs{x_{12}}\ll\abs{x_{13}},\abs{x_{23}}\;.
\ee 
\normalsize
With \equref{eq: conformal block convention}, this fixes the normalization of the conformal block as in the second row of Table 1 of \cite{Poland:2018epd}, i.e. $g(z,\zb)\sim\frac{\G(l+\half)}{\sqrt{\pi}\G(l+1)}z^h\zb^{\hb}$ for $0<z\ll\zb \ll1$.\footnote{For further details on the relation between $c_{ABO}$ and the conformal block normalization, one can refer to \cite{Karateev:2018oml}.} 

The global symmetry structures $\left(t^{ab}_o\right)_{mn}^{(i)s}$ in \equref{eq: three point structure} can be computed in various ways; for instance, one can compute them as explicit tensors \cite{Berkooz:2014yda}, or one can use index-free formalism to write them down as we did in \equref{eq: three point su(n) structures}. We will not dwell on the details here, but only present how reflection positivity fixes the overall signs of certain structures in our conventions. For this, we look at a reflection positive configuration of Hermitian operators $A$ and $B$; \equref{eq: conformal block expansion} becomes
\be 
\<A_{1m}B_{2n}B_{3p}A_{4r}\>\propto\sum\limits_{\substack{O\in A\times B\\O^\dagger\in B\times A\\i,j}}(-1)^{l_\cO}\lambda_{ABO}^{(i)}\lambda_{BAO^\dagger}^{(j)}\\\x\left(T_o^{abba}\right)_{mnpr}^{(ij)}g_O^{ABBA}(u,v)
\ee 
up to a positive proportionality constant. Via \equref{eq: symmetries of T general case}, this indicates
\small 
\be 
\hspace*{-1em}
\<A_{1m}B_{2n}|\cO|B_{3p}A_{4r}\>\propto\sum\limits_{\substack{i,j}}\lambda_{ABO}^{(i)}\lambda_{ABO^\dagger}^{(j)} \left(T_o^{abab}\right)_{mnrp}^{(ij)}g_O^{ABBA}(u,v)\;.
\ee 
\normalsize
For $A_{1m}=\left(A_{4r}\right)^\dagger$ and $B_{2n}=\left(B_{3p}\right)^\dagger$, the left hand side can be interpreted as the norm of a state in radial quantization, hence needs to be positive. We then conclude\footnote{Note that this relies on our choice that the conformal block is normalized to be positive.} $\sum\limits_{\substack{i,j}}\lambda_{ABO}^{(i)}\lambda_{ABO^\dagger}^{(j)}\left(T_o^{abab}\right)_{mnrp}^{(ij)}\ge 0$, or rather
\be 
\label{eq: positivity constraint for global symmetry tensor structures}
\begin{pmatrix}
	\left(T_o^{abab}\right)_{mnrp}^{(11)} & \left(T_o^{abab}\right)_{mnrp}^{(12)} &  \\
	\left(T_o^{abab}\right)_{mnrp}^{(21)} & \left(T_o^{abab}\right)_{mnrp}^{(22)} & \\
	&&\ddots
\end{pmatrix}\succeq 0 \;,\\\text{ for }A_{1m}=\left(A_{4r}\right)^\dagger\;,\; B_{2n}=\left(B_{3p}\right)^\dagger
\ee 
for real $\lambda$, which is the case for real scalars.

\subsection{Direct computation of the correlators in mean field theory limit}
In \secref{\ref{sec: Fermion bilinear scalar bootstrap}}, we discussed the importance of fermion bilinears in the exploration of QED$_3$ via nonperturbative methods. On the other hand, explicit computations in the mixed conformal bootstrap setup can be computationally demanding. One regime where computations can actually be done in a relatively straightforward manner is the mean field theory limit, where correlators can be computed via Wick contractions. Although such a MFT is expected to be rather unrelated to the physical QED$_3$, a better grasp of its correlators can nevertheless be useful. This is particularly true in the large spin limit, where the spectrum of any CFT approaches asymptotically to that of the MFT.

We start by considering the operators
\be 
\cO^m_i=&\overline\psi^m\psi_i-\frac{1}{N}\delta_i^m\overline\psi^k \psi_k\,,
\\
\cO'^{m}_i=&\frac{1}{\sqrt{N}}\left[
\left(\overline\psi^{[m}\psi_{[k}\right)\left(\overline\psi^{k]}\psi_{i]}\right)
-\frac{\delta_i^m}{N}
\left(\overline\psi^{[l}\psi_{[k}\right)\left(\overline\psi^{k]}\psi_{l]}\right)
\right]\,,
\\
\cO''^{m}_i=&\frac{1}{\sqrt{N}}(\overline\psi^k \psi_k) \cO^m_i\,,
\ee 
where $\psi$ is a Dirac fermion in the conventions of \secref{\ref{sec: group theory conventions}}.\footnote{
	These operators are also studied in \cite{Chester:2016ref}, except they work with
	\be 
	\tl\cO'^{m}_i=&\frac{1}{\sqrt{N}}\sum\limits_{k=3}^N(\psi_{(\a_1})_{[i}(\psi_{\a_2)})_{k]}(\overline\psi^{(\a_1})^{[m}(\overline\psi^{\a_2)})^{k]}- SU(N)\text{ traces}
	\ee 
	instead of $\cO'^{m}_i$. These two operators are equal if the summation range above is extended down to $k=1$.
} We now define the following operators
\small 
\be 
A_i^m=&\frac{i}{\sqrt{2}}\cO^m_i\,,\\
B_i^m=&-\frac{i}{\sqrt{2}}\sqrt{\frac{N}{N-1}}\cO''^m_i\,,\\
C_i^m=&-\sqrt{\frac{4(N-1) N}{3(N-2) (N+2)}}\left(
\cO'^m_i-i \frac{3 (N-2)}{2 \sqrt{2} (N-1)}  \cO''^m_i
\right)\,,
\ee 
\normalsize
which are orthonormal in the sense that
\be 
\hspace*{-1em}
\<A_1A_2\>=\frac{U_{12}}{x_{12}^{4\De}}\;,\quad
\<B_1B_2\>=\frac{U_{12}}{x_{12}^{8\De}}\;,\quad
\<C_1C_2\>=\frac{U_{12}}{x_{12}^{8\De}}\;,\\
\<X_1Y_2\>=0\; \text{ if } X\ne Y
\ee 
where we are using the index-free notation introduced in \secref{\ref{sec: index-free notation}}.\footnote{These equations follow from the normalization of the the Dirac field $\psi$ such that its real and imaginary parts are normalized as two independent Majorana fermions $\xi$ and $\chi$:
	\be 
	\<\chi^{\a,m}\left(x_1\right)\chi_{\b,i}\left(x_2\right)\>=\<\xi^{\a,m}\left(x_1\right)\xi_{\b,i}\left(x_2\right)\>=\frac{i}{2}\frac{(x_{12})^{\a}_{\;\;\b}}{x_{12}^{2\De+1}}\delta^m_i\;,\quad 
	\<\xi^{\a,m}\left(x_1\right)\chi_{\b,i}\left(x_2\right)\>=0
	\ee }

We can now treat $A^m_i$ as the lightest parity-odd adjoint bilinear scalar, whereas  $B^m_i$ and $C^m_i$ are the lightest parity-even adjoint bilinear scalars. Therefore, we can consider various correlators such as $\<AAB\>$ or $\<BBB\>$ and extract the OPE coefficients in the MFT limit. Performing the explicit computation, we find
\be 
\<A_1A_2X_3^{(1)}\>=\lambda_{AAX}\frac{T_{123}+T_{213}}{v^{2\De}}
\;,\\ 
\<X_1^{(1)}X_2^{(2)}X_3^{(3)}\>=\lambda_{X^{(1)}X^{(2)}X^{(3)}}\frac{T_{123}+T_{213}}{u^{2\De}v^{2\De}}\;,\\ 
\text{ for } X^{(i)}=B,C
\ee 
where $T_{i_1i_2\dots i_n}$ are defined in \equref{eq: four point structures in index free notation} and the operators are in the conformal frame of \equref{eq: conformal frame}. The OPE coefficients read as 
\footnotesize 
\be 
\lambda_{AAB}=&\frac{1}{2 \sqrt{N-1}}\,,
 &
\lambda_{AAC}=&
\frac{\sqrt{3} N}{2 \sqrt{(N-2) (N-1) (N+2)}}\,,
\\
\lambda_{BBB}=&\frac{3 N-4}{2 (N-1)^{3/2}}\,,
 &
\lambda_{BBC}=&
\frac{\sqrt{3} N
	\sqrt{\frac{N-2}{N^2+N-2}}}{2 (N-1)}\,,
\\
\lambda_{BCC}=&\frac{N^3-12 N+8}{2 (N-2) (N-1)^{3/2} (N+2)}\,,
 &
\lambda_{CCC}=&\frac{(N-4) \left(3 N^3+10 N^2+28 N-32\right)}{6 \sqrt{3}
	(N-2)^{3/2} (N-1)^{3/2} (N+2)^{3/2}}\,.
\ee 
\normalsize
For $N=4$, they become 
\be 
\lambda_{AAB}=\frac{1}{2 \sqrt{3}}\,,\quad
\lambda_{AAC}=\frac{1}{\sqrt{3}}\,,\quad
\lambda_{BBB}=\frac{4}{3\sqrt{3}}\,,\\
\lambda_{BBC}=\frac{2}{3 \sqrt{3}}\,,\quad
\lambda_{BCC}=\frac{1}{3 \sqrt{3}}\,.\quad
\lambda_{CCC}=0
\ee 

By using Wick contractions, we can also compute the four-point correlators and then compare them with the conformal block expansion in \equref{eq: conformal block expansion} to extract $F_{\pm,O}^{ABCD}(u,v)$ as defined in \equref{eq: definition of F and H}. For instance, for $\<AAAA\>$, if we define
\be 
\label{eq: via Wick rotation 2}
\<A_i^mA_j^nA_k^pA_l^r\>=\sum\limits_\a f_\a^{AAAA}(x_1,x_2,x_3,x_4)(\mathfrak{t}_\a)_{ijkl}^{mnpr}
\ee 
for various four-point tensor structures $(\mathfrak{t}_\a)_{ijkl}^{mnpr}$, we  can extract $F_{\pm,O}^{AAAA}(u,v)$ from the equation
\be 
\label{eq: extraction of F from MFT correlators}
\hspace*{-2em}
\left[\sum\limits_\a \left(u v\right)^{2\De}f_\a^{AAAA}(u,v)(\mathfrak{t}_\a)_{mnpr}\right]\pm \left[u\leftrightarrow v\right]=\\
\sum\limits_{\substack{O,O^\dagger \in A\times A\\i,j}}(-1)^{l_\cO}\lambda_{AAO}^{(i)}\lambda_{AAO}^{(j)} \left(T_o^{aaaa}\right)_{mnpr}^{(ij)} F_{\pm,O}^{AAAA}(u,v)
\ee 
by matching the structures $(\mathfrak{t}_\a)_{ijkl}^{mnpr}$ with different pieces of  $\left(T_o^{aaaa}\right)_{mnpr}^{(ij)}$.\footnote{Equation \eqref{eq: extraction of F from MFT correlators} generalizes to other correlators with the single modification that $(u v)^{2\De}$ is replaced by $(uv)^{4\De}$, $u^{2\De}v^{3\De}$, $(uv)^{3\De}$, and $u^{3\De}v^{2\De}$ for the correlators $\<BBBB\>$, $\<AABB\>$, $\<ABAB\>$, and $\<BAAB\>$ respectively (or any correlator with $C$ instead of $B$).
}

In this convention, we can explicitly compute that 
\small 
\be 
	\<A_1A_2A_3A_4\>=T_{13}T_{24}+\frac{1}{u^{2\De}}T_{12}T_{34}
	+
	\frac{1}{v^{2\De}}T_{14}T_{23}\\
	-\frac{u+v-1}{4 u^{\De+\half}v^{\De+\half}}\left(
	T_{1234}+T_{1432}
	\right)
	-\frac{1+u-v}{4 u^{\De+\half}}\left(
	T_{1243}+T_{1342}
	\right)\\
	-\frac{1-u+v}{4 v^{\De+\half}}\left(
	T_{1324}+T_{1423}
	\right)\,,
\ee 
\normalsize
where $T_{i_1i_2\dots i_n}$ are defined in \equref{eq: four point structures in index free notation}. We can now use 
\equref{eq: extraction of F from MFT correlators} and explicitly compute \mbox{$\mathcal{F}_{\pm,O}^{ij}\equiv
	(-1)^{l_\cO}\lambda_{AAO}^{(i)}\lambda_{AAO}^{(j)}  F_{\pm,O}^{AAAA}(u,v)$} as:
\footnotesize
\bea[eq: F and H in the MFT limit]
\mathcal{F}_{-,\mathsf{Adj}^+}^{11}=&\frac{N u^{\Delta } \left(N (u-1) u^{\Delta } v^{\Delta }+N \sqrt{u} v^{2 \Delta +\frac{1}{2}}+8 \sqrt{v} u^{\Delta }\right)}{16
	\left(N^2-4\right) \sqrt{v}}\;-\;\left(u\leftrightarrow v\right)\,,
\\
\mathcal{F}_{-,\mathsf{Adj}^-}^{22}=&\frac{1}{16} u^{\Delta } \left(-\frac{8 u^{\Delta }}{N}+(u-1) u^{\Delta } v^{\Delta -\frac{1}{2}}+\sqrt{u} v^{2 \Delta }\right)\;-\;\left(u\leftrightarrow v\right)\,,
\\
\mathcal{F}_{-,\mathsf{Singlet}}^{11}=&\frac{u^{\Delta } \left(-4 \left(N^2-2\right) \sqrt{v} u^{\Delta }+N (u-1) u^{\Delta } v^{\Delta }+N \sqrt{u} v^{2 \Delta +\frac{1}{2}}\right)}{4
	\left(N^2-1\right) \sqrt{v}}\nn\\&-\;\left(u\leftrightarrow v\right)\,,
\\
\mathcal{F}^{11}_{-,\mathsf{S\bar{S}}}=&\frac{1}{16} u^{\Delta } \left(4 u^{\Delta }-(u-1) u^{\Delta } v^{\Delta -\frac{1}{2}}-\sqrt{u} v^{2 \Delta }\right)\;-\;\left(u\leftrightarrow v\right)\,,
\\
\mathcal{F}^{11}_{-,\mathsf{ReA\bar{S}}}=&-\frac{u^{2 \Delta }}{2}\;-\;\left(u\leftrightarrow v\right)\,,
\\
\mathcal{F}^{11}_{-,\mathsf{A\bar{A}}}=&\frac{1}{16} u^{\Delta } \left(4 u^{\Delta }+(u-1) u^{\Delta } v^{\Delta -\frac{1}{2}}+\sqrt{u} v^{2 \Delta }\right)\;-\;\left(u\leftrightarrow v\right)\,,
\\
\mathcal{F}^{11}_{+,\mathsf{Adj}^+}=&\frac{N u^{2 \Delta }}{2 \left(N^2-4\right)}+\left(\frac{1}{16-4 N^2}+\frac{1}{16}\right) u^{2 \Delta } v^{\Delta -\frac{1}{2}}\nn\\&+\left(\frac{1}{4
	\left(N^2-4\right)}-\frac{1}{16}\right) u^{2 \Delta +1} v^{\Delta -\frac{1}{2}}
+\frac{N u^{2 \Delta } v^{2 \Delta }}{2
	\left(N^2-4\right)}\nn\\&+\left(\frac{1}{16-4 N^2}+\frac{1}{16}\right) u^{\Delta +\frac{1}{2}} v^{2 \Delta }-\frac{1}{16} u^{\Delta -\frac{1}{2}}
v^{\Delta -\frac{1}{2}}+\frac{1}{8} u^{\Delta +\frac{1}{2}} v^{\Delta -\frac{1}{2}}\nn\\&+\;\left(u\leftrightarrow v\right)\,,
\\
\mathcal{F}^{22}_{+,\mathsf{Adj}^-}=&-\frac{u^{2 \Delta }}{2 N}+\frac{u^{2 \Delta } v^{2 \Delta }}{2 N}+\frac{1}{16} u^{\Delta -\frac{1}{2}} v^{\Delta -\frac{1}{2}}+\frac{1}{16} u^{2
	\Delta } v^{\Delta -\frac{1}{2}}
-\frac{1}{8} u^{\Delta +\frac{1}{2}} v^{\Delta -\frac{1}{2}}\nn\\&-\frac{1}{16} u^{2 \Delta +1} v^{\Delta
	-\frac{1}{2}}+\frac{1}{16} u^{\Delta +\frac{1}{2}} v^{2 \Delta }\;+\;\left(u\leftrightarrow v\right)\,,
\\
\mathcal{F}^{11}_{+,\mathsf{Singlet}}=&\frac{N^2 u^{2 \Delta }}{N^2-1}+\frac{\left(2-N^2\right) u^{2 \Delta } v^{\Delta -\frac{1}{2}}}{4 N-4 N^3}-\frac{\left(2-N^2\right) u^{2 \Delta
		+1} v^{\Delta -\frac{1}{2}}}{4 N-4 N^3}+\frac{u^{2 \Delta } v^{2 \Delta }}{N^2-1}\nn\\&
+\frac{\left(2-N^2\right) u^{\Delta +\frac{1}{2}} v^{2 \Delta
}}{4 N-4 N^3}-\frac{u^{\Delta -\frac{1}{2}} v^{\Delta -\frac{1}{2}}}{4 N}+\frac{u^{\Delta +\frac{1}{2}} v^{\Delta -\frac{1}{2}}}{2 N}\;+\;\left(u\leftrightarrow v\right)\,,
\\
\mathcal{F}^{11}_{+,\mathsf{S\bar{S}}}=&\frac{u^{2 \Delta }}{4}+\frac{1}{16} u^{2 \Delta } v^{\Delta -\frac{1}{2}}-\frac{1}{16} u^{2 \Delta +1} v^{\Delta -\frac{1}{2}}+\frac{1}{4} u^{2
	\Delta } v^{2 \Delta }+\frac{1}{16} u^{\Delta +\frac{1}{2}} v^{2 \Delta }\;\nn\\&+\;\left(u\leftrightarrow v\right)\,,
\\
\mathcal{F}^{11}_{+,\mathsf{ReA\bar{S}}}=&\frac{1}{2} u^{2 \Delta } v^{2 \Delta }-\frac{u^{2 \Delta }}{2}\;+\;\left(u\leftrightarrow v\right)\,,
\\
\mathcal{F}^{11}_{+,\mathsf{A\bar{A}}}=&\frac{u^{2 \Delta }}{4}-\frac{1}{16} u^{2 \Delta } v^{\Delta -\frac{1}{2}}+\frac{1}{16} u^{2 \Delta +1} v^{\Delta -\frac{1}{2}}+\frac{1}{4} u^{2
	\Delta } v^{2 \Delta }-\frac{1}{16} u^{\Delta +\frac{1}{2}} v^{2 \Delta }\nn\\&+\;\left(u\leftrightarrow v\right)\,,
\eea
\normalsize 
where the representations $\mathsf{Adj}^\pm$ are those that come with the structures in \equref{eq: adjoint three-point structure} with the relative sign $\pm$. The computation can straightforwardly be extended to other correlators as explained above; however, we will not be providing explicit results as they are relatively lengthy.

The explicit forms in \equref{eq: F and H in the MFT limit} can be used to check the consistency of the crossing equations. Furthermore, one can use them to fix the overall signs of  the global symmetry tensor structures $\left(T_o^{abcd}\right)_{mnpr}^{(ij)}$ which cannot be fixed by group theory arguments. This is especially useful as the reflection positivity constraint in \equref{eq: positivity constraint for global symmetry tensor structures} is insufficient to fix the signs of \emph{all} tensor structures.

\section{Mixed correlator bootstrap of \texorpdfstring{$SU(4)$}{SU(4)} adjoint scalars with opposite parity charges}
\label{sec:rR-mixing}
In Fig.~\ref{multifbbd} we presented the fermion bilinear single correlator bootstrap results, which show interesting kinks in different channels after imposing gaps inspired by the perturbative QED$_3$ spectrum. One may expect to obtain stronger bootstrap results and even restrict the CFT data into a closed island by bootstrapping mixed correlators, reminiscent to the remarkable success in \cite{Kos:2014bka}. In addition to mixing with monopole operators, another simple candidate for the mixed correlator bootstrap study is the lowest scalar $R$ in the parity even $SU(4)$ adjoint representation. There are yet other interesting candidates for the mixed correlator bootstrap studies, such as the lowest scalar in the  $(422)$ representation of $SU(4)$ and the lowest spin 1 operator in the  real combination of $ ((310)+ (332))^-$ representation. Nevertheless, their mixed correlator bootstrap implementations are much more challenging. 

\begin{figure}
\centering
\hspace*{-1em}
$\begin{aligned}
\includegraphics[width=0.75\linewidth]{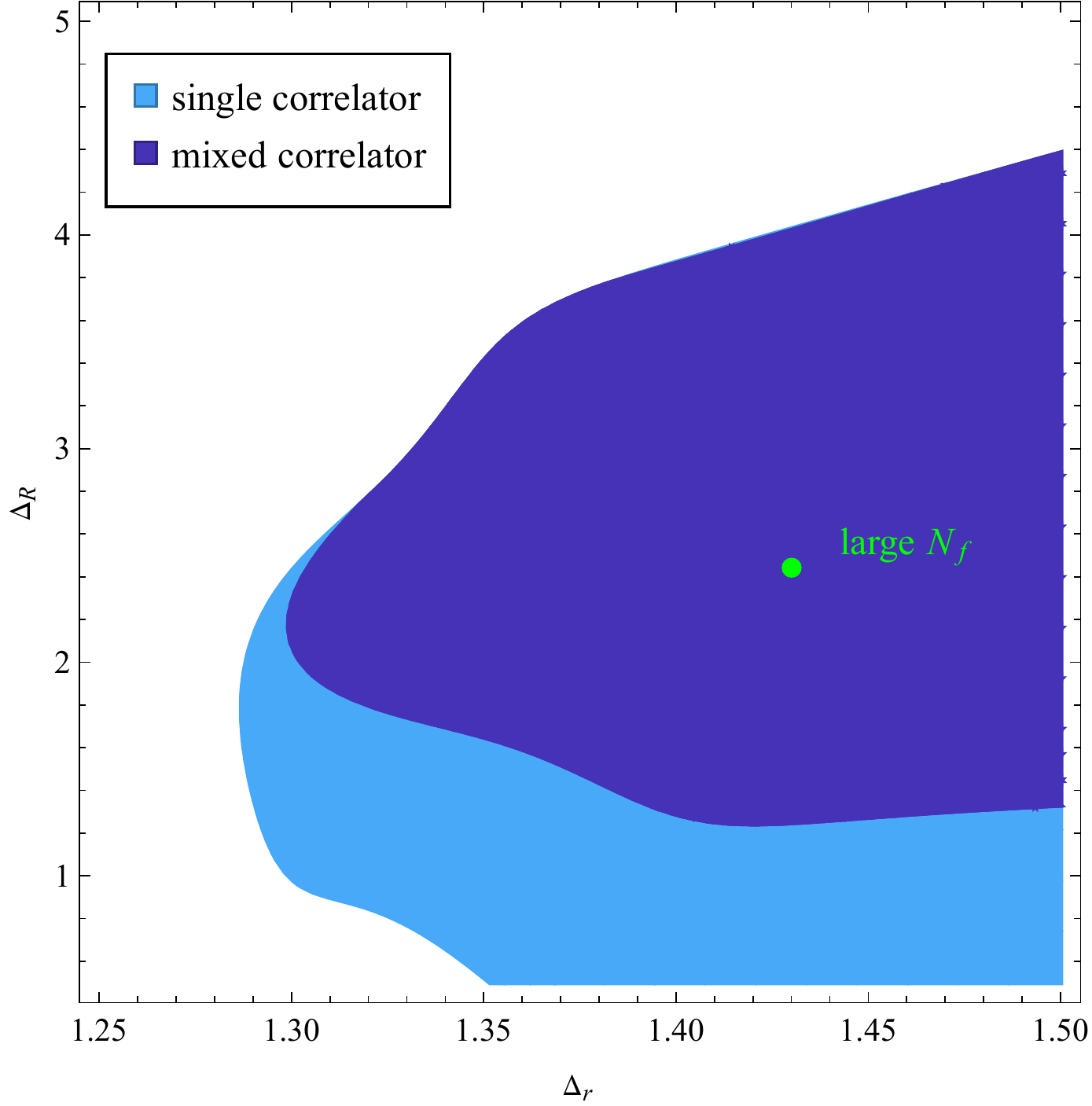}
\end{aligned}
\;\;
    \begin{aligned}
	\Delta_{r'}  &\geqslant 3.0 \\
	\Delta_{R'}  &\geqslant 4.0 \\
	\Delta_{{\rm Singlet},+}^{\ell=0} &\geqslant 3.0 \\
	\Delta_{{\rm Singlet},-}^{\ell=0} &\geqslant 2.0 \\
	\Delta_{A\bar{A},+}^{\ell=0} &\geqslant 2.8 \\
	\Delta_{A\bar{A},-}^{\ell=0} &\geqslant 3.0 \\
	\Delta_{S\bar{S},+}^{\ell=0} &\geqslant 3.5 \\
	\Delta_{S\bar{S},-}^{\ell=0} &\geqslant 3.0 \\
	\Delta_{S\bar{A},+}^{\ell=1} &\geqslant 3.0 \\
	\Delta_{S\bar{A},-}^{\ell=0} &\geqslant 3.0 \\
\end{aligned}
$
\caption{
Bootstrap bounds on the scaling dimensions of the fermion bilinear scalar $r$ and the lowest parity even $SU(4)$ adjoint scalar $R$ appearing in the $r\times r$ OPE.
The light shaded region represents the bound from the single correlator bootstrap with an external scalar $r$ at $\Lambda=19$, and the dark shaded region denotes the bound from the mixed correlator bootstrap with external scalars $r$ and $R$.
} \label{mp:adjmx}
\end{figure}

The results of our preliminary exploration of the mixed correlator bootstrap with external scalars $r$ and $R$ are shown in Fig.~\ref{mp:adjmx}. By introducing a gap $4.0$ for the second lowest scalar in the $(2,1,1)$ sector, there is a mild lower bound on the scaling dimension $\Delta_R$ from the single correlator bootstrap, which becomes stronger in the mixed correlator bootstrap results. This suggests the mixed correlator bootstrap indeed can help to generate a stronger bound. However, the lower bound on $\Delta_R$ obtained from the mixed correlator bootstrap is not close to the kink in the upper bound or the large $N_f$ perturbative result. The results suggest it is hard to further isolate the kinks in the single correlator bootstrap bound into a closed region using this mixed correlator bootstrap. This may not be surprising. As mentioned in our discussion for the bootstrap results in Fig.~\ref{T2J1}, it is hard to distinguish conformal QED$_3$ from QCD$_3$ in the bootstrap bounds on the scaling dimensions of fermion bilinear and 4-fermion operators, as both of them share a similar low-lying spectrum. However, they have significantly different central charges. It might be interesting to further explore the roles of conserved currents and their associated central charges in the bootstrap studies of conformal QED$_3$ in mixed correlator bootstraps involving 4-fermion operators.

\section{More details on the large \texorpdfstring{$N_f$}{Nf} mode construction}
\label{sec: large n mode construction}
In this section we give more details on the computations of the spectrum at large $N_f$, primarily following~\cite{Pufu:2013vpa,Chester:2017vdh}.
\subsection{Monopole harmonics}
\label{sec: monopole harmonics}
First, we review the spinor monopole spherical harmonics described in \cite{Pufu:2013vpa}. In terms of the scalar monopole spherical harmonics $Y_{q,lm}$ used in \equref{eq: scalar monopole spherical harmonics}, they read as
\be 
\label{eq:spinorHarmonics}
T_{q,lm}(\theta,\phi) \equiv
\colvec{
\sqrt{\frac{\ell+m+1}{2\ell+1} } Y_{q,lm}(\theta,\phi) \\
\sqrt{\frac{\ell-m}{2\ell+1} } Y_{q,l(m+1)}(\theta,\phi) 
}\;,\\ 
S_{q,lm}(\theta,\phi) \equiv \colvec{
-\sqrt{\frac{\ell-m}{2\ell+1} } Y_{q,lm}(\theta,\phi) \\
\sqrt{\frac{\ell+m+1}{2\ell+1} } Y_{q,l(m+1)}(\theta,\phi) 
} \,.
\ee 
The wave functions in 
\equref{eq:largeNModeExpansion}
are defined as
\begin{equation}
\label{eq:monopoleHarmonics}
\begin{aligned}
A_{qlm} &= \frac{q T_{q,lm}+(\lambda_l+l+1/2)S_{q,lm}}
{ \sqrt{(2l+1)(l+1/2+\lambda_l)} }\,, \\
B_{qlm} &= \frac{q T_{q,lm}+(\lambda_l-l-1/2)S_{q,lm}}
{ \sqrt{(2l+1)(l+1/2-\lambda_l)} }\,, \\
C_{q,q-1/2,m} &= S_{q,q-1/2,m}\,.
\end{aligned}
\end{equation}

The equation above along with \equref{eq: transformation of spinor monopole spherical harmonics} indicates that $A_{qlm}$ and $B_{qlm}$ does not transform nicely under space parity unless $q=0$. For $q=0$, we have\footnote{
The general relation reads as
\be 
\label{eq: transformation of spinor monopole spherical harmonics}
\text{Space Parity: }\quad
X_{q,lm}(\theta,\phi)\rightarrow(-1)^{l+m}e^{2iq\phi}X_{-q,lm}(\theta,\phi)\quad\text{ for }X=T,S
\ee 
which follows from the application of \equref{eq: transformation of scalar monopole spherical harmonics} for the spinor monopole spherical harmonics $T_{q,lm}^{\a}(x)$ and $S_{q,lm}^{\a}(x)$ defined in \cite{Pufu:2013vpa}.
} 
\small 
\be 
\label{eq: parity of A and B}
\hspace*{-1em}
\text{Space Parity: }\quad
X_{0lm}(\theta,\phi)\rightarrow(-1)^{l+m}X_{0lm}(\theta,\phi)\quad\text{ for }X=A,B
\ee 
\normalsize

Therefore, we can implement the parity transformation in the Hilbert space in a straightforward fashion for the $q=0$ sector. If we define
\be 
\cP \psi(x^0,x^1,x^2)\cP^{-1}=\g^2\psi(x^0,x^1,-x^2)
\ee 
we conclude via \equref{eq:largeNModeExpansion} and \equref{eq: parity of A and B} that for the $q=0$ sector,
\begin{equation}
	\label{eq:oscillatorReflectionRule}
	\begin{aligned}
		\cP a_{lm}^{i,\dagger} \cP^{-1} &= (-1)^{\ell+m} a_{lm}^{i,\dagger}\,, \\
		\cP b_{lm}^{i} \cP^{-1} &= (-1)^{\ell+m} b_{lm}^{i}\,,
	\end{aligned}
\end{equation}
where $c_{q-1/2,m}^{i,\dagger}$ does not show up in $q=0$ sectors.

\subsection{Construction of the large \texorpdfstring{$N_f$}{Nf} states}

Now we discuss how to construct the large $N_f$ states from the oscillator modes in (\ref{eq:largeNModeExpansion}). Schematically, this takes 3 steps:
\begin{itemize}
\item First, take all possible combinations of creation operators $a_{jm}^{i,\dagger}$, $b_{jm,i}^{\dagger}$, and $c_{q-1/2,m}^{i,\dagger}$ below a certain energy $E_{\max}$ that are charge neutral.
\item Then, for each string of creation operators, construct all $SU(N_f)$ and $SO(3)$ reps in the product of reps of individual operators.
\item Finally, for each representation, try anti-symmetrizing the identical fermion creation operators.
\end{itemize}

\subsubsection*{Selection of operators}
\label{sec:selecting-operators}
The first step is straightforward. The $c^\dagger$ operator has zero energy so there is always a ground state populated by $c^\dagger$ only. The operators $a_{jm}^{i,\dagger}$ and $b_{jm,i}^{\dagger}$ have energy
\begin{equation}
\lambda_j = \sqrt{(j+1/2)^2-q^2} \, .
\end{equation}
Since their spin is bounded from below $j \geq q+1/2$, the lowest energy of a single oscillator is $\lambda_{q+1/2} = \sqrt{2q+1}$, which sets an upper bound on the total number of $a_{jm}^{i,\dagger}$ and $b_{jm,i}^{\dagger}$
\begin{equation}
n_a + n_b \leq \frac{E_{\max}}{\sqrt{2q+1}} \, .
\end{equation}
After this, the gauge charge neutrality and $n_b\geq 0$ requires
\begin{equation}
n_a \leq \min\left\{ n_a + n_b, -q \pr{k-\frac{N}{2}} + \frac{n_a+n_b}{2} \right\} \, .
\end{equation}
Finally, the largest spin of each operator $a_{jm}^{i,\dagger}$ and $b_{jm,i}^{\dagger}$ is bounded by $E_{\max}$
\begin{equation}
j \leq \sqrt{E_{\max}^2 + q^2} -\frac{1}{2} \, .
\end{equation} 
These constraints leave us finitely many possible combinations. We can exhaust these possibilities and select those below $E_{\max}$.

\subsubsection*{Constructing \texorpdfstring{$SU(N_f)$}{SU(Nf)} and \texorpdfstring{$SO(3)$}{SO(3)} reps}
\label{sec:largeN-constructing-reps}
In this step we focus on the $SU(N_f)$ and $SO(3)$ states separately, and treat for now each creation operator as a distinct particle. An $SU(N_f)$ state corresponds to a tensor
\small 
\begin{equation}
| T \> \equiv 
T_{i_1,i_2,\cdots,i_{n_a}}^{k_1,k_2,\cdots,k_{n_b}}
a^{i_1,\dagger}a^{i_2,\dagger}\cdots a^{i_{n_a},\dagger} \,
b_{k_1}^\dagger b_{k_2}^\dagger\cdots b_{k_{n_b}}^\dagger 
| M_{\rm bare} \> \, ,
\end{equation}
\normalsize
where the spin indices are suppressed. To project to a certain representation, we diagonalize the quadratic Casimir operator
\begin{equation}
C_2 | T, r \> = c_2(r) | T, r \> \, .
\end{equation}
Similarly, we associate each $SO(3)$ state to a tensor
\begin{equation}
| U \> \equiv 
U^{m_1,m_2,\cdots,m_{n}}
| j_1, m_1 \> \otimes | j_2, m_2 \> \otimes \cdots \otimes 
| j_{n}, m_{n} \> \, ,
\end{equation}
where $-j_i \leq m_i \leq j_i$, 
and again diagonalize the $SO(3)$ quadratic Casimir 
\begin{equation}
L^2 | U, j \> = j(j+1) | U, j \> \, .
\end{equation}
We collect all eigenstates for the next step.

\subsubsection*{Anti-symmetrization}
\label{sec:anti-symmetrization}
Potentially, a state $| \Psi_{j,r} \>$ of spin $j$ transforming in an $SU(N)$ rep $r$ live in the linear space
\begin{equation}
| \Psi_{j,r} \> \in \underset{k,\ell}{\rm span} \big\{  \, | U_k, j \> \otimes | T_\ell, r \> \big\}\,,
\end{equation}
and we fully anti-symmetrize it to make it fermionic. If all creation operators $a_{jm}^{i,\dagger}$, $b_{jm,i}^{\dagger}$, and $c_{q-1/2,m}^{i,\dagger}$ have distinct quantum numbers, then the anti-symmetrization is trivial. However, if there are two or more operators of the same type $a,b,$ or $c$ having the same spin, we need to check if there is at least a state in the above space that is anti-symmetric under the permutation between those operators. We take as an example the states created by four identical $c^\dagger$ operators to explain the procedure to determine whether certain reps can show up.
\paragraph{Example: $q=1$ sector ground state representation}
\begin{equation}
| \Psi \> \sim \left( c_{1/2,m}^\dagger \right)^4 | M_{\rm bare} \> \,.
\end{equation}
This is also the ground state of the $q=1$ sector in the $N_f=4$ case. After brute-force diagonalizing the Casimir matrix, we obtain some number of eigenvectors in the reps listed below:
\small 
\begin{equation}
\label{eq:exampleLargeNRepsDimension}
\begin{array}{|c|c|c|c|c|}
\hline
c_2(r) & 0 & 4 & 6 & 8 \\
\hline
{\rm dimension} & 1 & 45 & 40 & 135 \\
\hline
\end{array}
\qquad 
\begin{array}{|c|c|c|c|c|}
\hline
j & 0 & 1 & 2 \\
\hline
{\rm dimension} & 2 & 9 & 5 \\
\hline
\end{array}\,\,.
\end{equation}
\normalsize
Note that the dimension of the eigenvector space is multiple times the dimension of the rep. This is because we may construct the same rep from different tensor contraction, and they mix when we permute the particles. We would like to study how the $c^\dagger$s' permutation group acts on the states. The generators of permutation group $\mathbb{Z}_n$ of $n$ particles are $(n-1)$ subsequent permutations, in our case $R_{12}$, $R_{23}$ and $R_{34}$. The matrix representation of these generators are, for example for $R_{12}$
\begin{align}
(R_{12}^{(r)})_{ik} \equiv \< T_i^{(2134)}, r | T_k^{(1234)}, r \>\,, \\
(R_{12}^{(j)})_{ik} \equiv \< U_i^{(2134)}, j | U_k^{(1234)}, j \> \, .
\end{align}
To show that an anti-symmetric state exists, we just need to find a common eigenvector of eigenvalue $(-1)$ for all three product matrices: $R_{12}^{(r)} \otimes R_{12}^{(j)}$, $R_{23}^{(r)} \otimes R_{23}^{(j)}$, and $R_{34}^{(r)} \otimes R_{34}^{(j)}$. Because $R_{12}$ and $R_{23}$ don't commute, generically we cannot simultaneously diagonalize them both, but the all-minus and all-plus sectors can be simultaneously diagonalized.

It may be tempting to try reducing this problem to individual matrices $R_{\alpha\beta}^{(r)}$ and $R_{\alpha\beta}^{(j)}$. The argument would sound like the following: The eigenvalue of the Kronecker product matrix $R_{\alpha\beta}^{(r)} \otimes R_{\alpha\beta}^{(j)}$ is the product of constituents, thus the eigenvalues of $R_{\alpha\beta}^{(r)}$ and $R_{\alpha\beta}^{(j)}$ individually must be either $(+1,-1)$ or $(-1,+1)$. But this implies that each eigenvector we find would be an eigenvector of all 6 matrices, which is in tension with the fact that the permutation operators don't commute. Indeed a straightforward check shows that this is not the case. What is wrong? The issue is that an eigenvector of the Kronecker product matrix doesn't necessarily factorize into a Kronecker product, so our target state may not have definitive permutation parity if projected to either $|T\>$ space or $|U\>$ space, but is anti-symmetric in the space of the product group representation. This makes the problem much harder because we are forced to run an eigenvalue problem on Kronecker product matrices which have huge dimension. 

To speed up the computation, we use the Lanczos method~\cite{lanczos1950iteration} to find the eigenvectors of eigenvalue $(-1)$. Lanczos method is a variational ansatz that aims at minimizing the expectation value of a matrix. Schematically, to diagonalize a Hermitian matrix $H$, we project $H$ to a basis spanned by
\begin{equation}
\{ \Psi, H\Psi, H^2\Psi, \cdots \}\,,
\end{equation} 
where $\Psi$ is the initial condition, and diagonalize the sub-matrix. The lowest eigenvalue of the sub-matrix is an approximation of the lowest eigenvalue of the whole matrix. If $H$ has big sparsity, which is the case in our example, the approximation will converge with a much smaller dimensional basis than the full dimension of $H$. Since we would like to find a state with eigenvalue $(-1)$ of all 3 matrices, we define
\begin{equation}
\hspace*{-1em}
H\Psi \equiv \left( 
\frac{3}{5} R_{12}^{(r)} \otimes R_{12}^{(j)}
+ \frac{5}{7} R_{23}^{(r)} \otimes R_{23}^{(j)}
+ \frac{7}{11} R_{34}^{(r)} \otimes R_{34}^{(j)}
\right) \Psi\,,
\end{equation}
where $\Psi$ has dimension $(\underset{k}{\dim}\{| U_k, j \> \} \times \underset{\ell}{\dim}\{| T_\ell, r \>\} )$. We use the Lanczos method to find the lowest eigenvalue of $H$. 
The 3 matrices will have eigenvalue $(-1)$ if and only if the eigenvalue of $H$ is $-\frac{3}{5}-\frac{5}{7}-\frac{7}{11} \approx -1.95065$, and that all other eigenvalues of $H$ are larger.
If the eigenvalue converges to $-1.95065$, then we conclude that an anti-symmetric state exists. Otherwise, the eigenvalue will converge to a greater value, and we conclude that a fermionic state that is constructed from 4 identical $c^\dagger$'s and transforms in reps $(r,j)$ does not exist. 

Using this method, we check the existence of anti-symmetric states in each pair of reps in (\ref{eq:exampleLargeNRepsDimension}). The result is the following:
\begin{equation}
\begin{array}{|c|c|c|c|}
\hline 
 & j=0 & j=1 & j=2 \\
\hline 
c_2 = 0 & {\rm No} & {\rm No} & {\rm Yes} \\
\hline 
c_2 = 4 & {\rm No} & {\rm Yes} & {\rm No} \\
\hline
c_2 = 6 & {\rm Yes} & {\rm No} & {\rm No} \\
\hline 
c_2 = 8 & {\rm No} & {\rm No} & {\rm No} \\
\hline
\end{array}\,\,.
\end{equation}
The lowest $q=1$ scalar monopole is indeed in the $SU(4)$ rep $\mathsf{A\bar{A}}$ which has $c_2=6$.

\subsection{Implications of the parity symmetry for uncharged sectors}
In \secref{\ref{sec: monopole harmonics}}, we discussed that the creation operators transform irreducibly under the space parity transformation in the $q=0$ sector. Indeed, $c_{q-1/2,m}^{i,\dagger}$ does not show up in this sector and $a_{jm}^{i,\dagger}$ and $b_{jm,i}^{\dagger}$ simply get a sign $(-1)^{j+m}$ under reflection. For an operator made of several $a^{\dagger}$'s and $b^{\dagger}$'s, $\CO \sim a_{j_1 m_1}^{\dagger} \cdots b_{j_n m_n}^{\dagger}$, the internal parity is the product of the signs of each constituents, factoring out the total $(-1)^{j_\CO+m_\CO}$; thus,
\begin{equation}
\label{eq:parityRuleNeutral}
\text{internal parity of }\CO = (-1)^{(\sum_i j_i) - j_{\CO}} \, .
\end{equation}

We can check this explicitly for several low-dimension operators. For instance, we know that $\widebar{\psi}\psi$ \& $\widebar{\psi}\g^\mu\slashed{\d} \psi$ are parity-odd whereas  $\widebar{\psi}\g^\mu \psi$, $\widebar{\psi}\slashed{\d} \psi$, and $\widebar{\psi}\slashed{\d}\g^\mu\slashed{\d} \psi$ are parity-even.\footnote{One can explicitly check this using $\psi\rightarrow\g^2\psi$ and $\widebar{\psi}\rightarrow-\widebar{\psi}\g^2$ along with some gamma algebra identities; however, we can see this more simply by group-theoretical arguments. Under the $\mathrm{Pin}(2,1)$ group, we label the representations as $j^p$ where $j$ is the usual spin and $p$ is the parity of the representation. We then have the branching $j_1^{p_1}\otimes j_2^{p_2}=\left(j_1+j_2\right)^{p_1 p_2}\oplus \left(j_1+j_2-1\right)^{-p_1 p_2}\oplus\cdots\oplus\abs{j_1-j_2}^{\pm}$ where parities alternate between representations. If we choose the fermions to have positive parity (this does not affect anything for operators containing an even number of fermions), we see that $\half^+\otimes\half^+=1^+\otimes 0^-$; hence, the scalar $\widebar{\psi}\psi$ has odd parity whereas the vector $\widebar{\psi}\g^\mu\psi$ has even parity.} In the large $N_f$ limit, $\Delta_\psi = 1$ and the operators have their engineering dimensions. The first two operators $\widebar\psi\psi$ and $\widebar\psi \gamma^\mu \psi$ have dimension 2, so they must be made of a pair of lowest spin creation operators, $a_{1/2}^\dagger b_{1/2}^\dagger$. Using (\ref{eq:parityRuleNeutral}) we determine that the scalar is parity odd and vector is parity even. Next we have two dimension-3 operators. In our construction the dimension-3 scalar does not exist, and the vector is made of $a_{3/2}^\dagger b_{1/2}^\dagger$ or $a_{1/2}^\dagger b_{3/2}^\dagger$. In either case, the parity is odd. Finally, the dimension-4 vector is made of $a_{3/2}^\dagger b_{3/2}^\dagger$, and has even parity.

Combining the parity rule with the large $N_f$ state construction discussed in the last subsection, we can write a summary of the $q=0$ sector as the table (up to 6 particles and energy level 6)
\footnotesize 
\begin{equation}
\begin{array}{|c|l|l|l|l|l|}
\hline
  & \mathsf{singlet} & \mathsf{Adj} & \mathsf{A\bar{A}} & \mathsf{S\bar{A}} & \mathsf{S\bar{S}} \\
\hline
 j=0 & 0^+,\,2^-,\,4^{\pm },\,5^-,\,6^{\pm } & 2^-,\,4^{\pm },\,5^-,\,6^{\pm } & 4^+,\,5^-,\,6^{\pm } & 5^-,\,6^{\pm } & 4^+,\,6^{\pm } \\
\hline
 j=1 & 2^+,\,3^-,\,4^{\pm },\,5^{\pm },\,6^{\pm } & 2^+,\,3^-,\,4^{\pm },\,5^{\pm },\,6^{\pm } & 4^-,\,5^+,\,6^{\pm } & 4^-,\,5^+,\,6^{\pm } & 5^+,\,6^{\pm } \\
\hline
 j=2 & 3^+,\,4^{\pm },\,5^{\pm },\,6^{\pm } & 3^+,\,4^{\pm },\,5^{\pm },\,6^{\pm } & 4^+,\,5^-,\,6^{\pm } & 5^-,\,6^{\pm } & 5^-,\,6^+ \\
\hline
 j=3 & 4^+,\,5^{\pm },\,6^{\pm } & 4^+,\,5^{\pm },\,6^{\pm } & 5^+,\,6^- & 5^+,\,6^- & 5^+,\,6^- \\
\hline
\end{array} \,\, ,
\end{equation}
\normalsize
where the number and superscript sign are the dimension and parity of the corresponding operator, respectively, and $\pm$ means both parity odd and even operators can be found at this dimension. The parity even operators appear in the $S$ sector of $M\times M$ OPE, and the parity odd operators appear in the $A$ sector.

\section{Mixed crossing equations between the lowest monopole \texorpdfstring{$\CM_{1/2}$}{M1/2} and the fermion bilinear \texorpdfstring{$r$}{r} } \label{rMmixing}

\begin{table}
	\centering
\caption{\label{tab:summarize-SU4}
	A summary of the conformal blocks and the OPE coefficients in the SU(4) mixed monopole-fermion-bilinear correlators.}

\begin{tabular}{ccccc}
\hline\hline\\[-1em]
$\SU(4)$ \textbf{name}& \textbf{Young tableaux} & $\SO(2)$ \textbf{rep} & \textbf{Spin} & \textbf{OPE}
\\[.1em]\hline\\[-1em]
		$\begin{array}{c}\mathsf{Singlet}\\{(000)}\end{array} $ 
	& $\bullet$ 
	& $\begin{array}{c}S\\A\\T\end{array}$
	&\begin{tabular}{l}even\\odd\\even\end{tabular} 
	&$\begin{array}{l}
		\lambda _{rrO},\lambda _{MMO} \\
		\lambda _{MMO} \\
		\lambda _{MMO} \\
		\end{array}$
\\[.1em]\hline\\[-1em]	
		$\begin{array}{c}\mathsf{Adj}\\{(211)}\end{array} $
	&$\raisebox{-1.2em}{\includegraphics[width=1.6em]{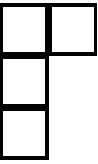}}$
	&$\begin{array}{c}
		S \\
		S \\
		A \\
		T \\
		\end{array}$
	&\begin{tabular}{l}odd\\even\\even\\odd\end{tabular}
	& $\begin{array}{l}
		\lambda _{rrO},\lambda _{MMO} \\
		\lambda _{rrO} \\
		\lambda _{MMO} \\
		\lambda _{MMO} \\
	\end{array}$
\\[.1em]\hline\\[-1em]	
		$\begin{array}{c}\mathsf{AA}\\{(220)}\end{array} $		
	&$\raisebox{-.8em}{\includegraphics[width=1.6em]{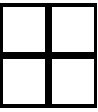}}$
	&$	\begin{array}{c}
		S \\
		A \\
		T \\
	\end{array}$
	&\begin{tabular}{l}even\\odd\\even	\end{tabular}
	&$	\begin{array}{l}
		\lambda _{rrO},\lambda _{MMO} \\
		\lambda _{MMO} \\
		\lambda _{MMO} \\
		\end{array}$
\\[.1em]\hline\\[-1em]	
		$\begin{array}{c}\mathsf{S \bar{A} }\\{(310)_R}\end{array}$
	&$\raisebox{-1.2em}{\includegraphics[width=2.4em]{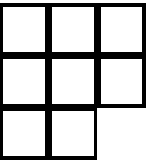}}$
	&$S$
	&\begin{tabular}{l}odd\end{tabular}
	&$\begin{array}{l}
		\lambda _{rrO}\color{white}{,\lambda _{MMO}}
		\end{array}$
\\[.1em]\hline\\[-1em]	
		$\begin{array}{c}\mathsf{ S \bar{S} }\\{(422)}\end{array} $
	&$\raisebox{-1.2em}{\includegraphics[width=3.2em]{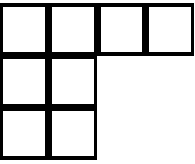}}$
	&$S$
	&\begin{tabular}{l}even	\end{tabular}
	&$\begin{array}{l}
		\lambda _{rrO}\color{white}{,\lambda _{MMO}}
		\end{array}$
\\[.1em]\hline\\[-1em]	
	$\begin{array}{c}\mathsf{Anti}\\{(110)}\end{array} $
	&$\raisebox{-.8em}{\includegraphics[width=.8em]{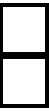}}$
	&$V$
	&\begin{tabular}{l}both	\end{tabular}
	& $\begin{array}{l}
		\lambda _{rMO}\color{white}{,\lambda _{MMO}}
	\end{array}$
\\[.1em]\hline\\[-1em]	
	$\begin{array}{c}\mathsf{Sym}\\{(200)}\end{array}$
	&$\raisebox{-.4em}{\includegraphics[width=1.6em]{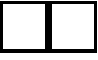}}$
	&$V$
	&\begin{tabular}{l}both	\end{tabular}
	& $\begin{array}{l}
		\lambda _{rMO}\color{white}{,\lambda _{MMO}}
	\end{array}$
\\[.1em]\hline\\[-1em]	
	$\begin{array}{c}\mathsf{AAdj}\\{(321)}\end{array} $
	&$	\raisebox{-1.2em}{\includegraphics[width=2.4em]{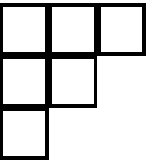}}$
	&$V$
	&\begin{tabular}{l}both	\end{tabular}
	& $\begin{array}{l}
		\lambda _{rMO}\color{white}{,\lambda _{MMO}}
	\end{array}$
\\[.1em]\hline\hline
\end{tabular}
\end{table}

We study the mixed correlator system of the lowest monopole $M\equiv\CM_{1/2} $ and the lowest parity odd adjoint scalar $r$. In addition to the $\<rrrr\>$ correlator discussed in (\ref{eq:rrrrCrossingEq}) and the $\<MMMM\>$ correlator discussed in (\ref{eq:repsMM}) and (\ref{eq:MMMMCrossingEq}), we further have the mixed correlators $\<MMrr\>$, $\<MrrM\>$ and $\<MrMr\>$. In the language of $SO(2)$ and $SU(4)$ representations, $M$ is in the $V,(110)$ representation and $r$ is in the $S,(211)$ representation. The additional tensor product of representation we have in the system is that of $r \times M$,
\begin{equation}
\begin{aligned}\label{eq:MrOPE}
SU(4):\quad& (110) \bigotimes (211)= (110) \bigoplus (200) \bigoplus (321)\, , \\
SO(2):\quad& V \bigotimes S= V \, .
\end{aligned}
\end{equation}
The full crossing equation system from all correlators is thus
\be 
0=\Vec{V}_{\rm \mathbb{1}}
+\sum_{\cO,i^+} 
	\left(\begin{matrix}
	\lambda_{MM\cO} & \lambda_{rr\cO}
	\end{matrix}\right)
	\Vec{V}_{\Delta,\ell}^{S,i^+}
	\left(\begin{matrix}
	\lambda_{MM\cO} \\ \lambda_{rr\cO}
	\end{matrix}\right)
	\\+ \lambda_{MMr}^2
	\Vec{V}_{MMr}
+\sum_{\cO,j} \lambda_{rr\cO}^2\Vec{V}_{\Delta,\ell}^{S,j}
+\sum_{\cO,i^-} \lambda_{MM\cO}^2\Vec{V}_{\Delta,\ell}^{A,i^-}
\\+\sum_{\cO,i^+} \lambda_{MM\cO}^2\Vec{V}_{\Delta,\ell}^{T,i^+}
+\sum_{\cO,k} \lambda_{rM\cO}^2\Vec{V}_{\Delta,\ell}^{V,k}
\ee 
where $i^{\pm}=(000)^\pm,(211)^\mp,(220)^\pm$, $j=(211)^+,(310)_R^-,(422)^+$, $k=(110),(200),(321)$ are the sets of representations and spins appearing in the summation. The $+(-)$ in the superscript of each representation means only even(odd) spins appear in the sum. The operators in representations $SO(2)$ $V$, $SU(4)$ $k$ can have any spin. 
The explicit forms of the vector blocks $\Vec{V}_{\rm \mathbb{1}}$, $\Vec{V}_{MMr}$, $\Vec{V}_{\Delta,\ell}^{S,i^+}$, 
etc. are given in the attached {\texttt Mathematica} notebook. There are in total 18 different channels and 24 crossing equations. Various selection rules from global symmetry representations, parity, and spin control the possible contributions to the OPE in each channel.
We summarize these selection rules in Table \ref{tab:summarize-SU4}. 

The OPE coefficients of the stress tensor $T^{\mu\nu}$, $SU(4)$ conserved current $J^{f\mu}$ and the topological $U(1)$ conserved current $J^{t\mu}$ are constrained by Ward identities in terms of the two-point coefficients $c_T$, $c_J$ and $c_J^t$, respectively. In our conventions, we have
\be 
\hspace*{-1em}
    c_T=\frac{9 \Delta _{M }^2}{4 \lambda _{MMT}^2}=\frac{9 \Delta _{r }^2}{4 \lambda
   _{rr T}^2} \,,
   \;
   \left(
    \begin{array}{c}
     \lambda _{MMJ}^{\text{mix}} \\
     \lambda _{rrJ}^{\text{mix}} \\
    \end{array}
    \right)=\frac{1}{\sqrt{c_J}}\left(
    \begin{array}{c}
     \sqrt{30} \\
     -\sqrt{60} \\
    \end{array}
    \right) \,,
    \\
    {\lambda _{MMJ^t}^{\text{mix}}}^2=\frac{6}{c_J^t} \, .
\ee 

\section{Numerical set-up and implementation}
\label{sec:software}
Our bootstrap computations are run with {\texttt SDPB} \cite{Simmons-Duffin:2015qma, Landry:2019qug} and set up using the packages \cite{hyperion-projects,scalar_blocks} and
 \cite{simpleboot}.
We also used {\texttt autoboot} to cross check the $r$ and $\cM_{1/2}$ mixed correlator crossing equation 
\cite{Go:2019lke}.

\begin{table}
	\begin{center}
		\caption{Parameters for the paper's computations. The sets $S_{19,27,31,39}$ are defined in (\ref{eq:spinsets}).}
		\footnotesize
		\hspace*{-1em} 
		\begin{tabular}{lcccc}
			\hline\hline\\[-1em]
			& \(\Lambda=19\) &  \(\Lambda=27\) & \(\Lambda=31\) & \(\Lambda=39\)\\[.1em]\hline\\[-1em]
			{\texttt{keptPoleOrder}}& 14 &  20 & 32 & 40 \\
			{\texttt{order}}& 60 & 60 & 80 & 90 \\
			{\texttt{spins}} & $S_{19}$ & $S_{27}$ & $S_{31}$ & $S_{39}$ \\
			{\texttt{precision}} & 640 &  640 & 768 & 1024 \\
			{\texttt{dualityGapThreshold}} & $10^{-30}$ & $10^{-30}$ & $10^{-30}$ & $10^{-30}$ \\
			{\texttt{primalErrorThreshold}}& $10^{-200}$& $10^{-200}$ & $10^{-200}$ & $10^{-200}$\\
			{\texttt{dualErrorThreshold}} & $10^{-200}$& $10^{-200}$ & $10^{-200}$& $10^{-200}$\\
			{\texttt{findPrimalFeasible}} & false & false & false & false \\
			{\texttt{findDualFeasible}} & false & false & false & false \\
			{\texttt{detectPrimalFeasibleJump}} & true & true & true & true \\
			{\texttt{detectDualFeasibleJump}} & true & true & true & true \\
			{\texttt{initialMatrixScalePrimal}} & $10^{40}$ & $10^{50}$ & $10^{50}$& $10^{60}$ \\
			{\texttt{initialMatrixScaleDual}} & $10^{40}$ & $10^{50}$ & $10^{50}$& $10^{60}$\\
			{\texttt{feasibleCenteringParameter}} & 0.1 & 0.1 & 0.1 & 0.1 \\
			{\texttt{infeasibleCenteringParameter}} & 0.3 & 0.3 & 0.3 & 0.3\\
			{\texttt{stepLengthReduction}} & 0.7 & 0.7 & 0.7 & 0.7 \\
			{\texttt{maxComplementarity}} & $10^{100}$ & $10^{130}$ & $10^{160}$ & $10^{200}$ \\
			\hline\hline
		\end{tabular}
		\label{islandparameter}
	\end{center}
\end{table}

The {\it interval positivity} condition plays an important role in our bootstrap study, which assumes the spectrum in the bootstrap equations satisfies the constraint: $$\Delta_{0}<\Delta\leqslant \Delta_1 ~~\text{or}~~ \Delta\geqslant\Delta_2,\qquad (\Delta_2>\Delta_1) \,.$$
It is less straightforward in {\texttt SDPB} to impose the positivity condition for the interval range of the scaling dimension ${\Delta_0}<\Delta\leqslant \Delta_1$.
 To do this requires a coordinate transformation to map the interval range to $(0,\infty)$, e.g.
    $$\Delta = {\Delta_0} + \frac{x}{1+x} (\Delta_1-{\Delta_0})\,,$$
    based on which the interval $\Delta\in({\Delta_0},  \Delta_1)$ is mapped to $x>0$. Then the positivity constraint in the whole range $x\in(0, \infty)$ can be effectively studied using {\texttt SDPB}.
    
An alternative setup for the interval positivity constraints is to simply sample the interval $({\Delta_0},  \Delta_1)$ with many isolated points, and refine the sampling until the bounds are well converged. We have done computations where we sample the interval range with step $\delta=0.005$, and find results consistent with the continuous formulation.

For the SDPB calculations, we provide a summary of the numerical parameters in Table \ref{islandparameter}. \footnote{The code used to run these calculations can be found so the reader may run these calculations themselves at the following link and commit: \url{https://gitlab.com/rajeev.erramilli/hyperion-projects/-/tree/fbcdd70673cf17cccc44f205a649fcfa8676130c}}

The spins used in the computations are:
\begin{align}
\label{eq:spinsets}
S_{19} &= \{0,\dots,26\}\cup \{49, 50\}\,, \nn\\
S_{27} &= \{0,\dots,26\}\cup \{29, 30, 33, 34, 37, 38, 41, 42, 45, 46, 49, 50\}\,, \nn\\
S_{31} &= \{0,\dots,44\}\cup \{47, 48, 51, 52, 55, 56, 59, 60, 63, 64, 67, 68\}\,,\nn\\
S_{39} &= \{0,\dots,64\}\cup \{67, 68, 71, 72, 75, 76, 79, 80, 83, 84, 87, 88\}\,.
\end{align}

\bibliography{ReferenceLibrary}{}

\providecommand{\href}[2]{#2}\begingroup\raggedright\begin{thebibliography}{100}

\bibitem{Appelquist:1988sr}
T.~Appelquist, D.~Nash, and L.~C.~R. Wijewardhana, ``{Critical Behavior in
  (2+1)-Dimensional QED},''
  \href{http://dx.doi.org/10.1103/PhysRevLett.60.2575}{{\em Phys. Rev. Lett.}
  {\bfseries 60} (1988) 2575}.

\bibitem{Nash:1989xx}
D.~Nash, ``{Higher Order Corrections in (2+1)-Dimensional QED},''
  \href{http://dx.doi.org/10.1103/PhysRevLett.62.3024}{{\em Phys. Rev. Lett.}
  {\bfseries 62} (1989) 3024}.

\bibitem{Polyakov:1975rs}
A.~M. Polyakov, ``{Compact Gauge Fields and the Infrared Catastrophe},''
  \href{http://dx.doi.org/10.1016/0370-2693(75)90162-8}{{\em Phys. Lett. B}
  {\bfseries 59} (1975) 82--84}.

\bibitem{Polyakov:1976fu}
A.~M. Polyakov, ``{Quark Confinement and Topology of Gauge Groups},''
  \href{http://dx.doi.org/10.1016/0550-3213(77)90086-4}{{\em Nucl. Phys. B}
  {\bfseries 120} (1977) 429--458}.

\bibitem{Pisarski}
R.~D. Pisarski, ``Chiral-symmetry breaking in three-dimensional
  electrodynamics,'' \href{http://dx.doi.org/10.1103/PhysRevD.29.2423}{{\em
  Phys. Rev. D} {\bfseries 29} (May, 1984) 2423--2426}.
  \url{https://link.aps.org/doi/10.1103/PhysRevD.29.2423}.

\bibitem{Appelquist:1986fd}
T.~W. Appelquist, M.~J. Bowick, D.~Karabali, and L.~C.~R. Wijewardhana,
  ``{Spontaneous Chiral Symmetry Breaking in Three-Dimensional QED},''
  \href{http://dx.doi.org/10.1103/PhysRevD.33.3704}{{\em Phys. Rev. D}
  {\bfseries 33} (1986) 3704}.

\bibitem{Dagotto:1989td}
E.~Dagotto, A.~Kocic, and J.~B. Kogut, ``{Chiral Symmetry Breaking in
  Three-dimensional {QED} With $N$(f) Flavors},''
  \href{http://dx.doi.org/10.1016/0550-3213(90)90665-Z}{{\em Nucl. Phys. B}
  {\bfseries 334} (1990) 279--301}.

\bibitem{Rantner:2000wer}
W.~Rantner and X.-G. Wen, ``{Electron spectral function and algebraic spin
  liquid for the normal state of underdoped high $T_c$ superconductors},''
  \href{http://dx.doi.org/10.1103/PhysRevLett.86.3871}{{\em Phys. Rev. Lett.}
  {\bfseries 86} no.~17, (2001) 3871},
  \href{http://arxiv.org/abs/cond-mat/0010378}{{\ttfamily
  arXiv:cond-mat/0010378}}.

\bibitem{Rantner:2002zz}
W.~Rantner and X.-G. Wen, ``{Spin correlations in the algebraic spin liquid:
  Implications for high-Tc superconductors},''
  \href{http://dx.doi.org/10.1103/PhysRevB.66.144501}{{\em Phys. Rev. B}
  {\bfseries 66} (2002) 144501},
  \href{http://arxiv.org/abs/cond-mat/0201521}{{\ttfamily
  arXiv:cond-mat/0201521}}.

\bibitem{Herbut:2002yq}
I.~F. Herbut, ``{QED(3) theory of underdoped high temperature
  superconductors},'' \href{http://dx.doi.org/10.1103/PhysRevB.66.094504}{{\em
  Phys. Rev. B} {\bfseries 66} (2002) 094504},
  \href{http://arxiv.org/abs/cond-mat/0202491}{{\ttfamily
  arXiv:cond-mat/0202491}}.

\bibitem{Franz:2002qy}
M.~Franz, Z.~Tesanovic, and O.~Vafek, ``{QED(3) theory of pairing pseudogap in
  cuprates. 1. From D wave superconductor to antiferromagnet via 'algebraic'
  Fermi liquid},'' \href{http://dx.doi.org/10.1103/PhysRevB.66.054535}{{\em
  Phys. Rev. B} {\bfseries 66} (2002) 054535},
  \href{http://arxiv.org/abs/cond-mat/0203333}{{\ttfamily
  arXiv:cond-mat/0203333}}.

\bibitem{Hermele_2005}
M.~Hermele, T.~Senthil, and M.~P.~A. Fisher, ``Algebraic spin liquid as the
  mother of many competing orders,''
  \href{http://dx.doi.org/10.1103/physrevb.72.104404}{{\em Physical Review B}
  {\bfseries 72} no.~10, (Sep, 2005) }.
  \url{http://dx.doi.org/10.1103/PhysRevB.72.104404}.

\bibitem{Wen2008}
M.~Hermele, Y.~Ran, P.~A. Lee, and X.-G. Wen, ``Properties of an algebraic spin
  liquid on the kagome lattice,''
  \href{http://dx.doi.org/10.1103/physrevb.77.224413}{{\em Physical Review B}
  {\bfseries 77} no.~22, (Jun, 2008) }.
  \url{http://dx.doi.org/10.1103/PhysRevB.77.224413}.

\bibitem{Senthil_2019}
T.~Senthil, D.~T. Son, C.~Wang, and C.~Xu, ``Duality between (2+1)d quantum
  critical points,''
  \href{http://dx.doi.org/10.1016/j.physrep.2019.09.001}{{\em Physics Reports}
  {\bfseries 827} (Sep, 2019) 1–48}.
  \url{http://dx.doi.org/10.1016/j.physrep.2019.09.001}.

\bibitem{Maris:1996zg}
P.~Maris, ``{The Influence of the full vertex and vacuum polarization on the
  fermion propagator in QED in three-dimensions},''
  \href{http://dx.doi.org/10.1103/PhysRevD.54.4049}{{\em Phys. Rev. D}
  {\bfseries 54} (1996) 4049--4058},
  \href{http://arxiv.org/abs/hep-ph/9606214}{{\ttfamily arXiv:hep-ph/9606214}}.

\bibitem{Aitchison:1997ua}
I.~J.~R. Aitchison, N.~E. Mavromatos, and D.~McNeill, ``{Inverse
  Landau-Khalatnikov transformation and infrared critical exponents of
  (2+1)-dimensional quantum electrodynamics},''
  \href{http://dx.doi.org/10.1016/S0370-2693(97)00447-4}{{\em Phys. Lett. B}
  {\bfseries 402} (1997) 154--158},
  \href{http://arxiv.org/abs/hep-th/9701087}{{\ttfamily arXiv:hep-th/9701087}}.

\bibitem{Appelquist:1999hr}
T.~Appelquist, A.~G. Cohen, and M.~Schmaltz, ``{A New constraint on strongly
  coupled gauge theories},''
  \href{http://dx.doi.org/10.1103/PhysRevD.60.045003}{{\em Phys. Rev. D}
  {\bfseries 60} (1999) 045003},
  \href{http://arxiv.org/abs/hep-th/9901109}{{\ttfamily arXiv:hep-th/9901109}}.

\bibitem{Kubota:2001kk}
K.-i. Kubota and H.~Terao, ``{Dynamical symmetry breaking in QED(3) from the
  Wilson RG point of view},'' \href{http://dx.doi.org/10.1143/PTP.105.809}{{\em
  Prog. Theor. Phys.} {\bfseries 105} (2001) 809--825},
  \href{http://arxiv.org/abs/hep-ph/0101073}{{\ttfamily arXiv:hep-ph/0101073}}.

\bibitem{Appelquist:2004ib}
T.~Appelquist and L.~C.~R. Wijewardhana,
  \href{http://dx.doi.org/10.1142/9789812702340_0022}{``{Phase structure of
  noncompact QED3 and the Abelian Higgs model},''} in {\em {3rd International
  Symposium on Quantum Theory and Symmetries}}.
\newblock 3, 2004.
\newblock \href{http://arxiv.org/abs/hep-ph/0403250}{{\ttfamily
  arXiv:hep-ph/0403250}}.

\bibitem{Franz_2003}
M.~Franz, T.~Pereg-Barnea, D.~E. Sheehy, and Z.~Tešanović, ``Gauge-invariant
  response functions in algebraic fermi liquids,''
  \href{http://dx.doi.org/10.1103/physrevb.68.024508}{{\em Physical Review B}
  {\bfseries 68} no.~2, (Jul, 2003) }.
  \url{http://dx.doi.org/10.1103/PhysRevB.68.024508}.

\bibitem{Fischer:2004nq}
C.~S. Fischer, R.~Alkofer, T.~Dahm, and P.~Maris, ``{Dynamical chiral symmetry
  breaking in unquenched QED(3)},''
  \href{http://dx.doi.org/10.1103/PhysRevD.70.073007}{{\em Phys. Rev. D}
  {\bfseries 70} (2004) 073007},
  \href{http://arxiv.org/abs/hep-ph/0407104}{{\ttfamily arXiv:hep-ph/0407104}}.

\bibitem{Kotikov:2016wrb}
A.~V. Kotikov, V.~I. Shilin, and S.~Teber, ``{Critical behavior of ( 2+1
  )-dimensional QED: 1/N$_f$ corrections in the Landau gauge},''
  \href{http://dx.doi.org/10.1103/PhysRevD.94.056009}{{\em Phys. Rev. D}
  {\bfseries 94} no.~5, (2016) 056009},
  \href{http://arxiv.org/abs/1605.01911}{{\ttfamily arXiv:1605.01911
  [hep-th]}}. [Erratum: Phys.Rev.D 99, 119901 (2019)].

\bibitem{Kaveh_2005}
K.~Kaveh and I.~F. Herbut, ``Chiral symmetry breaking in three-dimensional
  quantum electrodynamics in the presence of irrelevant interactions: A
  renormalization group study,''
  \href{http://dx.doi.org/10.1103/physrevb.71.184519}{{\em Physical Review B}
  {\bfseries 71} no.~18, (May, 2005) }.
  \url{http://dx.doi.org/10.1103/PhysRevB.71.184519}.

\bibitem{Giombi:2015haa}
S.~Giombi, I.~R. Klebanov, and G.~Tarnopolsky, ``{Conformal QED$_d$,
  $F$-Theorem and the $\epsilon$ Expansion},''
  \href{http://dx.doi.org/10.1088/1751-8113/49/13/135403}{{\em J. Phys. A}
  {\bfseries 49} no.~13, (2016) 135403},
  \href{http://arxiv.org/abs/1508.06354}{{\ttfamily arXiv:1508.06354
  [hep-th]}}.

\bibitem{DiPietro:2015taa}
L.~Di~Pietro, Z.~Komargodski, I.~Shamir, and E.~Stamou, ``{Quantum
  Electrodynamics in d=3 from the \ensuremath{\varepsilon} Expansion},''
  \href{http://dx.doi.org/10.1103/PhysRevLett.116.131601}{{\em Phys. Rev.
  Lett.} {\bfseries 116} no.~13, (2016) 131601},
  \href{http://arxiv.org/abs/1508.06278}{{\ttfamily arXiv:1508.06278
  [hep-th]}}.

\bibitem{Giombi:2016fct}
S.~Giombi, G.~Tarnopolsky, and I.~R. Klebanov, ``{On $C_{J}$ and $C_{T}$ in
  Conformal QED},'' \href{http://dx.doi.org/10.1007/JHEP08(2016)156}{{\em JHEP}
  {\bfseries 08} (2016) 156}, \href{http://arxiv.org/abs/1602.01076}{{\ttfamily
  arXiv:1602.01076 [hep-th]}}.

\bibitem{DiPietro:2017kcd}
L.~Di~Pietro and E.~Stamou, ``{Scaling dimensions in QED$_3$ from the
  $\epsilon$-expansion},''
  \href{http://dx.doi.org/10.1007/JHEP12(2017)054}{{\em JHEP} {\bfseries 12}
  (2017) 054}, \href{http://arxiv.org/abs/1708.03740}{{\ttfamily
  arXiv:1708.03740 [hep-th]}}.

\bibitem{Zerf:2018csr}
N.~Zerf, P.~Marquard, R.~Boyack, and J.~Maciejko, ``{Critical behavior of the
  QED$_3$-Gross-Neveu-Yukawa model at four loops},''
  \href{http://dx.doi.org/10.1103/PhysRevB.98.165125}{{\em Phys. Rev. B}
  {\bfseries 98} no.~16, (2018) 165125},
  \href{http://arxiv.org/abs/1808.00549}{{\ttfamily arXiv:1808.00549
  [cond-mat.str-el]}}.

\bibitem{Herbut:2016ide}
I.~F. Herbut, ``{Chiral symmetry breaking in three-dimensional quantum
  electrodynamics as fixed point annihilation},''
  \href{http://dx.doi.org/10.1103/PhysRevD.94.025036}{{\em Phys. Rev. D}
  {\bfseries 94} no.~2, (2016) 025036},
  \href{http://arxiv.org/abs/1605.09482}{{\ttfamily arXiv:1605.09482
  [hep-th]}}.

\bibitem{Gusynin:2016som}
V.~P. Gusynin and P.~K. Pyatkovskiy, ``{Critical number of fermions in
  three-dimensional QED},''
  \href{http://dx.doi.org/10.1103/PhysRevD.94.125009}{{\em Phys. Rev. D}
  {\bfseries 94} no.~12, (2016) 125009},
  \href{http://arxiv.org/abs/1607.08582}{{\ttfamily arXiv:1607.08582
  [hep-ph]}}.

\bibitem{Benvenuti:2018cwd}
S.~Benvenuti and H.~Khachatryan, ``{QED's in $2{+}1$ dimensions: complex fixed
  points and dualities},'' \href{http://arxiv.org/abs/1812.01544}{{\ttfamily
  arXiv:1812.01544 [hep-th]}}.

\bibitem{Christofi:2007ye}
S.~Christofi, S.~Hands, and C.~Strouthos, ``{Critical flavor number in the
  three dimensional Thirring model},''
  \href{http://dx.doi.org/10.1103/PhysRevD.75.101701}{{\em Phys. Rev. D}
  {\bfseries 75} (2007) 101701},
  \href{http://arxiv.org/abs/hep-lat/0701016}{{\ttfamily
  arXiv:hep-lat/0701016}}.

\bibitem{Janssen:2012pq}
L.~Janssen and H.~Gies, ``{Critical behavior of the (2+1)-dimensional Thirring
  model},'' \href{http://dx.doi.org/10.1103/PhysRevD.86.105007}{{\em Phys. Rev.
  D} {\bfseries 86} (2012) 105007},
  \href{http://arxiv.org/abs/1208.3327}{{\ttfamily arXiv:1208.3327 [hep-th]}}.

\bibitem{Hands:2020itv}
S.~Hands, M.~Mesiti, and J.~Worthy, ``{Critical behavior in the single flavor
  Thirring model in 2+1D},''
  \href{http://dx.doi.org/10.1103/PhysRevD.102.094502}{{\em Phys. Rev. D}
  {\bfseries 102} no.~9, (2020) 094502},
  \href{http://arxiv.org/abs/2009.02964}{{\ttfamily arXiv:2009.02964
  [hep-lat]}}.

\bibitem{Braun:2014wja}
J.~Braun, H.~Gies, L.~Janssen, and D.~Roscher, ``{Phase structure of
  many-flavor QED$_3$},''
  \href{http://dx.doi.org/10.1103/PhysRevD.90.036002}{{\em Phys. Rev. D}
  {\bfseries 90} no.~3, (2014) 036002},
  \href{http://arxiv.org/abs/1404.1362}{{\ttfamily arXiv:1404.1362 [hep-ph]}}.

\bibitem{Gukov:2016tnp}
S.~Gukov, ``{RG Flows and Bifurcations},''
  \href{http://dx.doi.org/10.1016/j.nuclphysb.2017.03.025}{{\em Nucl. Phys. B}
  {\bfseries 919} (2017) 583--638},
  \href{http://arxiv.org/abs/1608.06638}{{\ttfamily arXiv:1608.06638
  [hep-th]}}.

\bibitem{Hands:2002qt}
S.~J. Hands, J.~B. Kogut, L.~Scorzato, and C.~G. Strouthos, ``{The Chiral limit
  of noncompact QED in three-dimensions},''
  \href{http://dx.doi.org/10.1016/S0920-5632(03)01735-3}{{\em Nucl. Phys. B
  Proc. Suppl.} {\bfseries 119} (2003) 974--976},
  \href{http://arxiv.org/abs/hep-lat/0209133}{{\ttfamily
  arXiv:hep-lat/0209133}}.

\bibitem{Hands:2002dv}
S.~J. Hands, J.~B. Kogut, and C.~G. Strouthos, ``{Noncompact QED(3) with N(f)
  greater than or equal to 2},''
  \href{http://dx.doi.org/10.1016/S0550-3213(02)00869-6}{{\em Nucl. Phys. B}
  {\bfseries 645} (2002) 321--336},
  \href{http://arxiv.org/abs/hep-lat/0208030}{{\ttfamily
  arXiv:hep-lat/0208030}}.

\bibitem{Hands:2004bh}
S.~J. Hands, J.~B. Kogut, L.~Scorzato, and C.~G. Strouthos, ``{Non-compact
  QED(3) with N(f) = 1 and N(f) = 4},''
  \href{http://dx.doi.org/10.1103/PhysRevB.70.104501}{{\em Phys. Rev. B}
  {\bfseries 70} (2004) 104501},
  \href{http://arxiv.org/abs/hep-lat/0404013}{{\ttfamily
  arXiv:hep-lat/0404013}}.

\bibitem{Strouthos:2008kc}
C.~Strouthos and J.~B. Kogut, ``{The Phases of Non-Compact QED(3)},''
  \href{http://dx.doi.org/10.22323/1.042.0278}{{\em PoS} {\bfseries
  LATTICE2007} (2007) 278}, \href{http://arxiv.org/abs/0804.0300}{{\ttfamily
  arXiv:0804.0300 [hep-lat]}}.

\bibitem{Karthik:2015sgq}
N.~Karthik and R.~Narayanan, ``{No evidence for bilinear condensate in
  parity-invariant three-dimensional QED with massless fermions},''
  \href{http://dx.doi.org/10.1103/PhysRevD.93.045020}{{\em Phys. Rev. D}
  {\bfseries 93} no.~4, (2016) 045020},
  \href{http://arxiv.org/abs/1512.02993}{{\ttfamily arXiv:1512.02993
  [hep-lat]}}.

\bibitem{Karthik:2016ppr}
N.~Karthik and R.~Narayanan, ``{Scale-invariance of parity-invariant
  three-dimensional QED},''
  \href{http://dx.doi.org/10.1103/PhysRevD.94.065026}{{\em Phys. Rev. D}
  {\bfseries 94} no.~6, (2016) 065026},
  \href{http://arxiv.org/abs/1606.04109}{{\ttfamily arXiv:1606.04109
  [hep-th]}}.

\bibitem{Karthik:2019mrr}
N.~Karthik and R.~Narayanan, ``{Numerical determination of monopole scaling
  dimension in parity-invariant three-dimensional noncompact QED},''
  \href{http://dx.doi.org/10.1103/PhysRevD.100.054514}{{\em Phys. Rev. D}
  {\bfseries 100} no.~5, (2019) 054514},
  \href{http://arxiv.org/abs/1908.05500}{{\ttfamily arXiv:1908.05500
  [hep-lat]}}.

\bibitem{Karthik:2020shl}
N.~Karthik and R.~Narayanan, ``{QED$_3$-inspired three-dimensional conformal
  lattice gauge theory without fine-tuning},''
  \href{http://dx.doi.org/10.1103/PhysRevLett.125.261601}{{\em Phys. Rev.
  Lett.} {\bfseries 125} no.~26, (2020) 261601},
  \href{http://arxiv.org/abs/2009.01313}{{\ttfamily arXiv:2009.01313
  [hep-lat]}}.

\bibitem{Li:2021emd}
Z.~Li, ``{On conformality and self-duality of $N_f=2$ QED$_3$},''
  \href{http://arxiv.org/abs/2107.09020}{{\ttfamily arXiv:2107.09020
  [hep-th]}}.

\bibitem{Gies:2005as}
H.~Gies and J.~Jaeckel, ``{Chiral phase structure of QCD with many flavors},''
  \href{http://dx.doi.org/10.1140/epjc/s2006-02475-0}{{\em Eur. Phys. J. C}
  {\bfseries 46} (2006) 433--438},
  \href{http://arxiv.org/abs/hep-ph/0507171}{{\ttfamily arXiv:hep-ph/0507171}}.

\bibitem{Kaplan:2009kr}
D.~B. Kaplan, J.-W. Lee, D.~T. Son, and M.~A. Stephanov, ``{Conformality
  Lost},'' \href{http://dx.doi.org/10.1103/PhysRevD.80.125005}{{\em Phys. Rev.
  D} {\bfseries 80} (2009) 125005},
  \href{http://arxiv.org/abs/0905.4752}{{\ttfamily arXiv:0905.4752 [hep-th]}}.

\bibitem{Gorbenko:2018ncu}
V.~Gorbenko, S.~Rychkov, and B.~Zan, ``{Walking, Weak first-order transitions,
  and Complex CFTs},'' \href{http://dx.doi.org/10.1007/JHEP10(2018)108}{{\em
  JHEP} {\bfseries 10} (2018) 108},
  \href{http://arxiv.org/abs/1807.11512}{{\ttfamily arXiv:1807.11512
  [hep-th]}}.

\bibitem{Rattazzi:2008pe}
R.~Rattazzi, V.~S. Rychkov, E.~Tonni, and A.~Vichi, ``{Bounding scalar operator
  dimensions in 4D CFT},''
  \href{http://dx.doi.org/10.1088/1126-6708/2008/12/031}{{\em JHEP} {\bfseries
  12} (2008) 031}, \href{http://arxiv.org/abs/0807.0004}{{\ttfamily
  arXiv:0807.0004 [hep-th]}}.

\bibitem{Poland:2018epd}
D.~Poland, S.~Rychkov, and A.~Vichi, ``{The Conformal Bootstrap: Theory,
  Numerical Techniques, and Applications},''
  \href{http://dx.doi.org/10.1103/RevModPhys.91.015002}{{\em Rev. Mod. Phys.}
  {\bfseries 91} (2019) 015002},
  \href{http://arxiv.org/abs/1805.04405}{{\ttfamily arXiv:1805.04405
  [hep-th]}}.

\bibitem{Chester:2016wrc}
S.~M. Chester and S.~S. Pufu, ``{Towards bootstrapping QED$_{3}$},''
  \href{http://dx.doi.org/10.1007/JHEP08(2016)019}{{\em JHEP} {\bfseries 08}
  (2016) 019},
\href{http://arxiv.org/abs/1601.03476}{{\ttfamily arXiv:1601.03476 [hep-th]}}.

\bibitem{Chester:2017vdh}
S.~M. Chester, L.~V. Iliesiu, M.~Mezei, and S.~S. Pufu, ``{Monopole Operators
  in $U(1)$ Chern-Simons-Matter Theories},''
  \href{http://dx.doi.org/10.1007/JHEP05(2018)157}{{\em JHEP} {\bfseries 05}
  (2018) 157},
\href{http://arxiv.org/abs/1710.00654}{{\ttfamily arXiv:1710.00654 [hep-th]}}.

\bibitem{Li:2018lyb}
Z.~Li, ``{Solving QED$_3$ with Conformal Bootstrap},''
  \href{http://arxiv.org/abs/1812.09281}{{\ttfamily arXiv:1812.09281
  [hep-th]}}.

\bibitem{Li:2020bnb}
Z.~Li and D.~Poland, ``{Searching for gauge theories with the conformal
  bootstrap},'' \href{http://dx.doi.org/10.1007/JHEP03(2021)172}{{\em JHEP}
  {\bfseries 03} (2021) 172}, \href{http://arxiv.org/abs/2005.01721}{{\ttfamily
  arXiv:2005.01721 [hep-th]}}.

\bibitem{Li:2020tsm}
Z.~Li, ``{Symmetries of conformal correlation functions},''
  \href{http://arxiv.org/abs/2006.05119}{{\ttfamily arXiv:2006.05119
  [hep-th]}}.

\bibitem{He:2021xvg}
Y.-C. He, J.~Rong, and N.~Su, ``{A roadmap for bootstrapping critical gauge
  theories: decoupling operators of conformal field theories in $d>2$
  dimensions},'' \href{http://arxiv.org/abs/2101.07262}{{\ttfamily
  arXiv:2101.07262 [hep-th]}}.

\bibitem{Manenti:2021elk}
A.~Manenti and A.~Vichi, ``{Exploring $SU(N)$ adjoint correlators in $3d$},''
  \href{http://arxiv.org/abs/2101.07318}{{\ttfamily arXiv:2101.07318
  [hep-th]}}.

\bibitem{He:2021sto}
Y.-C. He, J.~Rong, and N.~Su, ``{Conformal bootstrap bounds for the $U(1)$
  Dirac spin liquid and $N=7$ Stiefel liquid},''
  \href{http://arxiv.org/abs/2107.14637}{{\ttfamily arXiv:2107.14637
  [cond-mat.str-el]}}.

\bibitem{Xu:2018wyg}
X.~Y. Xu, Y.~Qi, L.~Zhang, F.~F. Assaad, C.~Xu, and Z.~Y. Meng, ``{Monte Carlo
  Study of Lattice Compact Quantum Electrodynamics with Fermionic Matter: The
  Parent State of Quantum Phases},''
  \href{http://dx.doi.org/10.1103/PhysRevX.9.021022}{{\em Phys. Rev. X}
  {\bfseries 9} no.~2, (2019) 021022},
  \href{http://arxiv.org/abs/1807.07574}{{\ttfamily arXiv:1807.07574
  [cond-mat.str-el]}}.

\bibitem{Gracey:1993iu}
J.~A. Gracey, ``{Computation of critical exponent eta at O(1/N(f)**2) in
  quantum electrodynamics in arbitrary dimensions},''
  \href{http://dx.doi.org/10.1016/0550-3213(94)90257-7}{{\em Nucl. Phys. B}
  {\bfseries 414} (1994) 614--648},
  \href{http://arxiv.org/abs/hep-th/9312055}{{\ttfamily arXiv:hep-th/9312055}}.

\bibitem{Gracey:1993sn}
J.~A. Gracey, ``{Electron mass anomalous dimension at O(1/(Nf(2)) in quantum
  electrodynamics},''
  \href{http://dx.doi.org/10.1016/0370-2693(93)91017-H}{{\em Phys. Lett. B}
  {\bfseries 317} (1993) 415--420},
  \href{http://arxiv.org/abs/hep-th/9309092}{{\ttfamily arXiv:hep-th/9309092}}.

\bibitem{Rantner_2002}
W.~Rantner and X.-G. Wen, ``Spin correlations in the algebraic spin liquid:
  Implications for high-tcsuperconductors,''
  \href{http://dx.doi.org/10.1103/physrevb.66.144501}{{\em Physical Review B}
  {\bfseries 66} no.~14, (Oct, 2002) }.
  \url{http://dx.doi.org/10.1103/PhysRevB.66.144501}.

\bibitem{Xu_2008}
C.~Xu, ``Renormalization group studies on four-fermion interaction
  instabilities on algebraic spin liquids,''
  \href{http://dx.doi.org/10.1103/physrevb.78.054432}{{\em Physical Review B}
  {\bfseries 78} no.~5, (Aug, 2008) }.
  \url{http://dx.doi.org/10.1103/PhysRevB.78.054432}.

\bibitem{Kaul2008}
R.~K. Kaul and S.~Sachdev, ``Quantum criticality of u(1) gauge theories with
  fermionic and bosonic matter in two spatial dimensions,''
  \href{http://dx.doi.org/10.1103/PhysRevB.77.155105}{{\em Phys. Rev. B}
  {\bfseries 77} (Apr, 2008) 155105}.
  \url{https://link.aps.org/doi/10.1103/PhysRevB.77.155105}.

\bibitem{Chester:2016ref}
S.~M. Chester and S.~S. Pufu, ``{Anomalous dimensions of scalar operators in
  QED$_{3}$},'' \href{http://dx.doi.org/10.1007/JHEP08(2016)069}{{\em JHEP}
  {\bfseries 08} (2016) 069}, \href{http://arxiv.org/abs/1603.05582}{{\ttfamily
  arXiv:1603.05582 [hep-th]}}.

\bibitem{Huh:2013vga}
Y.~Huh, P.~Strack, and S.~Sachdev, ``{Conserved current correlators of
  conformal field theories in 2+1 dimensions},''
  \href{http://dx.doi.org/10.1103/PhysRevB.88.155109}{{\em Phys. Rev. B}
  {\bfseries 88} (2013) 155109},
  \href{http://arxiv.org/abs/1307.6863}{{\ttfamily arXiv:1307.6863
  [cond-mat.str-el]}}. [Erratum: Phys.Rev.B 90, 199902 (2014)].

\bibitem{Huh:2014eea}
Y.~Huh and P.~Strack, ``{Stress tensor and current correlators of interacting
  conformal field theories in 2+1 dimensions: Fermionic Dirac matter coupled to
  U(1) gauge field},'' \href{http://dx.doi.org/10.1007/JHEP01(2015)147}{{\em
  JHEP} {\bfseries 01} (2015) 147},
  \href{http://arxiv.org/abs/1410.1902}{{\ttfamily arXiv:1410.1902
  [cond-mat.str-el]}}. [Erratum: JHEP 03, 054 (2016)].

\bibitem{Borokhov:2002ib}
V.~Borokhov, A.~Kapustin, and X.-k. Wu, ``{Topological disorder operators in
  three-dimensional conformal field theory},''
  \href{http://dx.doi.org/10.1088/1126-6708/2002/11/049}{{\em JHEP} {\bfseries
  11} (2002) 049}, \href{http://arxiv.org/abs/hep-th/0206054}{{\ttfamily
  arXiv:hep-th/0206054}}.

\bibitem{Pufu:2013vpa}
S.~S. Pufu, ``{Anomalous dimensions of monopole operators in three-dimensional
  quantum electrodynamics},''
  \href{http://dx.doi.org/10.1103/PhysRevD.89.065016}{{\em Phys. Rev. D}
  {\bfseries 89} no.~6, (2014) 065016},
  \href{http://arxiv.org/abs/1303.6125}{{\ttfamily arXiv:1303.6125 [hep-th]}}.

\bibitem{Dyer:2013fja}
E.~Dyer, M.~Mezei, and S.~S. Pufu, ``{Monopole Taxonomy in Three-Dimensional
  Conformal Field Theories},'' \href{http://arxiv.org/abs/1309.1160}{{\ttfamily
  arXiv:1309.1160 [hep-th]}}.

\bibitem{Dupuis:2021flq}
E.~Dupuis, R.~Boyack, and W.~Witczak-Krempa, ``{Anomalous dimensions of
  monopole operators at the transitions between Dirac and topological spin
  liquids},'' \href{http://arxiv.org/abs/2108.05922}{{\ttfamily
  arXiv:2108.05922 [cond-mat.str-el]}}.

\bibitem{Berkooz:2014yda}
M.~Berkooz, R.~Yacoby, and A.~Zait, ``{Bounds on $\mathcal{N} = 1$
  superconformal theories with global symmetries},''
  \href{http://dx.doi.org/10.1007/JHEP01(2015)132}{{\em JHEP} {\bfseries 08}
  (2014) 008}, \href{http://arxiv.org/abs/1402.6068}{{\ttfamily arXiv:1402.6068
  [hep-th]}}. [Erratum: JHEP 01, 132 (2015)].

\bibitem{Nakayama:2017vdd}
Y.~Nakayama, ``{Bootstrap experiments on higher dimensional CFTs},''
  \href{http://dx.doi.org/10.1142/S0217751X18500367}{{\em Int. J. Mod. Phys. A}
  {\bfseries 33} no.~07, (2018) 1850036},
  \href{http://arxiv.org/abs/1705.02744}{{\ttfamily arXiv:1705.02744
  [hep-th]}}.

\bibitem{Iha:2016ppj}
H.~Iha, H.~Makino, and H.~Suzuki, ``{Upper bound on the mass anomalous
  dimension in many-flavor gauge theories: a conformal bootstrap approach},''
  \href{http://dx.doi.org/10.1093/ptep/ptw046}{{\em PTEP} {\bfseries 2016}
  no.~5, (2016) 053B03}, \href{http://arxiv.org/abs/1603.01995}{{\ttfamily
  arXiv:1603.01995 [hep-th]}}.

\bibitem{Poland:2011ey}
D.~Poland, D.~Simmons-Duffin, and A.~Vichi, ``{Carving Out the Space of 4D
  CFTs},'' \href{http://dx.doi.org/10.1007/JHEP05(2012)110}{{\em JHEP}
  {\bfseries 05} (2012) 110}, \href{http://arxiv.org/abs/1109.5176}{{\ttfamily
  arXiv:1109.5176 [hep-th]}}.

\bibitem{Dymarsky:2017xzb}
A.~Dymarsky, J.~Penedones, E.~Trevisani, and A.~Vichi, ``{Charting the space of
  3D CFTs with a continuous global symmetry},''
  \href{http://dx.doi.org/10.1007/JHEP05(2019)098}{{\em JHEP} {\bfseries 05}
  (2019) 098}, \href{http://arxiv.org/abs/1705.04278}{{\ttfamily
  arXiv:1705.04278 [hep-th]}}.

\bibitem{Dymarsky:2017yzx}
A.~Dymarsky, F.~Kos, P.~Kravchuk, D.~Poland, and D.~Simmons-Duffin, ``{The 3d
  Stress-Tensor Bootstrap},''
  \href{http://dx.doi.org/10.1007/JHEP02(2018)164}{{\em JHEP} {\bfseries 02}
  (2018) 164}, \href{http://arxiv.org/abs/1708.05718}{{\ttfamily
  arXiv:1708.05718 [hep-th]}}.

\bibitem{Reehorst:2019pzi}
M.~Reehorst, E.~Trevisani, and A.~Vichi, ``{Mixed Scalar-Current bootstrap in
  three dimensions},'' \href{http://dx.doi.org/10.1007/JHEP12(2020)156}{{\em
  JHEP} {\bfseries 12} (2020) 156},
  \href{http://arxiv.org/abs/1911.05747}{{\ttfamily arXiv:1911.05747
  [hep-th]}}.

\bibitem{Rattazzi:2010yc}
R.~Rattazzi, S.~Rychkov, and A.~Vichi, ``{Bounds in 4D Conformal Field Theories
  with Global Symmetry},''
  \href{http://dx.doi.org/10.1088/1751-8113/44/3/035402}{{\em J. Phys. A}
  {\bfseries 44} (2011) 035402},
  \href{http://arxiv.org/abs/1009.5985}{{\ttfamily arXiv:1009.5985 [hep-th]}}.

\bibitem{El-Showk:2012cjh}
S.~El-Showk, M.~F. Paulos, D.~Poland, S.~Rychkov, D.~Simmons-Duffin, and
  A.~Vichi, ``{Solving the 3D Ising Model with the Conformal Bootstrap},''
  \href{http://dx.doi.org/10.1103/PhysRevD.86.025022}{{\em Phys. Rev. D}
  {\bfseries 86} (2012) 025022},
  \href{http://arxiv.org/abs/1203.6064}{{\ttfamily arXiv:1203.6064 [hep-th]}}.

\bibitem{Agmon:2019imm}
N.~B. Agmon, S.~M. Chester, and S.~S. Pufu, ``{The M-theory Archipelago},''
  \href{http://dx.doi.org/10.1007/JHEP02(2020)010}{{\em JHEP} {\bfseries 02}
  (2020) 010}, \href{http://arxiv.org/abs/1907.13222}{{\ttfamily
  arXiv:1907.13222 [hep-th]}}.

\bibitem{Chester:2021aun}
S.~M. Chester, R.~Dempsey, and S.~S. Pufu, ``{Bootstrapping $\mathcal{N}=4$
  super-Yang-Mills on the conformal manifold},''
  \href{http://arxiv.org/abs/2111.07989}{{\ttfamily arXiv:2111.07989
  [hep-th]}}.

\bibitem{Iliesiu:2015qra}
L.~Iliesiu, F.~Kos, D.~Poland, S.~S. Pufu, D.~Simmons-Duffin, and R.~Yacoby,
  ``{Bootstrapping 3D Fermions},''
  \href{http://dx.doi.org/10.1007/JHEP03(2016)120}{{\em JHEP} {\bfseries 03}
  (2016) 120},
\href{http://arxiv.org/abs/1508.00012}{{\ttfamily arXiv:1508.00012 [hep-th]}}.

\bibitem{Iliesiu:2017nrv}
L.~Iliesiu, F.~Kos, D.~Poland, S.~S. Pufu, and D.~Simmons-Duffin,
  ``{Bootstrapping 3D Fermions with Global Symmetries},''
  \href{http://dx.doi.org/10.1007/JHEP01(2018)036}{{\em JHEP} {\bfseries 01}
  (2018) 036}, \href{http://arxiv.org/abs/1705.03484}{{\ttfamily
  arXiv:1705.03484 [hep-th]}}.

\bibitem{Albayrak:2019gnz}
S.~Albayrak, D.~Meltzer, and D.~Poland, ``{More Analytic Bootstrap:
  Nonperturbative Effects and Fermions},''
  \href{http://dx.doi.org/10.1007/JHEP08(2019)040}{{\em JHEP} {\bfseries 08}
  (2019) 040}, \href{http://arxiv.org/abs/1904.00032}{{\ttfamily
  arXiv:1904.00032 [hep-th]}}.

\bibitem{Albayrak:2020rxh}
S.~Albayrak, D.~Meltzer, and D.~Poland, ``{The Inversion Formula and 6j Symbol
  for 3d Fermions},'' \href{http://dx.doi.org/10.1007/JHEP09(2020)148}{{\em
  JHEP} {\bfseries 09} (2020) 148},
  \href{http://arxiv.org/abs/2006.07374}{{\ttfamily arXiv:2006.07374
  [hep-th]}}.

\bibitem{WU1976365}
T.~T. Wu and C.~N. Yang, ``Dirac monopole without strings: Monopole
  harmonics,''
  \href{http://dx.doi.org/https://doi.org/10.1016/0550-3213(76)90143-7}{{\em
  Nuclear Physics B} {\bfseries 107} no.~3, (1976) 365--380}.
  \url{https://www.sciencedirect.com/science/article/pii/0550321376901437}.

\bibitem{PhysRevD.16.1018}
T.~T. Wu and C.~N. Yang, ``Some properties of monopole harmonics,''
  \href{http://dx.doi.org/10.1103/PhysRevD.16.1018}{{\em Phys. Rev. D}
  {\bfseries 16} (Aug, 1977) 1018--1021}.
  \url{https://link.aps.org/doi/10.1103/PhysRevD.16.1018}.

\bibitem{onishchik2012lie}
A.~Onishchik, D.~Leites, and E.~Vinberg, {\em Lie Groups and Algebraic Groups}.
\newblock Springer Series in Soviet Mathematics. Springer Berlin Heidelberg,
  2012.
\newblock \url{https://books.google.com/books?id=TV7sCAAAQBAJ}.

\bibitem{Costa:2014rya}
M.~S. Costa and T.~Hansen, ``{Conformal correlators of mixed-symmetry
  tensors},'' \href{http://dx.doi.org/10.1007/JHEP02(2015)151}{{\em JHEP}
  {\bfseries 02} (2015) 151}, \href{http://arxiv.org/abs/1411.7351}{{\ttfamily
  arXiv:1411.7351 [hep-th]}}.

\bibitem{Costa:2016hju}
M.~S. Costa, T.~Hansen, J.~a. Penedones, and E.~Trevisani, ``{Projectors and
  seed conformal blocks for traceless mixed-symmetry tensors},''
  \href{http://dx.doi.org/10.1007/JHEP07(2016)018}{{\em JHEP} {\bfseries 07}
  (2016) 018}, \href{http://arxiv.org/abs/1603.05551}{{\ttfamily
  arXiv:1603.05551 [hep-th]}}.

\bibitem{Kos:2014bka}
F.~Kos, D.~Poland, and D.~Simmons-Duffin, ``{Bootstrapping Mixed Correlators in
  the 3D Ising Model},'' \href{http://dx.doi.org/10.1007/JHEP11(2014)109}{{\em
  JHEP} {\bfseries 11} (2014) 109},
  \href{http://arxiv.org/abs/1406.4858}{{\ttfamily arXiv:1406.4858 [hep-th]}}.

\bibitem{Giombi:2011rz}
S.~Giombi, S.~Prakash, and X.~Yin, ``{A Note on CFT Correlators in Three
  Dimensions},'' \href{http://dx.doi.org/10.1007/JHEP07(2013)105}{{\em JHEP}
  {\bfseries 07} (2013) 105}, \href{http://arxiv.org/abs/1104.4317}{{\ttfamily
  arXiv:1104.4317 [hep-th]}}.

\bibitem{Costa:2011mg}
M.~S. Costa, J.~Penedones, D.~Poland, and S.~Rychkov, ``{Spinning Conformal
  Correlators},'' \href{http://dx.doi.org/10.1007/JHEP11(2011)071}{{\em JHEP}
  {\bfseries 11} (2011) 071}, \href{http://arxiv.org/abs/1107.3554}{{\ttfamily
  arXiv:1107.3554 [hep-th]}}.

\bibitem{SimmonsDuffin:2012uy}
D.~Simmons-Duffin, ``{Projectors, Shadows, and Conformal Blocks},''
  \href{http://dx.doi.org/10.1007/JHEP04(2014)146}{{\em JHEP} {\bfseries 04}
  (2014) 146}, \href{http://arxiv.org/abs/1204.3894}{{\ttfamily arXiv:1204.3894
  [hep-th]}}.

\bibitem{Karateev:2018oml}
D.~Karateev, P.~Kravchuk, and D.~Simmons-Duffin, ``{Harmonic Analysis and Mean
  Field Theory},'' \href{http://dx.doi.org/10.1007/JHEP10(2019)217}{{\em JHEP}
  {\bfseries 10} (2019) 217}, \href{http://arxiv.org/abs/1809.05111}{{\ttfamily
  arXiv:1809.05111 [hep-th]}}.

\bibitem{lanczos1950iteration}
C.~Lanczos, ``An iteration method for the solution of the eigenvalue problem of
  linear differential and integral operators,''.

\bibitem{Simmons-Duffin:2015qma}
D.~Simmons-Duffin, ``{A Semidefinite Program Solver for the Conformal
  Bootstrap},'' \href{http://dx.doi.org/10.1007/JHEP06(2015)174}{{\em JHEP}
  {\bfseries 06} (2015) 174}, \href{http://arxiv.org/abs/1502.02033}{{\ttfamily
  arXiv:1502.02033 [hep-th]}}.

\bibitem{Landry:2019qug}
W.~Landry and D.~Simmons-Duffin, ``{Scaling the semidefinite program solver
  SDPB},'' \href{http://arxiv.org/abs/1909.09745}{{\ttfamily arXiv:1909.09745
  [hep-th]}}.

\bibitem{hyperion-projects}
D.~Simmons-Duffin, P.~Kravchuk, A.~Liu, and R.~S. Erramilli,
  ``hyperion-projects.''
\newblock \url{https://gitlab.com/pkravchuk/hyperion-projects}.

\bibitem{scalar_blocks}
\url{https://gitlab.com/bootstrapcollaboration/scalar_blocks}.

\bibitem{simpleboot}
N.~Su, ``simpleboot.''
\newblock \url{https://gitlab.com/bootstrapcollaboration/simpleboot}.

\bibitem{Go:2019lke}
M.~Go and Y.~Tachikawa, ``{autoboot: A generator of bootstrap equations with
  global symmetry},'' \href{http://dx.doi.org/10.1007/JHEP06(2019)084}{{\em
  JHEP} {\bfseries 06} (2019) 084},
  \href{http://arxiv.org/abs/1903.10522}{{\ttfamily arXiv:1903.10522
  [hep-th]}}.

\end{thebibliography}\endgroup


\providecommand{\href}[2]{#2}\begingroup\raggedright\endgroup
\bibliographystyle{utphys}
\end{document}